\documentclass[longauth]{aa}  

\usepackage{graphicx}
\usepackage[dvipsnames]{xcolor}
\usepackage{pdflscape}
\usepackage{longtable}     
\usepackage{caption}
\usepackage{placeins}
\usepackage{txfonts}
\usepackage[breaklinks,colorlinks,citecolor=blue,linkcolor=blue,urlcolor=blue]{hyperref}
\begin{document}

   \title{JADES NIRSpec initial data release for the {\em Hubble} Ultra Deep Field}
   
   \subtitle{Redshifts and line fluxes of distant galaxies from the deepest  JWST Cycle 1 NIRSpec multi-object spectroscopy}

\author{Andrew J. Bunker\thanks{These authors contributed equally to this work}
        \inst{1}
        \and
        Alex J. Cameron$^{\star}$
        \inst{1}
        \and
        Emma Curtis-Lake$^{\star}$
        \inst{2}
        \and
        Peter Jakobsen
        \inst{3}\,\inst{4}
        \and
        Stefano Carniani
        \inst{5}
        \and
        Mirko Curti
        \inst{6}\,\inst{7}\,\inst{8}
        \and
        Joris Witstok
        \inst{7}\,\inst{8}
        \and
        Roberto Maiolino
        \inst{7}\,\inst{8}\,\inst{9}
        \and
        Francesco D'Eugenio
        \inst{7}\,\inst{8}
        \and
        Tobias J. Looser
        \inst{7}\,\inst{8}
        \and
        Chris Willott
        \inst{10}
        \and
        Nina Bonaventura
        \inst{3}\,\inst{4}\,\inst{11}
        \and
        Kevin Hainline
        \inst{11}
        \and
        Hannah \"Ubler
        \inst{7}\,\inst{8}
        \and
        Christopher N. A. Willmer
        \inst{11}
        \and
        Aayush Saxena
        \inst{1}\,\inst{9}
        \and
        Renske Smit
        \inst{12}
        \and
        Stacey Alberts
        \inst{11}
        \and
        Santiago Arribas
        \inst{13}
        \and
        William M. Baker
        \inst{7}\,\inst{8}
        \and
        Stefi Baum
        \inst{14}
        \and
        Rachana Bhatawdekar
        \inst{15}
        \and
        Rebecca A. A. Bowler
        \inst{16}
        \and
        Kristan Boyett
        \inst{17}\,\inst{18}
        \and
        Stephane Charlot
        \inst{19}
        \and
        Zuyi Chen
        \inst{11}
        \and
        Jacopo Chevallard
        \inst{1}
        \and
        Chiara Circosta
        \inst{15}
        \and
        Christa DeCoursey
        \inst{11}
        \and
        Anna de Graaff
        \inst{20}
        \and
        Eiichi Egami
        \inst{11}
        \and
        Daniel J. Eisenstein
        \inst{21}
        \and
        Ryan Endsley
        \inst{22}
        \and
        Pierre Ferruit
        \inst{15}
        \and
        Giovanna Giardino
        \inst{23}
     \and
        Ryan Hausen
        \inst{24}
        \and
        Jakob M. Helton
        \inst{11}
        \and
        Raphael E. Hviding
        \inst{11}
        \and
        Zhiyuan Ji
        \inst{11}
        \and
        Benjamin D. Johnson
        \inst{21}
        \and
        Gareth C. Jones
        \inst{1}
        \and
        Nimisha Kumari
        \inst{25}
        \and
        Isaac Laseter
        \inst{26}
        \and
        Nora L\"utzgendorf
        \inst{25}
        \and
        Michael V. Maseda
        \inst{26}
        \and
        Erica Nelson
        \inst{27}
        \and
        Eleonora Parlanti
        \inst{5}
        \and
        Michele Perna
        \inst{13}
        \and
	Bernard J. Rauscher
	\inst{28}
	\and
        Tim Rawle
        \inst{15}
        \and
        Hans-Walter Rix
        \inst{29}
        \and
        Marcia Rieke
        \inst{11}
        \and
        Brant Robertson
        \inst{30}
        \and
        Bruno Rodr\'iguez Del Pino
        \inst{13}
        \and
        Lester Sandles
        \inst{7}\,\inst{8}
        \and
        Jan Scholtz
        \inst{7}\,\inst{8}
        \and
        Katherine Sharpe
        \inst{21}
        \and 
        Maya Skarbinski
        \inst{21}
        \and
        Daniel P. Stark
        \inst{11}
        \and
        Fengwu Sun
        \inst{11}
        \and
        Sandro Tacchella
        \inst{7}\,\inst{8}
        \and
        Michael W. Topping
        \inst{11}
        \and
        Natalia C. Villanueva
        \inst{21}
        \and
        Imaan E. B. Wallace
        \inst{1}
        \and
        Christina C. Williams
        \inst{31}
        \and
        Charity Woodrum
        \inst{11}
}

\institute{
   Department of Physics, University of Oxford, Denys Wilkinson Building, Keble Road, Oxford OX1 3RH, UK
   \and
   Centre for Astrophysics Research, Department of Physics, Astronomy and Mathematics, University of Hertfordshire, Hatfield AL10 9AB, UK
   \and
   Cosmic Dawn Center (DAWN), Copenhagen, Denmark 
   \and
   Niels Bohr Institute, University of Copenhagen, Jagtvej 128, DK-2200, Copenhagen, Denmark
   \and
   Scuola Normale Superiore, Piazza dei Cavalieri 7, I-56126 Pisa, Italy
   \and
   European Southern Observatory, Karl-Schwarzschild-Strasse 2, 85748 Garching, Germany
   \and
   Kavli Institute for Cosmology, University of Cambridge, Madingley Road, Cambridge, CB3 0HA, UK
   \and
   Cavendish Laboratory - Astrophysics Group, University of Cambridge, 19 JJ Thomson Avenue, Cambridge, CB3 0HE, UK. 
   \and
   Department of Physics and Astronomy, University College London, Gower Street, London WC1E 6BT, UK
   \and
   NRC Herzberg, 5071 West Saanich Rd, Victoria, BC V9E 2E7, Canada
   \and
   Steward Observatory University of Arizona 933 N. Cherry Avenue Tucson AZ 85721, USA
   \and
   Astrophysics Research Institute, Liverpool John Moores University, 146 Brownlow Hill, Liverpool L3 5RF, UK
   \and
   Centro de Astrobiolog\'ia (CAB), CSIC–INTA, Cra. de Ajalvir Km.~4, 28850- Torrej\'on de Ardoz, Madrid, Spain
   \and
   Department of Physics and Astronomy, University of Manitoba, Winnipeg, MB R3T 2N2, Canada
   \and
   European Space Agency (ESA), European Space Astronomy Centre (ESAC), Camino Bajo del Castillo s/n, 28692 Villanueva de la Cañada, Madrid, Spain; European Space Agency, ESA/ESTEC, Keplerlaan 1, 2201 AZ Noordwijk, NL
   \and
   Jodrell Bank Centre for Astrophysics, Department of Physics and Astronomy, School of Natural Sciences, The University of Manchester, Manchester, M13 9PL, UK
   \and
   School of Physics, University of Melbourne, Parkville 3010, VIC, Australia
   \and 
   ARC Centre of Excellence for All Sky Astrophysics in 3 Dimensions (ASTRO 3D), Australia
   \and
   Sorbonne Universit\'e, CNRS, UMR 7095, Institut d'Astrophysique de Paris, 98 bis bd Arago, 75014 Paris, France
   \and
   Max-Planck-Institut f\"ur Astronomie, K\"onigstuhl 17, D-69117, Heidelberg, Germany
   \and
   Center for Astrophysics $|$ Harvard \& Smithsonian, 60 Garden St., Cambridge MA 02138 USA
   \and
   Department of Astronomy, University of Texas, Austin, TX 78712, USA
   \and
   ATG Europe for the European Space Agency, ESTEC, Noordwijk,
The Netherlands
    \and
   Department of Physics and Astronomy, The Johns Hopkins University, 3400 N. Charles St., Baltimore, MD 21218
   \and
   European Space Agency, Space Telescope Science Institute, Baltimore, Maryland, USA
   \and
   Department of Astronomy, University of Wisconsin-Madison, 475 N. Charter St., Madison, WI 53706 USA
   \and
   Department for Astrophysical and Planetary Science, University of Colorado, Boulder, CO 80309, USA
   \and
   Observational Cosmology Laboratory, NASA Goddard Space Flight Center, 8800 Greenbelt Rd., Greenbelt, MD USA
   \and
   Max-Planck-Institut f\"ur Astronomie, 
    K\"onigstuhl 17, D-69117, 
    Heidelberg, Germany
\and
   Department of Astronomy and Astrophysics, University of California, Santa Cruz, 1156 High Street, Santa Cruz, CA 95064, USA
   \and
   NSF’s National Optical-Infrared Astronomy Research Laboratory, 950 North Cherry Avenue, Tucson, AZ 85719, USA
             }

   \date{Received June 4, 2023; accepted 21 May 2024}

  \abstract{We describe the NIRSpec component of the JWST Deep Extragalactic Survey (JADES), and provide deep spectroscopy of 253 sources targeted with the NIRSpec micro-shutter assembly in the {\em Hubble} Ultra Deep Field and surrounding GOODS-South. 
  The multi-object spectra presented here are the deepest so far obtained with JWST, amounting to up to 28 hours in the low-dispersion ($R\sim 30-300$) prism, and up to 7 hours in each of the three medium-resolution $R\approx 1000$ gratings and one high-dispersion grating, G395H ($R\approx2700$). Our low-dispersion and medium-dispersion spectra cover the wavelength range $0.6-5.3\,\mu$m. We describe the selection of the spectroscopic targets, the strategy for the allocation of targets to micro-shutters, and the design of the observations. We present the public release of the reduced 2D and 1D spectra, and a description of the reduction and calibration process. We measure spectroscopic redshifts for 178 of the objects targeted extending up to $z=13.2$. We present a catalogue of all emission lines detected at $S/N>5$, and our redshift determinations for the targets. Combined with the first JADES NIRCam data release, these public JADES spectroscopic and imaging
datasets provide a new foundation for discoveries of the infrared universe by the worldwide
scientific community.
}


   \maketitle
%

\section{Introduction}
\label{sec:intro}

JWST \citep{Gardner2023} is the largest programme in astrophysics to date, and is far more than simply the successor to the {\em Hubble} Space Telescope (HST). As well as having seven times the collecting area of HST, JWST operates over a wider range of wavelengths ($0.6-25\,\mu$m) in a lower-background environment (at L2), making it orders of magnitude more sensitive than previous observatories.
One of the major goals of the JWST mission is to study the formation and evolution of galaxies, in particular in the early universe through observations of high redshift galaxies. 

The  JWST Advanced Deep Extragalactic Survey (JADES, 
\citealt{Bunker2020}; \citealt{Rieke2020}; \citealt{eisenstein2023}) is 
the largest Cycle 1 programme aiming to study galaxy evolution out to the highest redshifts. 
JADES is a coordinated survey designed and executed by the NIRSpec and NIRCam Guaranteed Time Observation (GTO) teams. 
It provides NIRCam and MIRI imaging as well as NIRSpec spectroscopy over two fields.
An important aspect of JADES is the assembly of a large data set of spectroscopic observations spanning from cosmic noon to within the epoch of reionization, enabling confirmation of high-redshift candidates, accurate redshift measurements, and unprecedented constraints on the physical conditions in distant galaxies. With such spectroscopy, we can explore the mass-metallicity relation, dust attenuation, star formation rates and star formation histories in galaxies, as well as ionization parameters, ionizing photon escape fraction, and the presence of any active galactic nuclei. Spectroscopy is also key to understanding the physical states of the interstellar, circumgalactic and intergalactic media, and their evolution with cosmic time.
Crucially, assembling this data set is enabled by the new multi-object spectroscopy (MOS) capabilities of JWST with the near-infrared spectrograph (NIRSpec; \citealt{Jakobsen2022}).

NIRSpec operates in the range $0.6-5.3\,\mu$m, and has three spectral resolutions: a low-dispersion prism ($R\approx30-300$) which captures all the wavelength range with a single exposure, and medium- and high-resolution gratings ($R\approx1000$ and $R\approx2700$) which use three bands to cover the wavelength range. One unique feature of this spectrograph is its use of a micro-shutter assembly (MSA), developed specifically for NIRSpec to enable multi-object spectroscopy of hundreds of objects at once over a 3\farcm6 $\times$ 3\farcm4 field of view \citep{Ferruit2022}.

As described in \cite{eisenstein2023}, JADES has Deep and Medium tiers, where the Medium tier adds area to capture rarer objects, while the Deep tier allows for searches of the faintest, and most distant galaxies.  The survey covers two fields with huge legacy data sets thanks to the Great Observatories Origins Deep Survey \citep[GOODS][]{Dickinson2003, Giavalisco2004}, GOODS-South and GOODS-North.  As well as the multi-wavelength efforts of the original GOODS survey, there have been extensive observing efforts in the same area, including, but not limited to, the CANDELS survey with \textit{Hubble} \citep[Cosmic Assebly Near-Infrared Deep Extragalactic Legacy Survey, ][]{Grogin2011, Koekemoer2011} and the GREATS survey with \textit{Spitzer} \citep[GOODS Re-ionization Era wide-Area Treasury from \textit{Spitzer}, ][]{Stefanon2021}. In particular the GOODS-South field includes the region of the sky with the deepest \textit{Hubble} images ever taken, the {\em Hubble} Ultra Deep Field \citep[HUDF, ][]{Beckwith2006, Bouwens2010, Ellis2013}, where we focus the Deep portion of our survey. The Medium tier adds area with observations in GOODS-North, and the extended GOODS-South field, predominantly within the footprint of the CANDELS data.

In this paper, we present our deep spectroscopy of targets in the 
HUDF and surrounding GOODS-South field and outline our target selection strategy. We release the raw data and make our reduced data products available to the community\footnote{\url{https://archive.stsci.edu/hlsp/jades}}, and in a companion paper \citep{rieke2023} we present the complementary JADES NIRCam imaging of the HUDF. From the prism and medium-dispersion $R\approx1000$ spectra we derive redshifts and fluxes of prominent emission lines. The data from the single high-dispersion grating used (G395H, $R\approx2700$) also forms part of this data release, but we do not perform detailed on this analysis in this paper.

The structure of this paper is as follows. Section~\ref{sec:targets} describes how potential spectroscopic targets were selected from imaging data (primarily a combination of JADES NIRCam  and  HST), and how these were allocated to different priority classes so that the NIRSpec MSA configuration could be optimised for our science goals. The NIRSpec observations are described in Section~\ref{sec:obs} and the data processing is outlined in Section~\ref{sec:data_proc}. In Section~\ref{sec:emission_lines} we present our redshift measurements, and detected emission line fluxes of individual galaxies. Our conclusions are in Section~\ref{sec:conclusions}.
Throughout this work, we assume the Planck 2018 cosmology \citep{Planck18} and the AB magnitude system \citep{Oke1983}.

\section{Targets}
\label{sec:targets}

JADES observations take NIRCam imaging and NIRSpec spectroscopy in parallel. As the survey progresses, JADES aims to leverage NIRCam photometry to select targets for later NIRSpec observations where possible, as this will enable the identification of the highest-redshift objects and facilitates near mass-limited samples at lower redshifts.
However, for many of our early observations, we take spectroscopy in regions which have not yet been imaged by NIRCam.

\begin{figure*}
    \centering
    \includegraphics*[width=0.9\textwidth]{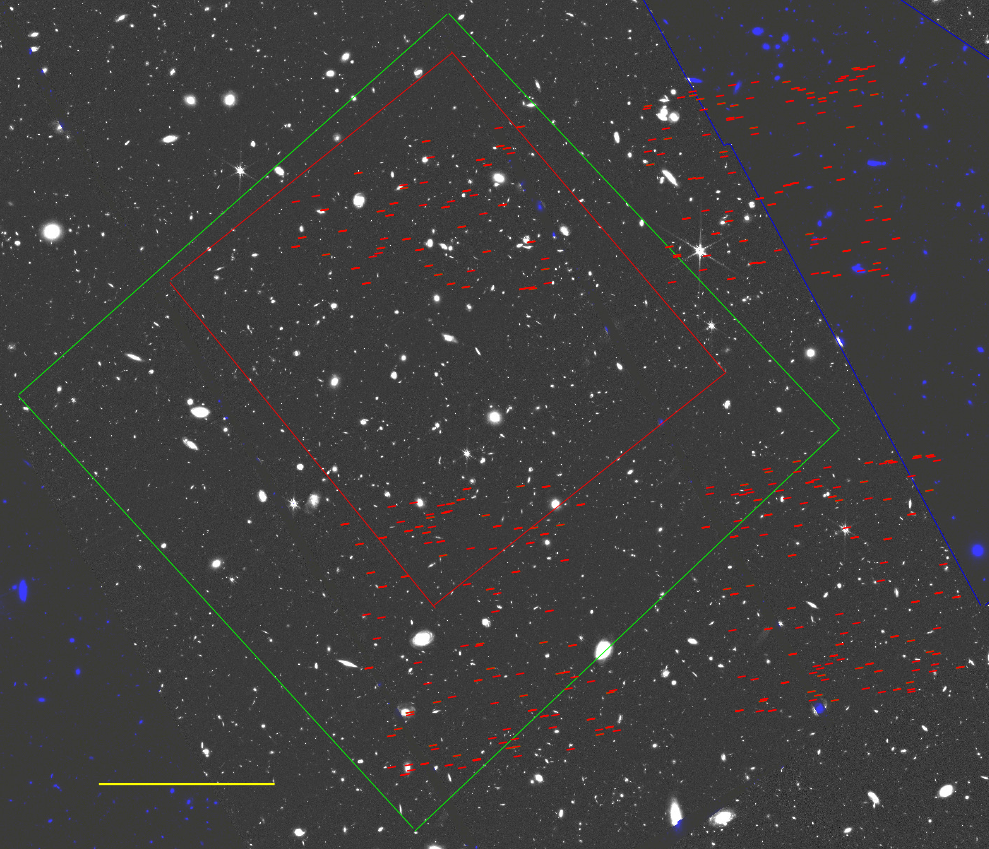}
    \caption{Field layout of the NIRSpec Deep-HST observations presented in this paper. The green rectange is the region covered by the original HST/ACS {\em Hubble} Ultra Deep Field. The red rectangle is the smaller area covered in the Ultra Deep HST/WFC\,3 imaging. The background image is the NIRCam F200W from JADES, except for the region to the right of the blue line which has not yet been observed by NIRCam; we show in this region the HST/WFC\,3 from GOODS-South/CANDELS in blue. The short red lines denote the five-shutter ($\approx 2\farcs 6$) extent observed (three open shutters each target per observation, nodded by $\pm 1$ shutter for background subtraction).  The the four quadrants of the NIRSpec MSA are clearly visible. More detailed views of each quadrant with the target ID numbers marked are shown in Figures~\ref{fig:open_shutters} and Figures~\ref{fig:quad1}, \ref{fig:quad2} \& \ref{fig:quad4}, which also show the sub-set of targets with grating spectra. The yellow scale bar at the bottom left is 1\,arcmin in length.}
    \label{fig:full_field}
\end{figure*}

In the initial planning phase, the Deep tier presented here was to be observed prior to NIRCam imaging and hence would comprise only targets previously identified (mostly from HST imaging).
However, scheduling changes meant that we ended up having NIRCam data available shortly before our final MSA configuration needed to be set\footnote{After the original APT file for PID 1210 had been submitted and scheduled by STScI, an electrical short appearing in a column of the MSA \citep{rawle22} necessitated that the three MSA configurations be modified and substituted.}. This unforeseen opportunity was exploited scientifically to refine the target selection by making use of the additional JADES photometry from NIRCam images, with observations completed 16 days before the NIRSpec observations. 
Thus, our selected targets represent a NIRCam-based selection, supplemented with some HST-based targets compiled from the literature. We note that the NIRCam images available when drawing up our target list did not cover the full region of the NIRSpec MSA (see Figure~\ref{fig:full_field}).

We used NIRCam data taken between 29th September and the 5th October 2022 described in \cite{rieke2023}, which added nine photometric bands, potentially improving photometric redshifts over previous HST-based studies as well as identifying HST-dark sources. We used a very early reduction of the data and describe the limitations of this in Appendix~\ref{section:caveats}. 
We measured the HST and NIRCam photometry using $0\farcs 3$-arcsec diameter apertures and applying aperture corrections for each filter appropriate for compact sources. We estimated photometric redshifts from two different SED-fitting codes with very different template sets and underlying assumptions, \textsc{eazy} \citep{Brammer2008} and \textsc{beagle} \citep{Chevallard2016}, and these were used in our target selection.

In this Section, we first describe our over-arching prioritisation system for allocating targets for spectroscopy. We then describe the assembly of these NIRCam- and HST-based catalogues, which formed the source material for our target allocation.

\subsection{Priority class system}
\label{sec:priority_class_system}

The target selection for the NIRSpec MOS observations was designed to prioritise rare targets, either at high redshift, or with low number density, while building up a statistical sample spanning from cosmic noon to within the epoch of reionization.  This was achieved by sorting the potential targets 
into a limited number of priority classes and employing the NIRSpec team's eMPT software suite \citep{Bonaventura2023} to optimise the placement of targets within each class in sequence on the MSA.
The priority class criteria employed are presented in Table~\ref{tab:priorities}.  The science goals for the JADES survey as a whole are diverse (Section~\ref{sec:intro}, see also \citealt{eisenstein2023}), and the first deep pointings presented here represent the initial step in building up the entire sample.

We emphasise that the JADES NIRSpec survey does not employ a single selection function, but within each priority class there is a well defined set of criteria.
The highest priority targets (Class 1) are used to set and optimise the NIRSpec pointing centres (see Section \ref{sec:eMPT}) and are the bright, robust highest redshift candidates ($z>8.5$).  Classes 2 and 3 allow for less robust candidates and fainter candidates, respectively, at similarly high redshifts.  Progressing down the priority classes predominantly represents a progression in decreasing redshift, as the number counts then increase.  A notable departure from this is Class 5 in which we include bright objects to achieve a few high signal-to-noise continuum spectra per pointing.  

In Class 4, we aim to target galaxies in the redshift interval $5.7<z<8.5$ which are expected to have sufficiently bright rest-frame optical emission lines to enable emission line ratio work and exploration of interstellar medium (ISM) conditions
\citep{Cameron2023,Curti2023}. Our goal is to achieve $S/N > 25$ in the H$\alpha$ line (available at $z < 7$) or, when not available (i.e.\ $z>7$), [O {\sc iii}] $\lambda$5007. This would gives an expected $S/N \approx 8$ or more  for the H$\beta$ line,  and the resulting uncertainty of $10-15$\% on the Balmer decrement, $f(H\alpha)/f(H\beta)$, allows for an estimate of the attenuation due to dust (e.g.\ \citealt{sandles2023}). To achieve these target emission line fluxes in Class 4, we select on the rest-frame UV magnitude around 1500\AA .
At $z\approx 6$, a galaxy with a star formation rate of $2.5\,M_{\odot}\,{\mathrm yr}^{-1}$ has  a magnitude in the F115W filter of $AB= 27.5$ for the rest-UV longward of the Lyman-$\alpha$ break (assuming a \citealt{Salpeter1955} initial mass function and no dust extinction), and an expected H$\alpha$ flux of $3 \times 10^{-19}\,\mathrm{erg\,cm}^{-2}\,\mathrm{s}^{-1}$ at $4.5\,\mu$m (adopting the \citealt{Kennicutt1998} conversion from star formation rate to H$\alpha$ flux). This should be detectable at $S/N=25$ in the prism spectroscopy of the Deep tier of JADES (duration $\approx 100$\,ksec), using the STScI Exposure Time Calculator\footnote{\url{https://jwst.etc.stsci.edu/}}. Hence for Class 4 we adopt a magnitude cut of $AB=27.5$ in the broad-band filters just above the Lyman-$\alpha$ break. Fainter targets in the same redshift interval appear in Class 6.

Class 7 represents the statistical sample spanning $1.5<z<5.7$, which will be built up over multiple tiers, spanning from cosmic noon to the epoch of reionization. 
Within Class 7, we paid particular attention to placing unusual objects first before the more common star-forming galaxy population.
Specifically, galaxies which exhibited colours in the rest-frame UVJ colour--colour plane consistent with being passive or quenched galaxies  were identified following the criteria for specific star formation rate (sSFR), $\log(\textrm{sSFR} /\textrm{yr}^{-1})<-9.5$ given in \cite{Leja2019}.
We also prioritised ALMA sources which had a match to sources in the HST or NIRCam images \citep[e.g.][]{Aravena2016, Decarli2016, Rujopakarn2016, Dunlop2017, Franco2018, Yamaguchi2019,Hodge2013}, 
along with AGN including those selected from the IR, from variability, or from X-ray selection with an optical/near-IR counterpart \citep{Alonso-Herrero2006,Castello-Mor2013,Del-Moro2016,Luo2011,Luo2017,Sarajedini2011,Treister2006,Treister2009a,Treister2009b,Young2012}

Any unused areas on the MSA following the placement of sources in Classes 1--7 (described above) were filled with very low priority targets in Class 8 and 9, which comprised: fainter targets which did not pass the brightness to be in Class 7; targets at lower redshifts than $z\approx 1.5$; and targets for which the astrometry was unreliable. Blank sky shutters were also added.

\subsection{Establishing input catalogue of possible spectroscopic targets}
\label{sec:input_catalog}

The {\em Hubble} Ultra Deep Field (HUDF) and surrounding GOODS-South are very well studied fields. To provide an input target list for potential observation with the NIRSpec MSA, we compiled a large list of galaxies from the literature, which we cross-matched with our NIRCam-derived catalogues after first correcting the coordinates of the literature sources onto the same Gaia DR2 astrometric frame (see Appendix~\ref{sub:astrometry}), leveraging the CHArGE re-reduction of the GOODS-S HST imaging which has been registered to the GAIA DR2 astrometric frame \citep{Kokorev2022_CHARGE, grizli}\footnote{\url{https://s3.amazonaws.com/grizli-stsci/Mosaics/index.html}}. Where no match to an HST-detected object was identified within $0\farcs 3$ with the NIRCam-based catalogue (with co-ordinates defined as target centres), the target catalogue was supplemented with the HST-detected object. In regions of MSA footprint with no NIRCam coverage, all objects are taken from the HST-based catalogues.
We later impose selection criteria to populate the various priority classes which dictated the allocation of observed sources to the MSA micro-shutters. The priority classes for each object eventually observed are presented in Table~\ref{tab:targets}, where we list the final priorities allocated on the basis of NIRCam photometry (where available), and we also give the initial priority allocations on the basis of HST data alone.

A main driver of the JADES survey is to observe the highest redshift targets, for which
we compiled a sample of galaxies which had been identified as $z>5.6$ candidates by one or more studies in the literature, or from our NIRCam+HST analysis. The lower end of this redshift range corresponds to where the $i'$-band drop-out technique using the HST/ACS filter set becomes effective.
The list of galaxies compiled from the literature includes any studies that have previously selected $z\gtrsim6$ candidates based on the Lyman break technique and/or photometric redshifts \citep{Bunker2004, YanWindhorst2004, Oesch2010, Oesch2013, Lorenzoni2011, Lorenzoni2013, Yan2010, Ellis2013, McLure2013, Schenker2013, Bouwens2015,Bouwens2021, Finkelstein2015, Harikane2016}.
The sample from the literature was largely based on Lyman-break drop-out selection (e.g.\ \citealt{Bunker2004,Bouwens2015} ) although some are more generally based on photometric redshifts (e.g.\ \citealt{Finkelstein2015}). We cross-matched different samples in the literature which present high redshift candidates, and we note that while many galaxies were in common (using a matching tolerance of $0\farcs2$), there was a significant fraction which appeared in only one selection. This may be due to those papers using earlier reductions of HST data, perhaps not including all the data now available, or slightly different colour cuts, $S/N$ thresholds, and photometric aperture choices by the various research groups. Hence, to refine this selection of potential high-redshift targets, we inspected all the $z>5.6$ candidates from the literature. We used a slightly lower redshift cut for this inspection of candidates 
than ultimately adopted for Classes 4 \& 6 ($z>5.7$, Table~\ref{tab:priorities}) so as to allow slight changes in photometric redshift due to our remeasured photometry. For each $z>5.6$ candidate, we re-measured the aperture photometry from the HST images (with $0\farcs 36$-diameter apertures and appropriate aperture corrections) and ran photometric redshift fits with \textsc{eazy} \citep{Brammer2008} and \textsc{beagle} \citep{Chevallard2016}. For those galaxies which also appeared in our NIRCam-based catalogue, we also calculated the photometric redshift including both the NIRCam and HST
 photometry.

We also visually inspected the HST images (and NIRCam images where available) in all wavebands, using a co-addition of all the HST data taken in the GOODS-South field from the {\em Hubble} Legacy Field  v2.0 images \citep{Whitaker2019,Illingworth2016}, which goes deeper than many of the images used in the past to construct the early catalogues of Lyman break galaxies. For the sources selected from NIRCam photometry, we also removed spurious high redshift candidates due to artifacts and deblending issues.

We retained only the most robust candidates in our highest priority classes, those which were clearly detected at longer wavelengths, had a strong spectral break and were undetected at short wavelengths, and where the photometric redshifts strongly favoured a high redshift solution. Some objects were either only faintly detected or had spectral energy distributions where the photometric redshift was unclear (with both high and low redshift solutions possible). These were placed in a class for more marginal targets, which were allocated at lower priority than the more robust candidate high-redshift galaxies. 
In the case of the highest redshifts ($z_{phot}>8.5$) from the previous literature, the most robust candidates with HST F160W magnitudes brighter than $AB=29$ were placed in the top priority `Class 1' (Table~\ref{tab:priorities}), with those judged to be less robust placed in Class 2. From our NIRCam-based selection, we added two targets not appearing in the literature to Class 1 which had robust photometric redshifts $z>9$ and were brighter than $AB=29.5$\,mag in filters just longward of the putative Lyman-$\alpha$ break, and a further two NIRCam-selected targets which were judged to be less robust were added to Class 2.
Galaxies with $z_{phot}>8.5$ and fainter than F160W $AB=29$ in the literature-based selection appear in Class 3.

Galaxies with redshifts in the interval $5.7<z<8.5$ span the epoch of reionization and are also potentially selected by the Lyman-break technique using drop-outs in the F775W, F850LP and F105W filters on HST. For these targets, we impose a magnitude cut on the broad-band filter longward of the Lyman-$\alpha$ break, sampling the rest-frame UV (a proxy for star formation). Those galaxies brighter than $AB=27.5$ in that filter were allocated to Class 4 (with this magnitude cut justified in Section~\ref{sec:priority_class_system}), with less robust candidates and  slightly fainter galaxies ($27.5<{AB}<29$) in Class 6.1. Candidates fainter than $AB=29$ appear in Class 6.2. 
Some objects were up-weighted in this visual inspection exercise from Class 6 to Class 4 if they showed signs of strong line emission in the NIRCam photometry ($<20\%$).
Our input sample, after visual inspection and photometric checks, comprised about 300 galaxies at $z>5.7$ within the total NIRSpec MSA footprint.
There were other cases (about 5\% of the sample drawn from the literature) where we identified targets which seem to have flux below the putative Lyman-$\alpha$ break, and these were demoted to lower redshift classes based on the our revised photometric redshifts  (including the NIRCam photometry where available).  A number of high-redshift candidates from the literature were essentially undetected in the full co-added HST imaging, and these were removed from our sample (about 20\%, but we note that many of these would not have passed the magnitude cuts to place them in our very highest priority classes).

Galaxies with photometric redshifts below $z=5.7$ formed our lower-priority classes, in particular Class 5 (bright objects), and Class 7 (a magnitude-limited sample prioritised in redshift slices). 
When assigning priorities in Class 7, we used the opportunity to base the magnitude limit of $AB=29$\,mag on the longest wavelength NIRCam filter available (F444W), to make the selection as close to a mass-selected sample as possible, and to homogenise the selection with that planned for other tiers of JADES. Where NIRCam imaging was not available (or a literature source did not have a match in our NIRCam catalogue), we imposed a magnitude cut in HST/WFC\,3 F160W of $AB=29$, as this $H$-band filter is the reddest available HST data. This HST photometry for each galaxy was drawn from 
the latest available catalogue in which it appeared out of: \citet{Whitaker2019}, \citet{Rafelski2015_UVUDF}, \citet{Skelton2014_3dhst} or \citet{Guo2013}. If the source did not appear in any of these large catalogues, then we adopted the HST $H$-mag from the discovery paper if available (e.g.\ Lyman break catalogues) or we remeasured the photometry.

In Class 7, photometric redshifts are used to assign objects to four different redshift bins, with the smaller number of objects in the higher redshift slice $4.5<z<5.7$ 
being allocated to MSA shutters before the next slice ($3.5<z<4.5$)
and then those with $2.5<z<3.5$ 
and finally $1.5<z<2.5$. 
In the MSA target allocation in Class 7, we first placed the unusual targets (quiescent galaxies, AGN and ALMA sources) descending through the four redshift bins in order (sub-classes 7.1--7.4) before then placing shutters on the more common star-forming galaxies, again working down the four redshifts bins in turn to allocate targets.  Where available, the photometric redshifts were drawn from the high-redshift catalogues of \citet{Bouwens2021}, \citet{Finkelstein2015} or \citet{Bouwens2015}, which generally utilised the Lyman break in HST filters extending to the UV. We supplemented Class 7 with photometric redshifts from the UVUDF survey \citep{Rafelski2015_UVUDF}, or, if unavailable, from  the 3DHST survey \citep{Brammer2012_3dhst, Skelton2014_3dhst}, which in particular extended to lower redshifts than the Lyman break selected catalogues. 
For some lower redshift objects, the additional NIRCam photometry was not guaranteed to improve the photometric redshifts due to the small aperture used compared to the size of the objects, and the differences in point spread function (PSF).  We therefore only replaced the HST-derived photometric redshift with the HST+NIRCam-derived photometric redshift when the \textsc{beagle} and \textsc{eazy} photometric redshifts agreed.  Specifically, the redshift bin was assigned first using the HST-based photometric redshift (where available).  This was then adjusted only if the range between the \textsc{eazy} and \textsc{beagle} (primary or secondary redshift solution) 95\% credible regions overlapped, or the redshift solutions agreed within $\Delta z=0.1$. 
 As with Classes 1$-$6, all the Class 7 sources were visually inspected on the HST and NIRCam images, and a few eliminated as being unreliable.

\begin{table*}
\tiny
\caption{Target prioritisation categories.}
\label{table:1}      
\centering          
\begin{tabular}{c c c c c c c }   
\hline\hline       
 Priority & Criteria & Total targets & Targets  & JWST-only & Success\tablefootmark{4} & interloper \\ 
         &          & /MSA footprint & allocated & targets & rate & fraction \\
\hline                    
      & $z$\tablefootmark{1}$ > 8.5$, $F160W < 29$ (HST)\\
   1  & OR F115W dropout or higher $|$ robust $z_{phot}>9$ \&\\
   & \& detection band $< 29.5$  (JWST) & 6 & 6 & 2 & 83\% & 0\% \\
      & AND identified as robust from visual inspection\\
\hline
2 & As for Class 1 but lower visual inspection score & 6 & 2 & 2 & 50\% & 0\% \\
\hline
  & $z > 8.5$, $F160W > 29$ (HST)\\
3  & AND not rejected at visual inspection& 9 & 3 & 2 & $>$33\%\tablefootmark{a} & $<$33\%\tablefootmark{a}\\
  & Some objects not identified as robust originally in \\
  & Classes 1 and 2 are demoted to this class\\
\hline
  & $5.7 < z < 8.5$, pass rest-UV cut (HST) \\
4 & OR $6 < z < 8.5$, detection band $< 27.5$ (JWST) & 76 & 20 & 1 & 85\% & 5\%\tablefootmark{b} \\
  & AND identified as robust from visual inspection\\
\hline 
  & $2<z_{phot}<5.7$, $F160W<23$ (HST)\\
5 & OR $z_{phot} > 2$, any filter $< 22.5$ (JWST) & 18 & 5 & 1\\
\hline
  & $5.7<z<8.5$, F160W < 29 (HST) \\
6.1 & $5.7<z<8.5$, (F105W $< 29 | $F150W $< 29$) (JWST) & 100 & 9 & 3 & 89\% & 11\% \\
  & AND not rejected at visual inspection\\
  \\
  & $5.7 < z< 8.5$, F160W $> 29$ (HST) \\
6.2  & OR $5.7<z<6.5$, F444W\tablefootmark{2} $< 27.5$ (JWST) & 102 & 7 & 1 & 57\% & 29\% \\
  & AND not rejected at visual inspection \\
\hline
  & $4.5\leq z<5.7$, F160W $<29$ (HST)\\
7.1  & OR $4.5\leq z<5.7$, F444W\tablefootmark{2} $<27.5$ (JWST) & 4 & 1 & 0 \\
  & AND rare galaxy up-weighting\tablefootmark{3}\\
  \\
  & $3.5\leq z<4.5$, F160W $<29$ (HST)\\
7.2  & OR $3.5\leq z<4.5$, F444W\tablefootmark{2} $<27.5$ (JWST) & 5 & 1 & 0 \\
  & AND rare galaxy up-weighting\tablefootmark{3}\\
  \\
  & $2.5\leq z<3.5$, F160W $<29$ (HST)\\
7.3  & OR $2.5\leq z<3.5$, F444W\tablefootmark{2} $<27.5$ (JWST) & 12 & 0 & 0\\
  & AND rare galaxy up-weighting\tablefootmark{3}\\
  \\
  & $1.5\leq z<2.5$, F160W $<29$ (HST)\\
7.4  & OR $1.5\leq z<2.5$, F444W\tablefootmark{2} $<27.5$ (JWST) & 18 & 1 & 0\\
  & AND rare galaxy up-weighting\tablefootmark{3}\\
  \\
  & $4.5\leq z<5.7$, F160W $<29$ (HST)\\
7.5  & OR $4.5\leq z<5.7$, F444W\tablefootmark{2} $<27.5$ (JWST) & 246 & 23 & 2 & 83\% & 4\% \\
  \\
  & $3.5\leq z<4.5$, F160W $<29$ (HST)\\
7.6  & OR $3.5\leq z<4.5$, F444W\tablefootmark{2} $<27.5$ (JWST) & 570 & 31 & 1 & 71\% & 3\% \\
  \\
  & $2.5\leq z<3.5$, F160W $<29$ (HST)\\
7.7  & OR $2.5\leq z<3.5$, F444W\tablefootmark{2} $<27.5$ (JWST) & 1015 & 45 & 4 & 78\% & 4\% \\
  \\
  & $1.5\leq z<2.5$, F160W $<29$ (HST)\\
7.8  & OR $1.5\leq z<2.5$, F444W\tablefootmark{2} $<27.5$ (JWST) & 1565 & 47 & 4 & 64\% & 4\% \\
\hline
8.1 & F160W $ > 28.5$, $1.5\leq z<5.7$ AND has GAIA2 coords (HST) & 2119 & 20 & 3\\ 
    & OR $1.5\leq z<5.7$, $27.5 <$ F444W\tablefootmark{2} $<29$ (JWST)\\
    \\
8.2 & $24.5 < $F160W$ < 29$, $z<1.5$ AND has GAIA2 coords (HST) & 836 & 17 & 1\\
    & OR $z<1.5$, F444W\tablefootmark{2} $ < 29$ (JWST)\\
    \\
8.3 & F160W $>29$, $z<1.5$ AND has GAIA2 coords (HST) & 361 & 3& 0\\\\
    \hline
9 & fillers (not deliberately rejected) & 1569 & 12 & 0\\
\hline
\label{tab:priorities}
\end{tabular}
\tablefoot{
The HST and JWST entries for each class denote the different priority criteria whether the source was primarily selected from JWST or HST(see text for details).  The number of targets per MSA footprint were estimated from the full 3\farcm6 $\times$ 3\farcm4 field of view.\\
\tablefoottext{1}{In this table, $z$ denotes a redshift estimate either from a photometric redshift, or from dropout criteria.}
\tablefoottext{2}{Denotes photometry derived from Kron apertures.}
\tablefoottext{3}{``Rare galaxy up-weighting'' was applied to targets that were identified as candidates for being either quiescent, hosting an active galactic nucleus (AGN), or Lyman-continuum leakers.}
\tablefoottext{4}{The success rate is the fraction of galaxies targeted who had a spectroscopic redshift measured within $\Delta z=0.1$ of the predicted redshift interval for that priority class. Galaxies lying outside this range are classed as interlopers.}
\tablefoottext{a}{The spectrum of object 9992 is Class 3 is ambiguous and may show two sources, a low-redshift galaxy at $z=1.962$ and hints of a second galaxy at $z>9$.}
\tablefoottext{b}{One target in Class 4 for which we did not get a good spectrum, 10035328, is a star (with a proper motion of 0\farcs16 between 
HST/WFC3 and NIRCam) and we class it as an interloper.}
}
\end{table*}

\subsection{Target assignment}
\label{sec:eMPT}

The NIRSpec multi-object spectroscopic observations presented in this paper were carried out in MOS mode with the MSA \citep{Ferruit2022} with NIRCam operating in parallel. The MSA configurations employed were designed using the NIRSpec GTO team's so-called e\textsc{MPT} software suite \citep{Bonaventura2023}, and then imported into the STScI Astronomers Proposal Tool (APT) for execution. For a given choice of disperser and assigned roll angle, the e\textsc{MPT} is capable of identifying the pointings of the MSA on the sky that capture the largest possible number of high priority targets whose images fall within the open areas of operational shutters to a specified accuracy without their spectra overlapping on the detector. Three shutter tall slitlets were assigned to each target, and the telescope was nodded by one shutter facet (529~mas) along the spatial direction such that the targets were observed in each shutter in sequence. An `acceptance zone' spanning 184~mas in the dispersion direction and 445~mas in the spatial direction was employed throughout, corresponding to the full open area of a shutter with $\simeq9$~mas shaved off the edges. This was to prevent targets leaving the open shutter areas during any of the nods due to differential optical distortion arising in the telescope and NIRSpec optics. For all targets, only shutters whose low resolution prism spectra avoid truncation by the gap between the two detector arrays of NIRSpec \citep{Jakobsen2022,Ferruit2022} were employed. 

As described in \cite{Bonaventura2023}, the e\textsc{MPT} approach to designing the MSA masks starts by exercising its so-called `initial pointing algorithm' (IPA) module. This identifies the ensemble of candidate pointings within a specified range of the nominal pointing that provide the largest possible coverage of the targets designated as Priority Class 1 in the input catalogue at the roll angle assigned to the observation. Other e\textsc{MPT} modules are then employed to fill up the remainder of the MSA mask at each pointing with additional targets in decreasing order of scientific priority. For the observations presented here, three separate `dithered' pointings were planned with the goal of smoothing out detector defects in the dispersed spectra beyond that achieved by the three nods performed at each pointing. For the highest priority targets the objective was to achieve the largest possible total exposure time by observing these targets at all three dithers, while for the brighter lower priority targets the desire was to observe as many targets as possible, especially considering that these targets are placed  on the MSA last and therefore become progressively more difficult to accommodate. For any given trial of three pointings drawn from the set of all optimal Priority Class 1 covering pointings identified by the IPA module, the e\textsc{MPT} distinguishes between targets that can be observed at all three pointings, in only two of the pointings, and in only a single pointing, and gives the user complete control over the order in which targets in each subset are placed on the MSA. This process is  carried out for all candidate triple pointings that constitute reasonable dithers of the spectra on the detector, and the triple pointing achieving the best overall target coverage was selected as the final one. 

Another important consideration when using the MSA is to avoid targets being contaminated by the  unintended light from nearby targets entering any of the (nodded) slitlets. The e\textsc{MPT} automatically eliminates such targets in a `point source' manner based on the input catalogue, but since the contamination due to extended sources is difficult to automate, the  candidate MSA masks produced by the e\textsc{MPT} were subjected to a final visual inspection. Remaining undesirable targets were flagged and subsequently removed from the MSA masks. However, since the removal of a higher priority target can significantly change the placement of all lower priority targets that are placed after it, rather than start the process again from scratch, bespoke software was employed that allowed the process to converge after one or two  iterations by optimally filling the gaps in the MSA opened up by removed contaminated targets with other non-overlapping ones, while leaving all others in place.

The grating exposures taken at each pointing employed as the starting point the same MSA masks as the prism exposures, but were modified using  bespoke software to protect the grating spectra of the first five priority class targets by closing all shutters containing lower priority targets whose spectra collided with those of the higher priority targets. 

Through the above process, prism spectra of a total of 253 unique objects were obtained in the three pointings and series of exposures described in this paper. Of these, 27\% were observed in all three pointings, 24\% in two pointings, and 49\% in a single pointing. The three pointings cover 145, 155 and 149 individual prism targets each. In comparison, the grating observations cover a total of 198 unique targets, of which 28\% are observed at three pointings, 21\% at two pointings and 51\% in a single pointing. The three grating pointings cover 119, 121 and 111 individual targets each.

In practice, the late addition of high priority NIRCam sources, and re-prioritisation of the catalogue (see Section \ref{sec:input_catalog}) were incorporated without being able to re-optimise the pointings themselves. As a consequence, only four of the six highest priority targets in Table~\ref{tab:targets} were observed in three dithers as opposed to all of them as would have been the case if the pointing could have been tweaked. One of the added Priority 1 NIRCam targets (ID 2773) was observed in two pointings and a second one (ID 17400) in only a single pointing.

\begin{figure*}
    \centering
    \includegraphics*[width=0.95\textwidth]{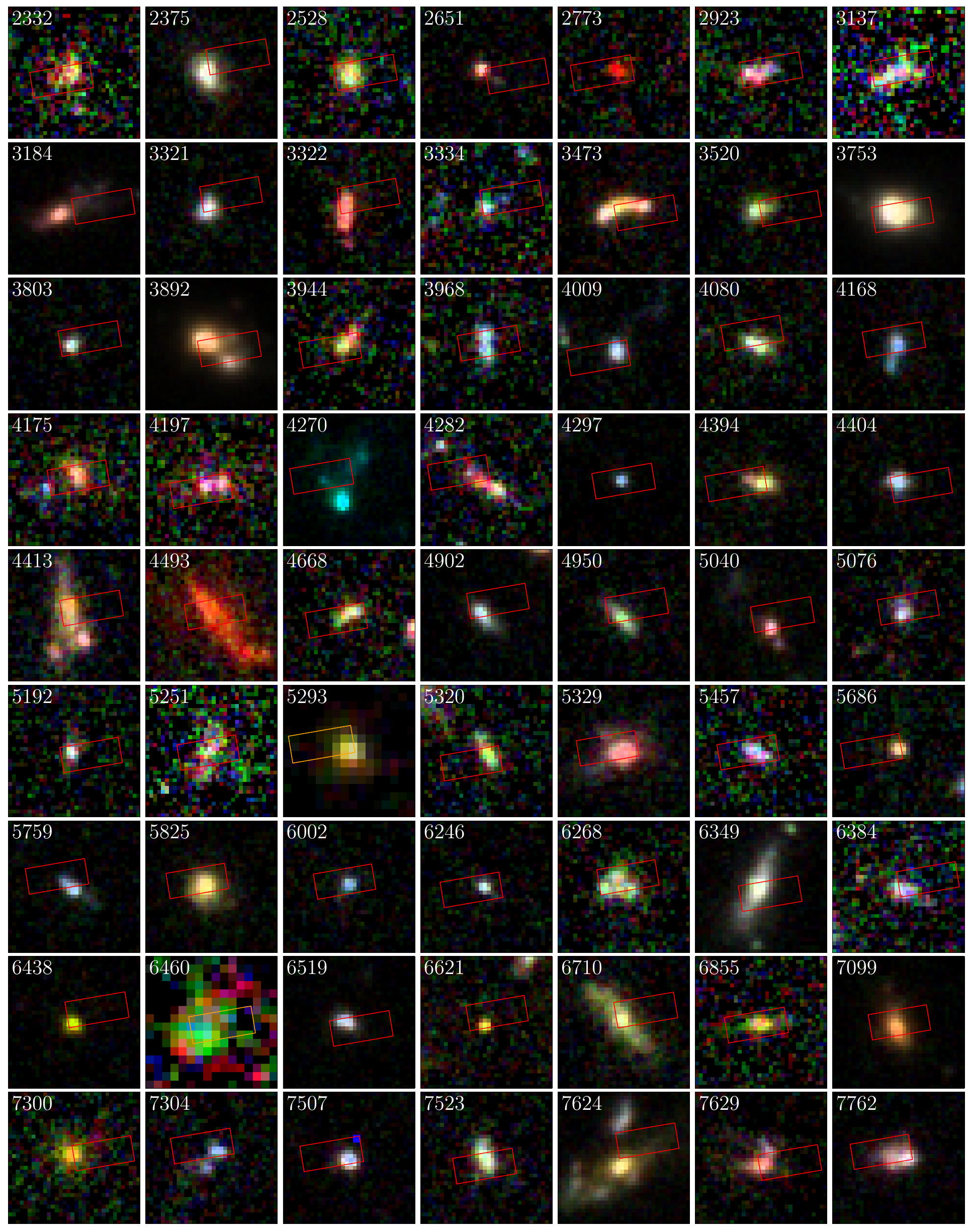}
    \caption{Overlay of target shutter positions onto the images, with the illuminated shutter regions outlined ($0\farcs46\times 0\farcs20$). The first 63 targets sorted by NIRSpec ID number (IDs 2333--7762) are shown here, starting at the top left, with the other 190 targets shown in Appendix~\ref{sec:postage_stamps}. 
    A red 
    outline indicates that the image is derived from the JWST/NIRCam F115W/F150W/F200W images from JADES (blue/green/red channels), and an orange outline denotes HST ACS-F850LP/WFC3-F125W/WFC3-F160W images. The individual images are $1\farcs 0$ on a side, and are centred on the input coordinate of the target. North is up and East is to the left. 
}
    \label{fig:postage_mosaic}
\end{figure*}

\begin{figure*}
    \centering
    \includegraphics*[width=0.9\textwidth]{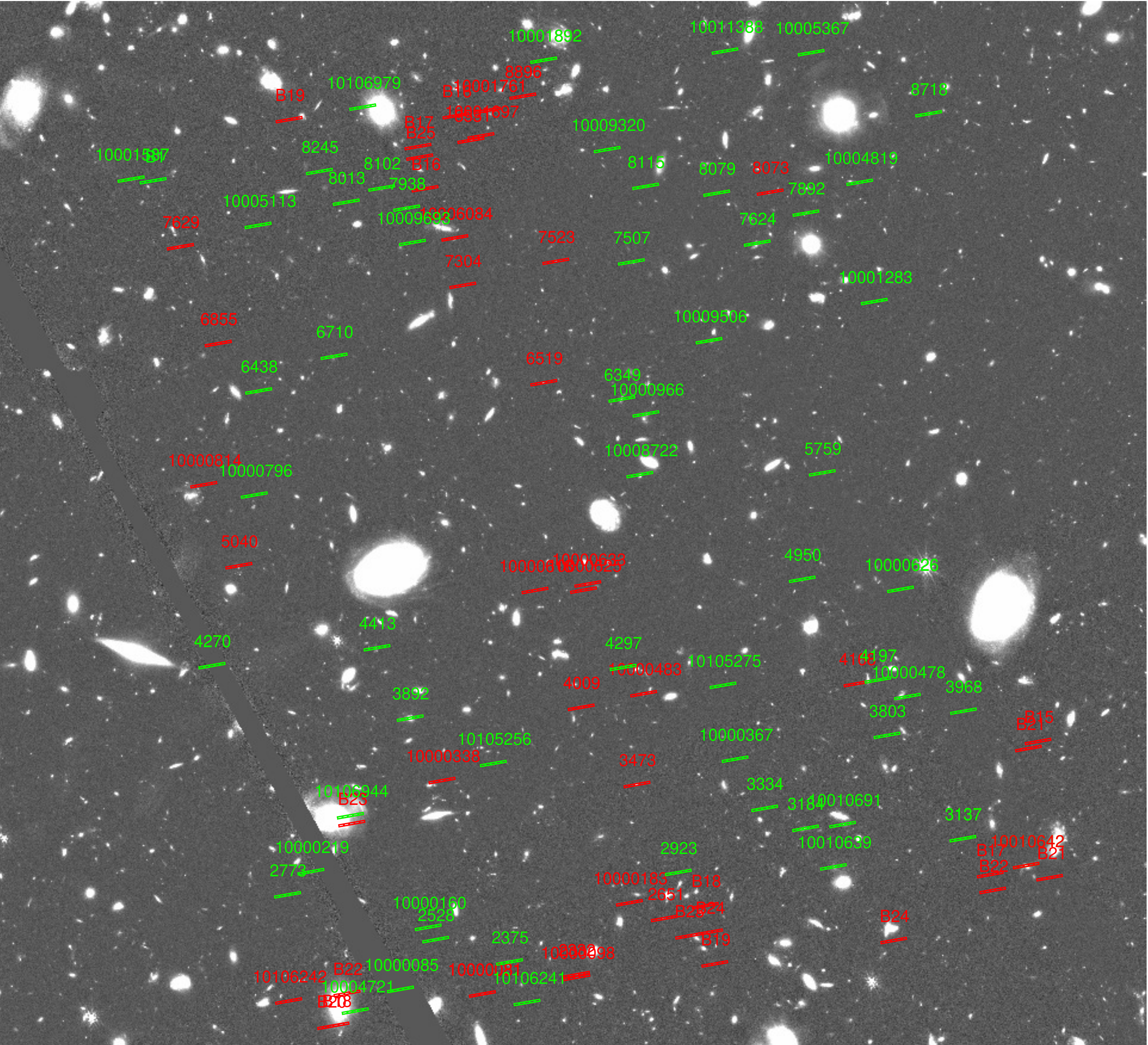}
    \caption{One of the four MSA quadrants (Q3), showing allocation of micro-shutters to targets. The other quadrants are shown in Appendix~\ref{sec:MSAquads}. Those shutters in green are covered by both the grating configurations and the low-dispersion prism. The red shutters are open only in the prism observations, as they would lead to overlapping spectra for our high priority targets in the grating configuration. Three micro-shutters are opened for each target, but the nodding by $\pm 1$ shutter means that spectra are obtained over the areas covered by five shutters (including background) which are displayed. The field displayed is  the NIRCam F200W image, and is $1.8$\,arcmin on a side. North is up and East is to the left. Shutters with the prefix `B' are empty sky background.}
    \label{fig:open_shutters}
\end{figure*}

\section{Observations}
\label{sec:obs}

The NIRSpec MSA observations of JADES Deep/HST were taken on UT 21-25 October 2022 as the JWST Program ID: 1210 (PI: N.\ L\"utzgendorf). The observations were split into three visits, which differed in their pointings by $<1$\,arcsec and employed separate MSA masks (see Section~\ref{sec:eMPT}) but identical exposure sequences.
The three pointings were selected such that they  shift the spectra on the detector by a sufficient amount  in order to smooth out detector defects; one shutter (268 mas or 2.6 pixels) in the dispersion direction plus one shutter (529~mas or 5.0~pixels) in the spatial direction for the second pointing, and three shutters (804~mas or 7.8~pixels) in the dispersion direction for the third pointing.

At each pointing we took observations with the low-resolution prism and four grating/filter combinations (G140M/F070LP, G235M/F170LP, G395M/F290LP and G395H/F290LP). We used the NRSIRS2 readout pattern \citep{Rauscher2017} with 19 groups for an integration time of 1400 seconds, with two integrations per exposure. The MSA configurations opened three adjacent shutters for each target and the targets were `nodded’ between these shutters (perpendicular to the dispersion direction) with an exposure at each position. For the prism only, this sequence was repeated four times to obtain very deep observations. At each one of the three pointings the total integration time (number of exposures) was 33.6\,ks (24) for the prism and 8.4\,ks (6) for each grating. Thus the sources observed in all three pointings attained total integration time of 100\,ks for the prism and 25\,ks for each grating.

\section{Data processing}
\label{sec:data_proc}

In processing this data, the NIRSpec GTO Team used a custom pipeline derived from the pipeline originally developed by the ESA NIRSpec Science Operations Team (SOT) described in section 4.3 of \cite{Ferruit2022} and based on the workflow and algorithms described in \cite{Oliveira2018}. This custom pipeline will be presented in a future paper (Carniani et al., in preparation).
We briefly describe here the main data reduction steps. The two NIRSpec detectors were read non-destructively multiple times using the NRSIRS2 readout mode. The master bias frame and dark current were subtracted, and we also corrected artefacts such as snowballs \citep{Ferruit2022,Giardino2019}. For each exposure we fit the slope (i.e.\ the count rate) for each pixel, identifying and removing jumps due to cosmic ray strikes, and flagging when saturation occurred. 
We background-subtracted the 2D spectrum in each shutter by taking the average of the two other exposures in the three-nod pattern. In some cases of spatially-extended objects, or those falling close to one end of a shutter, we excluded the adjacent shutter containing light from the target object (or in some cases a contaminating source) from the background subtraction. We note that very extended sources (a small minority of our targets) may be prone to some self-subtraction using this local background subtraction approach.

The individual 2D spectra from each shutter were then flat fielded and corrected for illumination by the spectrograph optics and the wavelength-dependent throughput of the dispersing element. The wavelength and flux calibration was then applied, with each pixel of the 2D spectrum having an associated wavelength and distance along the shutter, accounting for the slight tilt of the shutters relative to the dispersion direction, along with optical distortions. At each stage of the data reduction process we also propagated noise and data quality arrays.

The position of the object within the micro-shutter along the dispersion direction was also taken into account when applying the wavelength calibration — many of our targets are compact (Figure~\ref{fig:postage_mosaic})
with intrinsic sizes smaller than the $0\farcs 2$ shutter width, so making wrong assumptions about the slit being uniformly illuminated or that each object is well centered would lead to wavelength offsets.
We applied a path-loss correction to account for flux falling outside the micro-shutter; given the large wavelength range covered by NIRSpec ($0.6<\lambda<5.3\,\mu$m) it was critical to account for the considerable PSF variation with wavelength.
We took into account the position of the object within the micro-shutter (see the ``intra-shutter offset” columns in Table~\ref{tab:targets}), and calculated the slit loss as a function of wavelength for a point source at this location;
this was a reasonable approximation for many of our targets 
which are often compact (Figure~\ref{fig:postage_mosaic}), particularly at high redshift and also potentially for the star-forming regions giving rise to emission lines within more extended galaxies.

The spectra are curved on the detector due to optical distortions, and we rectify the 2D spectrum (transforming such that the wavelength and distance along the microshutter in the cross-dispersion direction lie along the $x$ and $y$ axes respectively), re-sampling the 2D spectrum onto a finer wavelength grid in the process.
For the gratings, the re-sampled pixel scale was $6.36$\,\AA , $10.68$\,\AA\ and $17.95$\,\AA\ for the G140M, G235M and G395M gratings, respectively.
For the prism, where the resolving power varies in a non linear way between $R\approx30-330$ \citep{Jakobsen2022}, we used an irregularly-gridded wavelength sampling with intervals between $26-122$\,\AA , with the coarsest sampling (largest wavelength interval per pixel) around $1.5\,\mu$m where the resolving power is at its lowest.
The 1D spectra for the three nod positions from each of the (up to) three pointings were then combined by a weighted average into a single 1D spectrum for each target, masking pixels previously flagged as bad in the data quality files, and rejecting outliers using a sigma clipping algorithm.
We also separately combined all the 2D spectra for each target from the different nods and pointings, although the 1D combined spectrum comes from a combination of the 1D individual spectra rather than an extraction of the combined 2D spectrum.
The resulting 1D and 2D spectra reduced data products for all targets are made available as part of this data release, along with the raw data. Example spectra covering a range of redshifts are shown in Appendix~\ref{section:example_spectra} (Figures~\ref{fig:example_spectra1}--\ref{fig:example_spectra10}).

\section{Redshift determination and emission line fluxes}
\label{sec:emission_lines}

\begin{figure}
    \centering
    \includegraphics[width=\columnwidth]{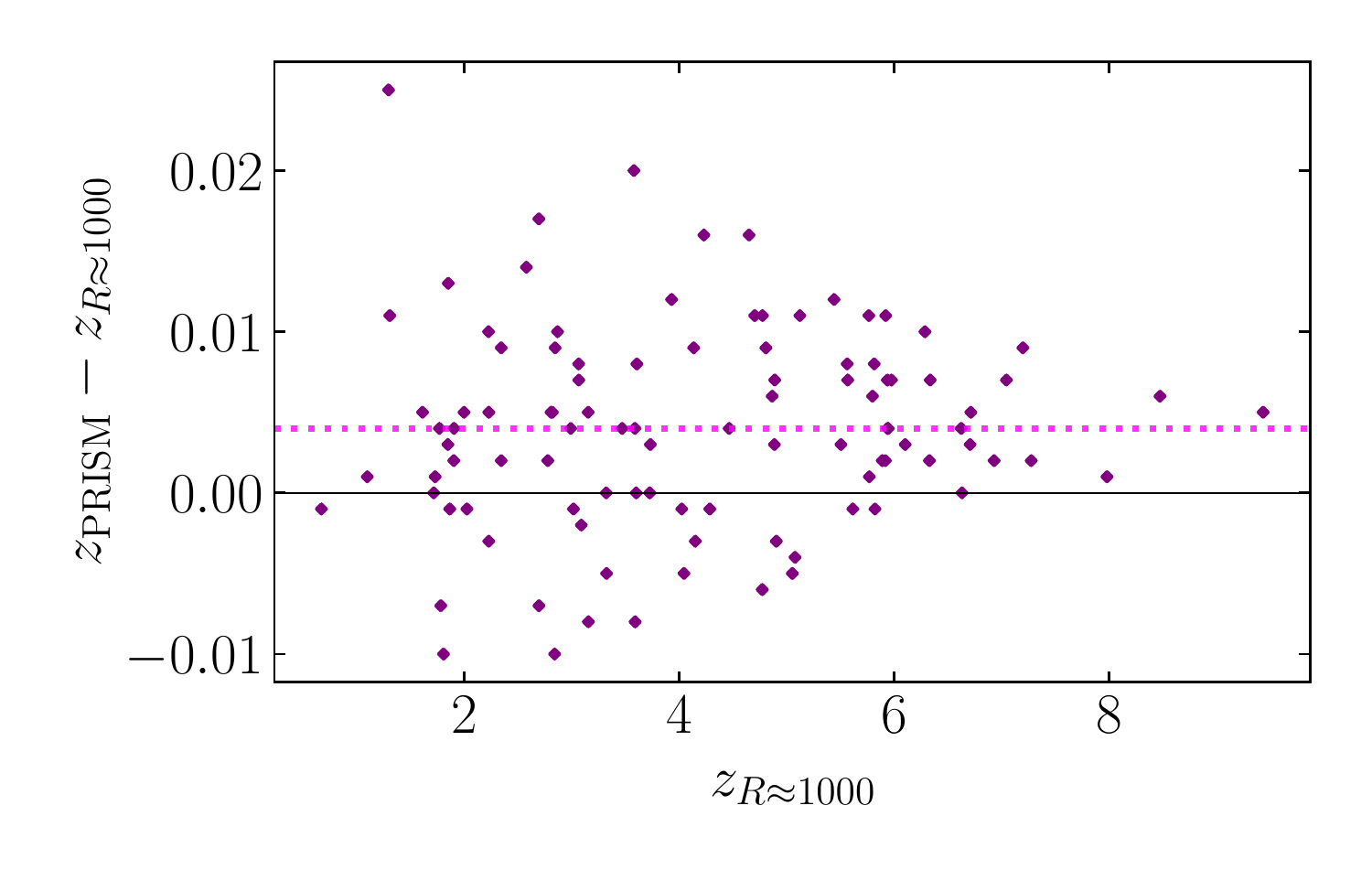}
    \includegraphics[width=\columnwidth]{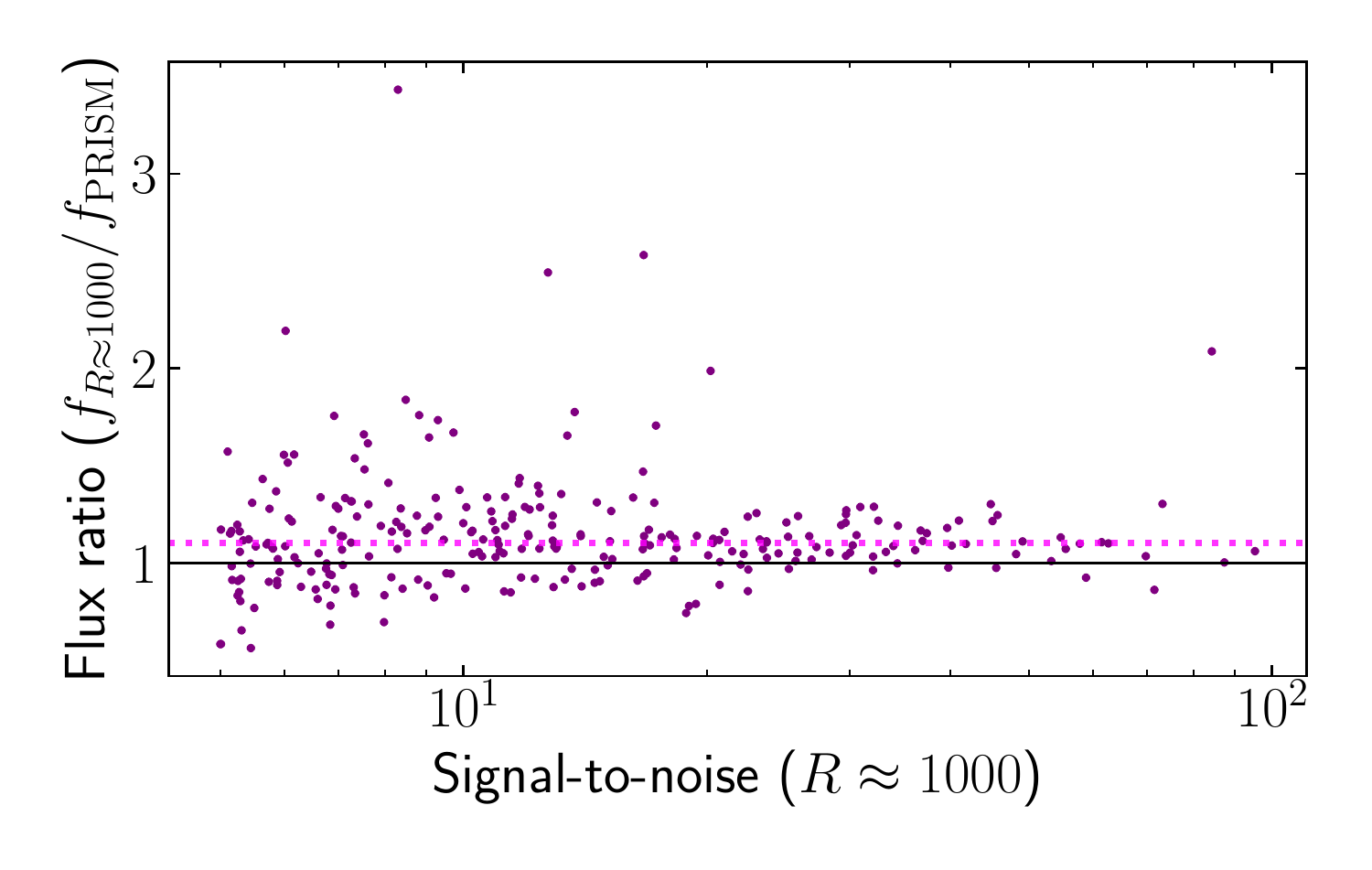}
    \caption{Comparison of spectral measurements between the low-dispersion prism and medium-dispersion gratings. \textit{Upper:} Comparison of redshift as determined from Prism/Clear and $R\approx1000$ grating observations for targets with emission lines clearly detected in both modes. There is a systematic offset of $\Delta z=0.0039$, with the prism yielding systematically higher redshifts. \textit{Lower:} Comparison of emission lines fluxes measured from prism and $R\approx1000$ grating. Measurements derived from the grating are systematically higher with a median value of $f_{R\approx1000}/f_{\rm PRISM}$ = 1.105 and a standard deviation of 0.298}
    \label{fig:redshift_comparison}
\end{figure}

 \begin{figure}
    \centering
    \includegraphics[width=\columnwidth]{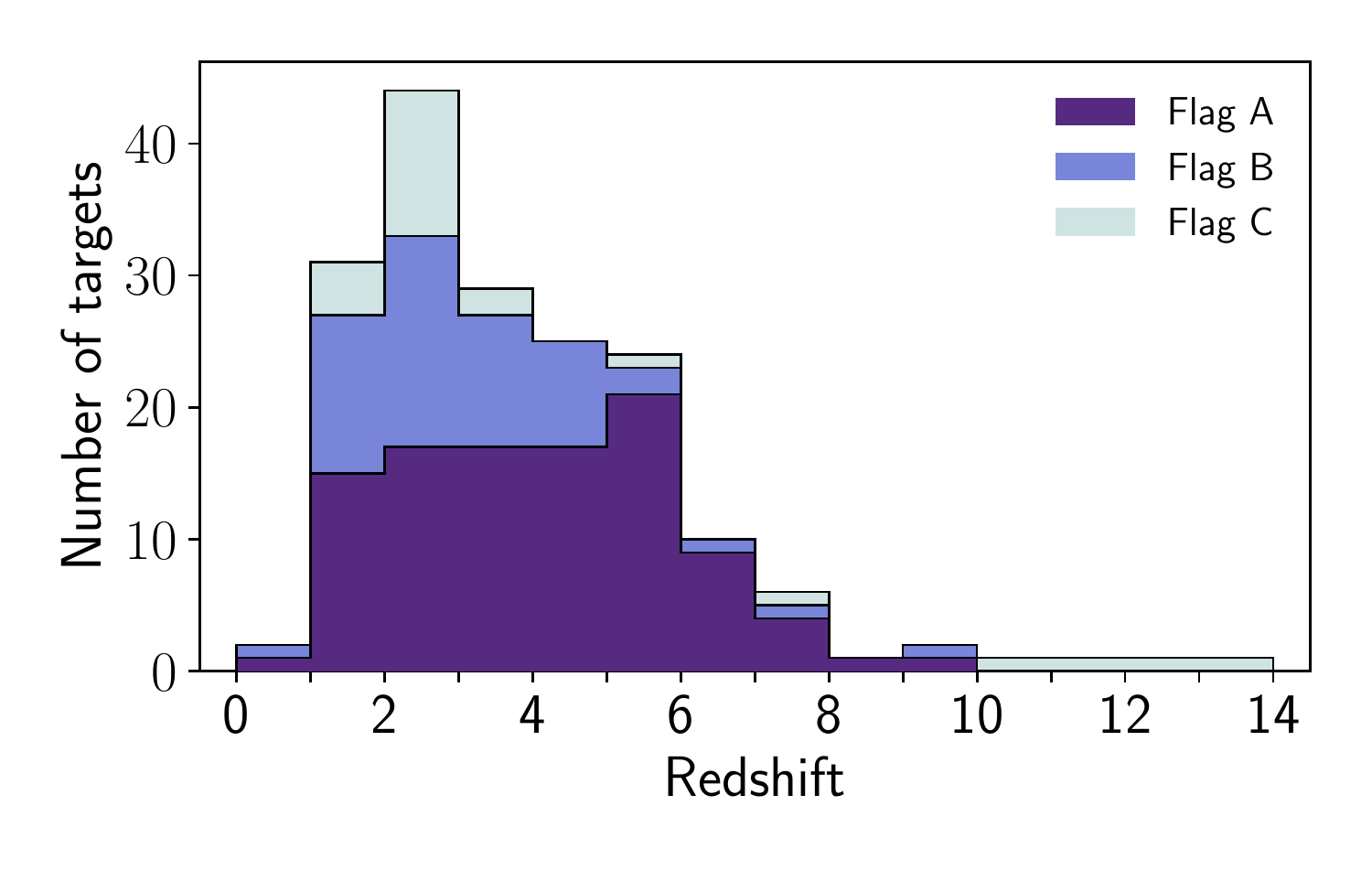}
    \caption{Histogram of spectroscopic redshifts obtained from $S/N>5$ emission lines. The separate histograms are for the medium-dispersion $R\approx1000$ gratings (flag A, darkest purple), additional galaxies with $S/N>5$ emission lines detected with the low-dispersion $R\approx30-300$ prism (flag B, lighter purple histogram) and galaxies with more marginal redshifts (flag C, lightest histogram).}
    \label{fig:redshift_historgram}
\end{figure}

In this section, we report spectroscopic redshifts and emission line fluxes determined from our spectroscopic observations, and assess the success rate of our priority class system for target selection. 

\subsection{Visual inspection and emission line fitting}
\label{sec:visual_redshifts}

The 1D and 2D spectra of all spectral configurations were visually inspected as a first pass on the redshift determination. The SED fitting code {\sc bagpipes} \citep{Carnall2018, Carnall2019} was run on the 1D Prism/CLEAR spectra and the redshifts arising from this fitting were used a starting point for the inspection, but ultimately the assessment of the human inspector would overrule this value if necessary.
In many cases, several clear emission lines were observed and the redshifts were unambiguous. Sometimes spectral breaks were visible, most notably the Lyman-$\alpha$ break (e.g.\ \citealt{CurtisLake2023, Looser2023}), and sometimes the Balmer break or 4000\,\AA\ break. In fainter targets, the $S/N$ of individual features were sometimes low, but the coincidence of more than one of these led to a tentative redshift.

We then performed emission line fitting to further refine the redshifts, and obtain measurements of the fluxes of significant emission lines.
For the $R\approx1000$ grating data, the continuum was typically only marginally detected and we subtracted this by fitting a spline to the spectrum after masking out any regions which could be contaminated by prominent emission lines. We then performed a single-component Gaussian fit to each line individually, allowing the flux, redshift and line-width to vary independently. In the case of unresolved doublets, such as [O {\sc ii}] $\lambda\lambda$3726, 3729, we simply fit the entire doublet as a single component. In the case of H$\alpha$ and [N {\sc ii}] $\lambda$6583, although these lines are never blended in our $R\approx1000$ grating data, these lines were fit simultaneously and had their line centroids fixed relative to one another. The line flux was obtained as the integrated area under the best-fit Gaussian, and the formal uncertainty on this Gaussian fit was taken as the noise on the line flux. We retained only emission lines which were measured with $S/N>5$, and visually inspected each fit to ensure the measurement was robust. The emission line fluxes arising from this are reported in Table~\ref{tab:grating_lines}. We note that there were two cases (ID 10013704 and ID 8083) where a broad component under H$\alpha$ meant that a single component fit was not appropriate. In these cases, the reported H$\alpha$ flux is obtained by integrating the whole line between the zero-power points in the spectrum.

In the case of the Prism/CLEAR data, the much lower spectral resolution ($R\approx30-300$) means that blending of emission lines is much more common in these data. Furthermore, which emission lines are blended changes with galaxy redshift due to the wavelength-dependent nature of the resolution.
For this reason, although the approach to emission line fitting on the Prism/CLEAR spectra largely followed the same process as described above for the $R\approx1000$ mode, some redshift-dependent modifications were implemented. As such, 
which line fluxes are reported as blends changes 
with redshift.
At all redshifts for the prism data, H$\alpha$+[N {\sc ii}] was fit as a single component, as were close doublets such as [O {\sc ii}] $\lambda\lambda$3726, 3729 and [S {\sc ii}] $\lambda\lambda$6716, 6731. The flux of H$\alpha$+[N {\sc ii}] and [S {\sc ii}] $\lambda\lambda$6716, 6731 were fit for simultaneously, with fixed centroids.
In all cases, the fit to H$\beta$ and [O {\sc iii}] $\lambda\lambda$4959, 5007 was performed simultaneously with the centroids fixed relative to one another. Above $z>5.3$, the fluxes of all three components were fit (and reported) independently. At lower redshifts, the ratio of [O {\sc iii}] $\lambda$5007/$\lambda$4959 was fixed to 2.98, but the H$\beta$ flux could still vary independently. Between $2<z<5.3$, we report the flux of the [O {\sc iii}] $\lambda\lambda$4959, 5007 as a blend.
For the [O {\sc iii}] $\lambda$4363 and H$\gamma$ complex, above $z>7.5$, the resolution allowed for a two-component fit to yield fluxes that are reported separately in Table~\ref{tab:prism_lines}. Between $5.3<z<7.5$, this flux is measured with a two-component fit, but is reported as a blend. At lower redshifts this blend was fit with a single component.

Below $z<2$ 
the reported fluxes of lines with rest-frame wavelengths blue-ward of 7000~\AA{} ($\lambda_{\rm obs}\lesssim2 \mu$m) are no longer obtained from Gaussian fitting, but instead are measured simply by integrating the continuum-subtracted spectrum of the specified blend. Lines red-ward of this are measured with Gaussian fitting and are fit independently with the exception of He{\sc i} 10830 and Pa-$\gamma$ which are fit simultaneously.

We note that there are many cases where emission lines were identified visually in the data that did not meet our $S/N>5$ threshhold to be included in Tables~\ref{tab:grating_lines}~-~\ref{tab:prism_lines}, however we opted not to report these fluxes. Particularly in the case of the Prism/CLEAR spectra, which generally speaking have significant continuum detections, reported fluxes for fainter lines become highly sensitive to how the continuum is modelled.
We also do not fit for Lyman-$\alpha$ in the Prism/Clear spectra here as the flux measurement is highly sensitive to how the continuum and Lyman-$\alpha$ break is modelled. Lyman-$\alpha$ measurements are however reported in \cite{jones2024} and \cite{saxena2024}.
We also note, there may be cases where reported lines are blended with other faint lines, despite this not being explicitly reported as such here. For example, [Ne {\sc iii}] $\lambda$3869 can be blended with He{\sc i} $\lambda$3889 emission. The reported flux in such cases where it appears as a single-peaked feature will reflect the whole complex.

For galaxies which had at least one emission line detected in the $R\approx1000$ data, we calculate $z_{R\approx1000}$ as the $S/N$-weighted average of the redshifts arising from the measured centroids of detected, non-blended lines and adopt this as our preferred redshift (flag `A' in Table~\ref{tab:targets}).
There were 150 cases where a galaxy did not yield a grating redshift in this way (either due to low $S/N$, or lack of a grating spectrum), but in 52 of these a $z_{\rm PRISM}$ could be derived analogously from the Prism/CLEAR fits (flag `B' in Table~\ref{tab:targets}).
This accounted for 155 highly confident redshift determinations.
Of the remaining 98 cases, we report a further grade `C' redshift for 23 targets where the redshift had been determined as being secure from visual inspection (either based on a spectral break and/or one or more low $S/N$ emission lines), and in Table~\ref{tab:marginal} we simply report the redshift obtained from this original visual inspection.
This leaves 75 targets for which the redshift is speculative, ambiguous or unable to be determined. These are heavily weighted toward our lowest priority classes.
As can be seen from the slit overlays on the JWST/NIRCam or HST images in Figure~\ref{fig:postage_mosaic}, some targets fall on the edge of the micro-shutter which will reduce the flux. Some shutters do appear empty, and these are largely targets based on catalogues from the literature which are either spurious or whose astrometry is less accurate (e.g.\ not from HST).

The medium-dispersion gratings yield more accurate redshifts than the low-dispersion prism, with the typical uncertainty in the centroid for a $S/N=10$ line being $1$\,\AA\ for the G140M grating, rising to $2$\,\AA\ for the G395M, compared to $16-50$\,\AA\ for the prism.
Hence for flag A redshifts, the typical uncertainty is $\Delta z / (1+z) \approx 10^{-4}$ and for flag B redshift the typical uncertainty is $\Delta z / (1+z) \approx 0.0003 - 0.003$.
Flag C redshift were determined visually and so are less precise.
A histogram of redshifts determined from these spectra is shown in Figure~\ref{fig:redshift_historgram}.

\subsection{Comparison of prism and grating observations}

We note that our Prism/CLEAR spectra are all non-overlapping, and thus cannot contain contamination from targets placed elsewhere on the MSA. This is not true for the grating data, for which spectra can be overlapping (although our highest priority targets are protected, see Section~\ref{sec:eMPT}). Thus, these spectra occasionally show spurious emission lines. However, given that the Prism/CLEAR observations were significantly deeper than the $R\approx1000$ grating data, targets which are observed with significant emission lines in the grating always show the same significant emission in the low-resolution data. Thus, all the grating redshift measurements here can be confirmed to be robust.
We note that there were three targets (IDs 8880, 9343, and 10013545) for which the the reduced Prism/CLEAR spectrum is sufficiently corrupted that, beyond simply confirming the presence of emission lines, we did not measure the emission line fluxes to be reported in Table~\ref{tab:prism_lines}. However, in all three cases, secure redshifts and line fluxes were already measured from the $R\approx1000$ data.

We have 100 galaxies for which we have robust measurements of both $z_{R\approx1000}$ and $z_{\rm PRISM}$. 
In Figure~\ref{fig:redshift_comparison} we compare the redshift determinations from the Prism/CLEAR spectra and the $R\approx1000$ gratings for each galaxy, and find a small systematic offset (with the grating determination of redshift slightly lower than that from the prism) with a median offset of 0.00388 and standard-deviation 0.00628.

We also compare the flux ratio for the same lines where these are detected in both the Prism/Clear and the $R\approx 1000$ gratings, excluding lines which are significantly blended in the prism, and these ratios are shown in Figure~\ref{fig:redshift_comparison}. We note that the grating fluxes are on average 10\% higher than the prism. \cite{Bunker2023} found that the flux in the prism agreed well with the NIRCam magnitudes (where the spectrum was integrated over the NIRCam filter bandpass), with the grating spectra showing less good agreement, suggesting that the flux calibration in the prism is more accurate.

\subsection{Comments on individual targets}
\label{sub:individual_targets}

A few MSA shutters exhibited unusual spectral features, often due to more than one source in the shutter. We briefly discuss these below, along with objects which are likely to be stars where proper motion can be seen between the HST/WFC3 images and the {\em JWST}/NIRCam images taken $\sim 13$\, years later.

\subsubsection{ID 5293 -- star}

Proper motion can be identified between HST/WFC3 F775W imaging and {\em JWST}/NIRCam 
F277W 
imaging for this object. Furthermore the spectrum looks visually like a brown dwarf star. No proper motion was clearly seen when comparing different-epoch observations from HST alone -- possibly due to the motion being comparable to the spatial resolution of HST/WFC3. However, with the better spatial resolution of NIRCam/SW, we detect a motion of $\approx 0\farcs05$.

\subsubsection{ID 7624 -- two sources in shutter}

Two sources can clearly be seen in the HST imaging (Figure~\ref{fig:postage_mosaic}), and the slit falls between the two sources. Object 7624 in Class 7.7 was the intended target (to the south of the slit), but a Lyman-break galaxy (a F435W $b$-band dropout) lies just to the north. We observe line detections consistent with [O {\sc iii}] $\lambda$5007 and H$\alpha$ at $z=2.665$ (from the target object 7624) and also at $z=4.854$ from the $b$-band drop-out. 

\subsubsection{ID 8896 -- possible double source}

This micro-shutter was originally targeted on a low redshift galaxy (Class 7.8).
We detect at least three compelling emission lines, and one more marginal line. 
There are emission lines that are consistent with [O {\sc iii}] and H$\alpha$ at $z=1.984$, and this is reported in Table~\ref{tab:prism_lines}.
However, we note that there are two robust lines that are consistent with a $z=6.287$ galaxy seen with [O {\sc iii}] and H$\alpha$. 
The imaging does not obviously reveal the presence of two objects (Figure~\ref{fig:postage_mosaic}), however there does not seem to be a plausible redshift solution that matches all of these lines simultaneously for a single object.

\subsubsection{ID 9992 -- possible double source}
\label{sec:9992}

This object was targeted as a $z>8.5$ candidate (Class 3). The Prism/CLEAR spectrum reveals a number of emission lines. The two most significant emission features are consistent with [O {\sc iii}] and H$\alpha$ at $z=1.962$. However, an emission line at 5.1 $\mu$m could be H$\beta$ or [O {\sc iii}] $\lambda$5007 at $z>9$, and this would be consistent with a tentative spectral break observed at 1.25 $\mu$m being a Lyman-$\alpha$ break. 
The NIRCam photometry reveals two components separated by only $0\farcs 15$ (Figure~\ref{fig:postage_mosaic}). One of these is photometrically consistent with a drop-out galaxy at $z\sim10$, while the other (over which the central shutter is better placed) is more consistent with lower redshift solutions.
We do not consider our high redshift solution from the spectrum to be highly robust, and Table~\ref{tab:targets} reports the low-redshift solution.

\subsubsection{ID 10040 - multiple sources in shutter}

Imaging clearly shows multiple sources with flux in the shutter. The spectrum has clear detections of emission lines consistent with [O {\sc iii}] and H$\alpha$ at $z=3.14$, which we report in Table~\ref{tab:prism_lines}. There is also, however,  continuum detected in the 2D spectrum, which appears to be spatially offset from the emission lines that we detect and which may arise from another source.

\subsubsection{ID 10035328 - star}
\label{sec:10035328}

Proper motion can be identified between the HST/WFC3 images and the {\em JWST}/NIRCam images.

\subsection{Quantifying the success of target selection}
\label{sec:redshift_success}

To measure the success of the class-based allocation, we looked at which targets from which classes actually ended up having the redshift expected (and desired line flux $S/N$ in the case of Class 4).

In our highest-priority Class 1 (predicted redshifts $z>8.5$ and  ${AB}<~29$),  we targeted six galaxies, five of which were robustly confirmed to be at high redshift: three at $z>11$ have previously been reported in \cite{CurtisLake2023} (GSz12-0=2773 at $z=12.63$, GSz11-0=10014220 at $z=11.58$, GSz13-0=17400 at $z=13.20$) and have strong Lyman-$\alpha$ breaks but no significant line emission. Galaxy 10058975 at $z=9.43$ exhibits many strong emission lines (see Figure~\ref{fig:example_spectra1}), as does galaxy 8013 (which falls just below the targeted redshift cut at $z=8.47$). One galaxy, ID 10014170, did not have obvious features in its spectrum and its redshift is ambiguous. Hence, we have a success rate of 83\% in pre-selecting Priority Class 1 targets which are then spectroscopically confirmed to be at high redshift. 

In Class 2 (candidates at $z>8.5$ with ${AB}<29$ which are more marginal), of the two targets one has a robust redshift of $z=9.68$ (galaxy 6438), and the second target (galaxy 7300) has an inconclusive spectrum. Class 3 has three $z>8.5$ candidates fainter than $AB>29$, but even here we are successful in confirming the high redshift nature of some of these: of the three targets, ID 10014177 was previously reported in \cite{CurtisLake2023} as GSz10-0 at $z=10.38$. Galaxy 9992 was discussed in Section~\ref{sec:9992}; there is clearly a low-redshift interloper at $z=1.962$, but inspection of the imaging reveals that there is a second galaxy, and the spectrum provides hints of other lines which may be consistent with a second source at $z>9$. We regard this spectrum as inconclusive. The third galaxy (ID 6621) has no strong emission lines but may exhibit a spectral break consistent with a Lyman-$\alpha$ break at a tentative redshift of $z=9.6$.
Hence, for all 11 of the $z>8.5$ candidates targeted, seven were clearly at high redshift (a fraction of 64\%), and four had inconclusive redshifts (36\%).

We now discuss the success of the selection in Class 4, where galaxies at $5.7<z<8.5$ were targeted which were sufficiently bright (${AB}<27.5$ in the wavebands corresponding to the rest-UV) that high $S/N$ emission lines are expected.
Of the 20 sources in Class 4 which were targeted, 17 were confirmed to be at high redshift (including galaxy 16745 at $z=5.57$, which fell just below the targeted redshift range). The emission line fluxes for H$\alpha$ were typically brighter than  $7.5\times 10^{-19}\,{\rm erg\,cm}^{-2}\,{\rm s}^{-1}$ (except for one object, ID 6384) as expected from our rest-UV pre-selection (Section~\ref{sec:targets}).
The uncertainty on the line flux from the prism is about $0.3\times 10^{-19}\,{\rm erg\,cm}^{-2}\,{\rm s}^{-1}$ as predicted, so we met our requirement of $S/N>25$ in H$\alpha$ for the sample in Class 4, enabling the physics of the ISM and the metal enrichment to be explored \citep{Cameron2023,Curti2023}.
Our overall success rate in Class 4 is 85\%, although we note that one of the three objects (out of 20) for which we did not obtain a good spectrum is ID 10035328, which is a likely star (see Section~\ref{sec:10035328}).

Fainter candidates than in Class 4 but in the same redshift interval $5.7<z<8.5$ were placed in Class 6.1 if they were brighter than ${AB}=29$ in HST/F160W (or the NIRCam filter around rest-frame 1500\AA ), and in Class 6.2 if they were fainter than that. The success rate in Class 6.1 is very good, with eight of nine galaxies having redshifts in the targeted range. We note that object 8115 does not have emission lines but does a strong Lyman-$\alpha$ break and a weaker Balmer break at $z=7.3$, and the JADES spectrum has been discussed in \cite{Looser2023} as a potential quiescent galaxy. The NIRSpec spectrum of ID 3334 at $z=6.71$ has been presented in \cite{Witstok2023} and shows evidence of broad rest-frame UV absorption around 2175\AA . Object 3137 is a low-redshift interloper with $z=1.91$, so our overall success rate in Class 6.1 is $89\%$. In the fainter Class 6.2, four of the seven objects have spectroscopic redshifts 
within the target  range, with two galaxies also at redshifts slightly below this (object 8113 at $z=4.90$ and object 17260 at $z=4.89$). One spectrum (object 10014117) had no significant features from which a redshift could be determined. The NIRSpec spectrum of object 10013682 shows very strong Lyman-$\alpha$ emission at the systemic redshift of $z=7.28$, as discussed in \cite{Saxena2023}. Our success rate for Class 6.2 is 57\%, rising to 86\% if the two galaxies at $z=4.9$ are included.

The success rate for Class 7 is recorded in Table~\ref{tab:priorities} for the sub-classes 7.5--7.8, where there are significant numbers ($>20$) of galaxies targeted. Taking the metric for `success' as being a measured spectroscopic redshift within $\Delta z=0.1$ of the intended redshift range, we have success rates of 83\% in the highest redshift slice ($4.5<z<5.7$, Class 7.5), around 80\% for Classes 7.6 and 7.7, and 64\% for the lowest redshift bin ($1.5<z<2.5$, Class 7.8). The interloper fractions were $\approx 3-4$\%, but these comprised galaxies only slightly outside the desried redshift bin (e.g.\ object 3892 has a redshift of $z=2.80$ and was selected to be in Class 7.6 at $3.5<z<4.5$). Those galaxies in Class 7 for which a reliable redshift could not be inferred amounted to $<$20\% of those targeted, and these tended to be the sources which were less well centred within the shutters, resulting in large slit losses.

Overall our priority class pre-selection strategy seemed successful; for the more robust galaxies, 80\% or more of the time the spectroscopic redshift fell in the anticipated range, and the line fluxes for Class 4 (which had been pre-selected on the basis of the rest-UV) were also as anticipated.

\section{Conclusions}
\label{sec:conclusions}

We have presented very deep spectroscopy obtained with {\em JWST}/NIRSpec in its multi-object MSA mode. In all,  253 targets were observed  in this JADES Deep/HST spectroscopy covering the {\em Hubble} Ultra Deep Field, and the surrounding GOODS-South, with total integrations times of up to 28 hours for the low-dispersion prism ($R\approx30-300$), and up to 7 hours in the three medium dispersion gratings ($R\approx1000$) and one high dispersion grating (G395H, $R\approx2700$). We detected emission lines with $S/N>5$ in 155 targets with the low-dispersion prism, 103 of which also had emission lines detected at this significance in $R\approx1000$ gratings. The robust redshifts determined for these galaxies spanned a range from $z=0.66$ to $z=13.2$, with 18 lying at $z>6$. A further 23 galaxies has more tentative redshifts.
We are able to detect emission lines at $S/N>5$ as faint as $\approx 10^{-19}\,{\rm erg\,cm}^{-2}\,{\rm s}^{-1}$ in our deepest prism spectra, and we have been able to confirm redshifts for some sources fainter than ${AB}=29$. Our selection of targets preferentially places the rarer high redshift targets on the MSA at higher priority, with more numerous lower-redshift galaxies filling unused regions, so that we can probe a large redshift range from `cosmic noon' ($z\sim 2$)  to within the epoch of reionzation ($z>6$) with reasonable numbers of galaxies in several redshift slices. We have demonstrated that our pre-selection of targets from HST and JWST imaging, based on broad-band magnitudes and photometric redshifts (including many Lyman break galaxy candidates) is highly effective, with $\sim 80$\% of galaxies targeted having spectroscopic confirmation within the expected redshift bin. Hence our target selection and the quality and depth of the NIRSpec MSA spectroscopy means that our science goals for the JADES project can be met.

\begin{figure*}[h]
    \centering
    \includegraphics[width=0.98\linewidth]{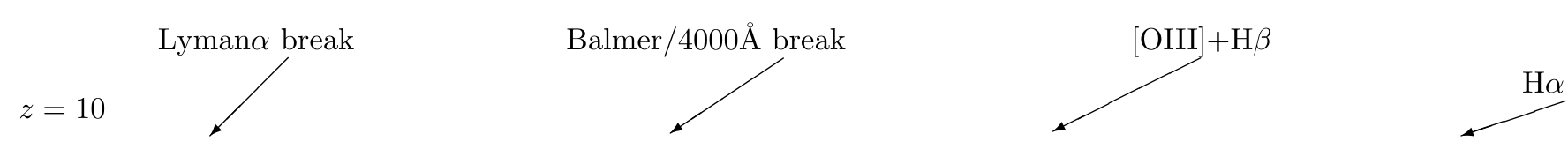}
    \vspace{-0.15cm}
    
    \includegraphics[width=0.98\linewidth]{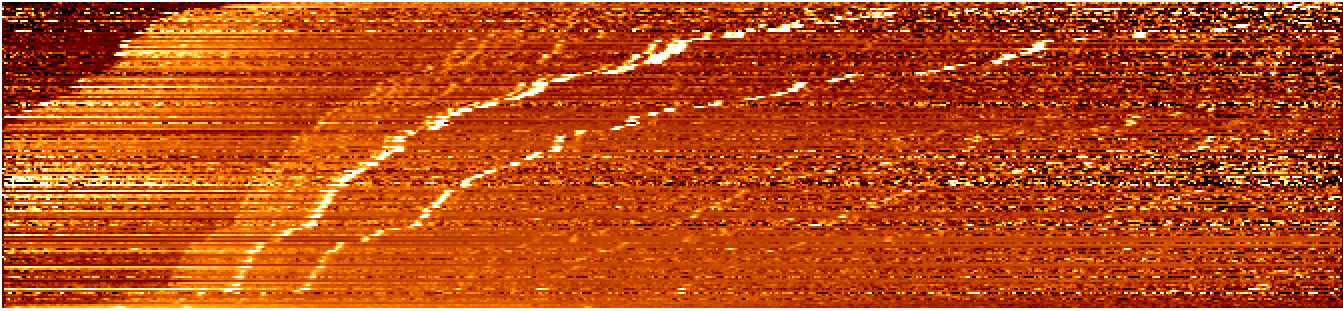}
    \vspace{-0.2cm}
    
    \includegraphics[width=0.98\linewidth]{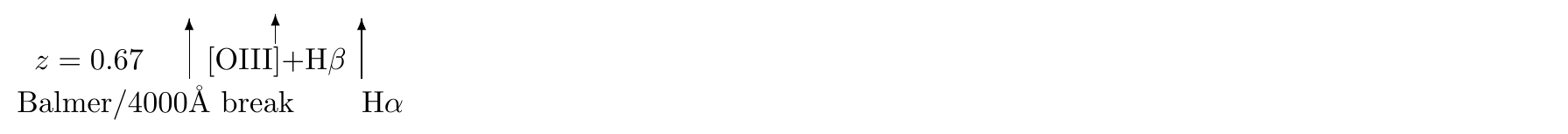}

    \caption{Extracted 1D NIRSpec prism spectra of galaxies with good redshifts in the GOODS-South/HUDF field, ordered by redshift (with highest redshifts at the top). Each spectrum is plotted in flux units of $f_{\lambda}\,\lambda^{1.5}$ (i.e.\  a galaxy with a spectral slope of $\beta=-1.5$ will have constant brightness with wavelength in this plot) and normalised by the mean intensity at $2.3<\lambda<4.45\,\mu$m. Wavelength increases to the right, from $0.6-5.3\,\mu$m. The Lyman-$\alpha$ break and Balmer/4000\,\AA\ break are clearly visible, as are the prominent emission lines H$\alpha$ and the H$\beta$+[OIII]\,4959,5007 complex (which is blended at low redshift but resolved at higher redshift).}
    \label{fig:DeepHSTstack}
\end{figure*}

\begin{acknowledgements}
    We sincerely thank Gabe Brammer for his work aligning HST imaging to the GAIA DR2 frame. In the absence of that work, this data release may well have amounted to 253 spectacularly deep spectra of empty sky. 
    We thank the referee for helpful comments on this manuscript.
    The JADES Collaboration thanks the Instrument Development Teams and the instrument teams at the European Space Agency and the Space Telescope Science Institute for the support that made this program possible. We also thank our program coordinators at STScI for their help in planning complicated parallel observations. We thank all the members of the NIRSpec and NIRCam Instrument Science Teams for making these observations possible.
AJB, AJC, AS, JC, GCJ, IW acknowledge funding from the ``FirstGalaxies" Advanced Grant from the European Research Council (ERC) under the European Union’s Horizon 2020 research and innovation programme (Grant agreement No. 789056). ECL acknowledges support of an STFC Webb Fellowship (ST/W001438/1). The Cosmic Dawn Center (DAWN) is funded by the Danish National Research Foundation under grant no.140. SC acknowledges support by European Union’s HE ERC Starting Grant No. 101040227 - WINGS. RM, JW, FDE, TJL, WB, LS, JS acknowledges support by the Science and Technology Facilities Council (STFC) and by the ERC through Advanced Grant 695671 ``QUENCH".  JW also acknowledges support from the Fondation MERAC. RS acknowledges support from a STFC Ernest Rutherford Fellowship (ST/S004831/1). SA, BRP acknowledges support from Grant PID2021-127718NB-I00 funded by the Spanish Ministry of Science and Innovation/State Agency of Research (MICIN/AEI/ 10.13039/501100011033). RB acknowledges support from an STFC Ernest Rutherford Fellowship [grant number ST/T003596/1]. This research is supported in part by the Australian Research Council Centre of Excellence for All Sky Astrophysics in 3 Dimensions (ASTRO 3D), through project number CE170100013. EE, BJD, MR, FS acknowledges the JWST/NIRCam contract to the University of Arizona NAS5-02015. DJE is supported as a Simons Investigator and by JWST/NIRCam contract to the University of Arizona, NAS5-02015. RH acknowledges funding provided by the Johns Hopkins University, Institute for Data Intensive Engineering and Science (IDIES). REH acknowledges acknowledges support from the National Science Foundation Graduate Research Fellowship Program under Grant No. DGE-1746060. MP acknowledges support from the research project PID2021-127718NB-I00 of the Spanish Ministry of Science and Innovation/State Agency of Research (MICIN/AEI/ 10.13039/501100011033), and the Programa Atracci\'on de Talento de la Comunidad de Madrid via grant 2018-T2/TIC-11715. BER acknowledges support from the NIRCam Science Team contract to the University of Arizona, NAS5-02015. The research of CCW is supported by NOIRLab, which is managed by the Association of Universities for Research in Astronomy (AURA) under a cooperative agreement with the National Science Foundation. CW is supported by the National Science Foundation through the Graduate Research Fellowship Program funded by Grant Award No. DGE-1746060.

This study made use of the Prospero high performance computing facility at Liverpool John Moores University.  This work was performed using resources provided by the Cambridge Service for Data Driven Discovery (CSD3) operated by the University of Cambridge Research Computing Service (www.csd3.cam.ac.uk), provided by Dell EMC and Intel using Tier-2 funding from the Engineering and Physical Sciences Research Council (capital grant EP/T022159/1), and DiRAC funding from the Science and Technology Facilities Council (www.dirac.ac.uk). The authors acknowledge use of the lux supercomputer at UC Santa Cruz, funded by NSF MRI grant AST 1828315.
\end{acknowledgements}

%
%

\bibliographystyle{aa}
\bibliography{paper}

\newpage

\begin{appendix}

\section{Astrometry of HST-based targets}
\label{sub:astrometry}

Astrometric offsets have previously been noted between previous reductions of HST imaging and other datasets registered to the GAIA DR2 astrometric frame \citep{Dunlop2017, Franco2018, Whitaker2019}. 
These studies have corrected the astrometry of CANDELS and 3DHST onto the Gaia DR2 frame \citep{GaiaMission, GaiaDR2} with a bulk offset of 0.26 arcsec.

We used the Complete {\em Hubble} Archive for Galaxy Evolution (CHArGE) re-reduction of the HST imaging in GOODS-S, where all input frames have been carefully registered to the GAIA DR2 astrometric frame \citep{Kokorev2022_CHARGE, grizli}\footnote{\url{https://s3.amazonaws.com/grizli-stsci/Mosaics/index.html}}.
In order to attain the astrometric accuracy required to ensure light is captured by an MSA shutter, we revisited the astrometry of the literature photometric candidates.
For the large catalogues from 3DHST and UVUDF, and for the Lyman break surveys of \cite{Finkelstein2015} and \cite{Harikane2016}, we matched the reported position of all brighter than F160W$<26$ to the GAIA-DR2-registered catalogy=ue using $0\farcs 5$  tolerance.
Based on this, we identified that, in addition to the bulk offset previously identified, there was also a plate scale difference amounting to about 1 part in 5000.
Across the full GOODS-S field ($10'\times 15'$ in size) this amounts to systematic errors larger than the width of a micro-shutter, highlighting the need for correcting this effect.

We fit a simple tangent-plane astrometric transformation allowing for a bulk offset, plate scale, and rotation to these offsets. We favoured the approach of fitting an astrometric transformation to the coordinates over re-measuring the centroids of objects in the CHArGE images because this new reduction had 100 mas drizzled pixels, and was not optimised for the selection of $z\gtrsim6$ targets in the HUDF, where the HST data were deepest. Thus, many of of the targets that ended up in our highest priority classes were not clearly detected in these reductions.

Fitting this simple transformation to the \citet{Skelton2014_3dhst} catalogue, we found that we reduced the residual RMS scatter on the positional offsets to 34 mas across the entire GOODS-South field. This corresponds to 17\% of the illuminated NIRSpec slit width of $0\farcs2$.

Given many of our highest priority targets were taken from Lyman-break catalogues, and did not necessarily appear in \citet{Skelton2014_3dhst}, we also constructed separate astrometric transformations for catalogues from \citet{Bouwens2015, Bouwens2021}, \citet{Finkelstein2015} and \citet{Harikane2016}.
We found that very similar astrometric offsets were present in these catalogues, however the exact magnitude of each component of the correction varied slightly.
The residuals on transformed coordinates were similarly $\sim 30-50$ mas after applying the relevant transformation.

We considered the corrections applied to \citet{Skelton2014_3dhst} to be the most robust, since that catalogue had more entries than the Lyman-break catalogues. Thus, to obtain updated `GAIA DR2' coordinates, we used the correction derived from \citet{Skelton2014_3dhst} for all targets that had a counterpart in this catalogue.

In some cases, high-priority targets were not matched to a counterpart in the \citet{Skelton2014_3dhst} catalogue, in which case we used a correction from the astrometric fit to one of the catalogues of \citet{Finkelstein2015}, \citet{Harikane2016} or \citet{Bouwens2021} to obtain updated coordinates.

There were some cases where we placed high-priority targets that were not in one of the catalogues discussed above (e.g.\ \citealt{Bouwens2011b} and the $z\sim10$ candidate from \citealt{McLure2013}). In these case we remeasured the centroid using the latest HLF ({\em Hubble} Legacy Field) v2.0 reduction of the GOODS-South images, which we assessed  as having astrometry in sufficiently good agreement with the GAIA DR2 frame.

Finally, we retained some targets in our catalogue for which we did not have a reliable conversion of the reported coordinates to the GAIA DR2 frame. However, these targets were flagged such that they could not appear any higher than Priority Class 9, and were only placed at the expense of extra sky shutters (Table~\ref{tab:priorities}).

\section{Caveats of early data release}
\label{section:caveats}

At the time of target selection two weeks after the images were obtained, several issues present in the image reduction and analysis may have affected the prioritisation of the targets.  The flat fields that were available at the time introduced spurious small-scale structure in the background at faint levels in some areas, leading to a large number of spurious detections close to the detection limit in those areas.  Improvements to the flat fields since then will allow us to push to deeper limits in future targeting.  In particular, Class 3 was designed to include fainter, less secure high redshift targets, but we retained those allocated from the HST pre-selection, and did not supplement with JWST-based sources fainter than $AB=29.5$\,mag.
Improvements in the flat fielding, in the small and large scale background subtraction including wisps, and in the object deblending, especially near large bright sources, means that the current photometric measurements and source positions may differ from the very early estimates available at the time of target selection. In particular the Kron-based aperture measurements that were used for the flux cut in Class 7, which was designed to approach a total magnitude cut, were significantly impacted by these improvements.

\section{Slit Overlays of Targets}
\label{sec:postage_stamps}

\begin{figure*}
    \centering
    \includegraphics*[width=0.95\textwidth]{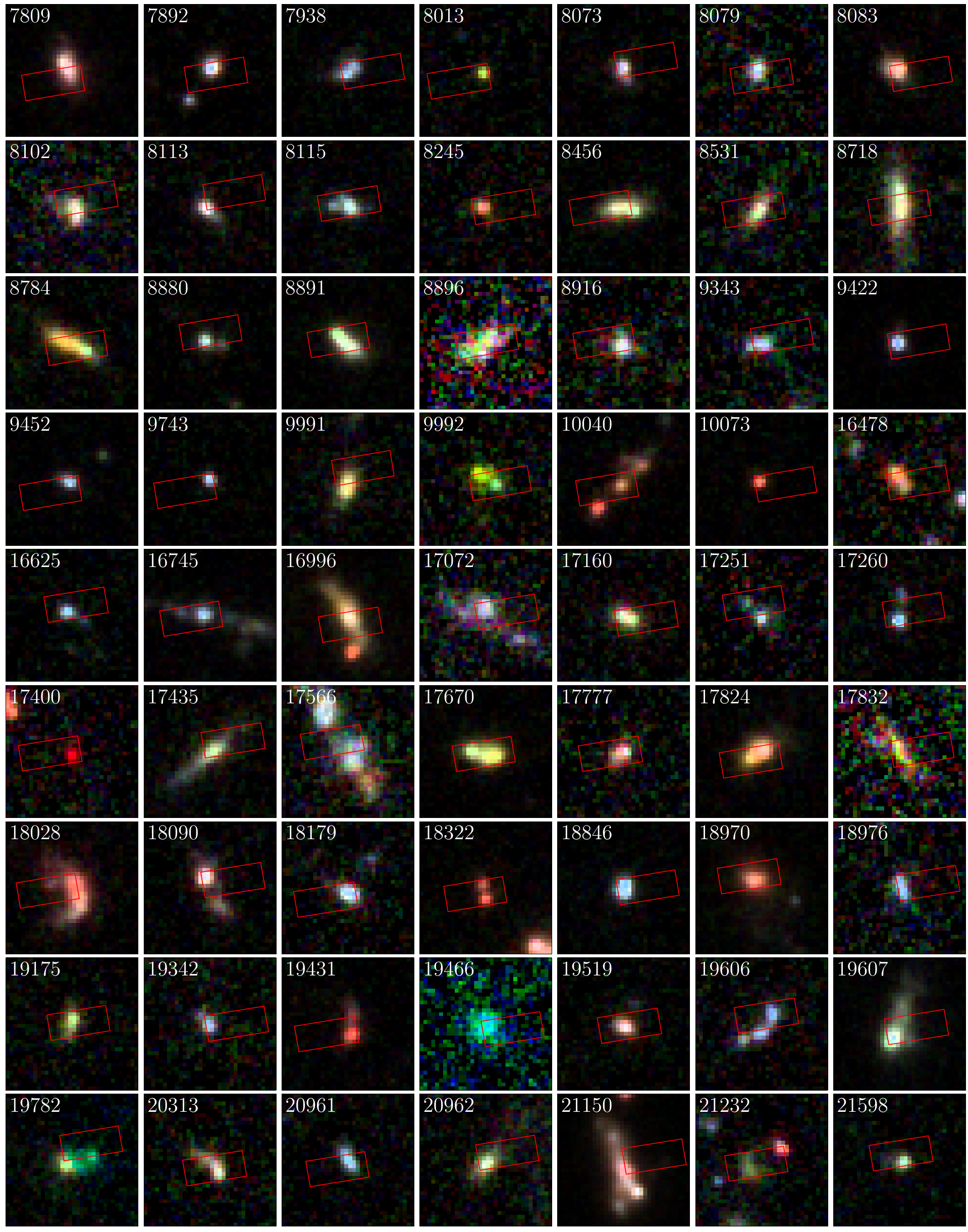}
    \caption{Overlay of target shutter positions onto the images for target IDs 7809--21598 sorted by NIRSpec ID number, starting at the top left. The illuminated shutter regions are outlined ($0\farcs46\times 0\farcs20$).
     The image is derived from the JWST/NIRCam F115W/F150W/F200W images from JADES (blue/green/red channels).
     The individual images are $1\farcs 0$ on a side, and are centred on the input coordinate of the target. North is up and East is to the left. 
}
    \label{fig:postage_mosaic1}
\end{figure*}

\begin{figure*}
    \centering
    \includegraphics*[width=0.95\textwidth]{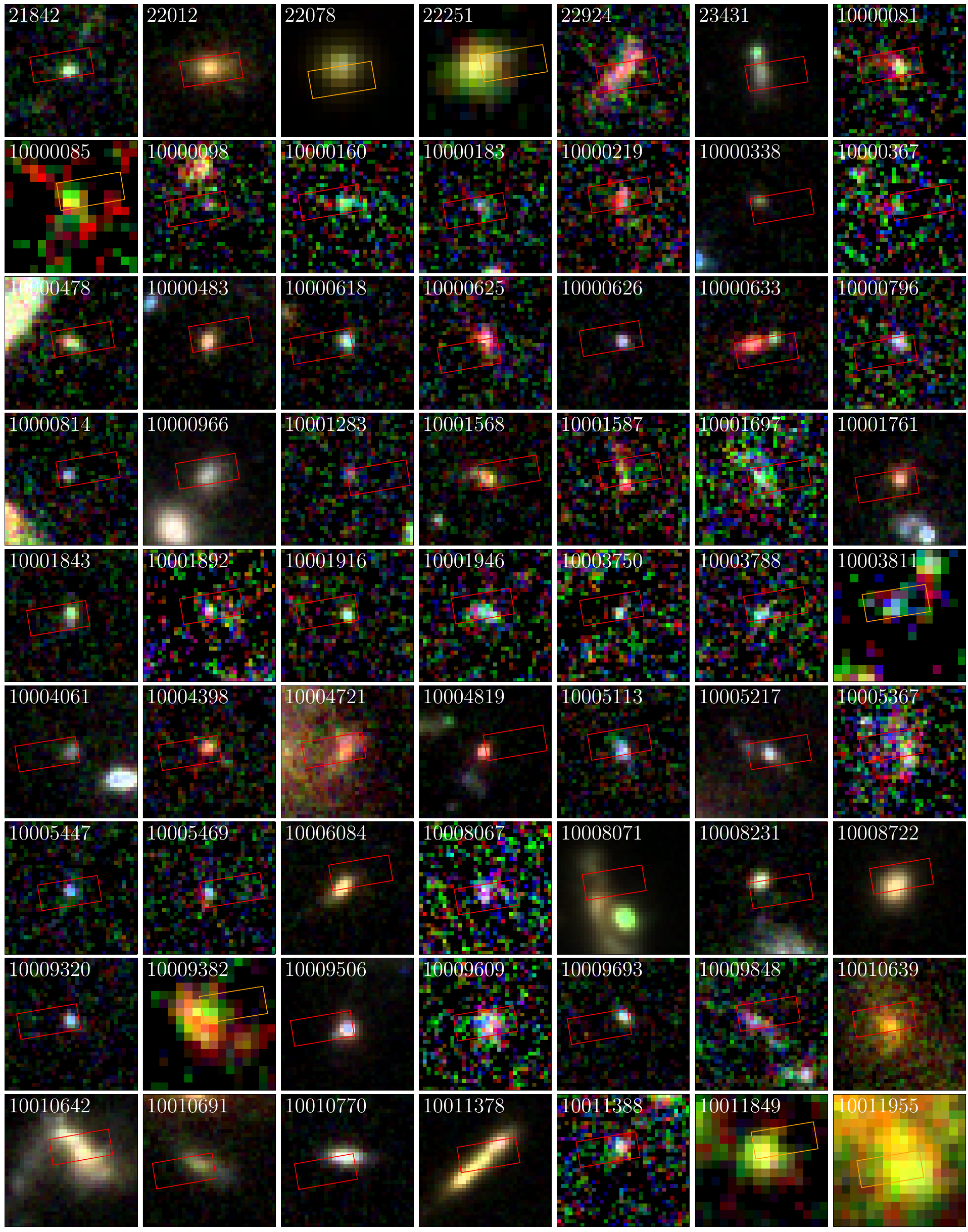}
    \caption{Overlay of target shutter positions onto the images for target IDs 21842--10011955 sorted by NIRSpec ID number, starting at the top left. The illuminated shutter regions are outlined ($0\farcs46\times 0\farcs20$). A red 
    outline indicates that the image is derived from the JWST/NIRCam F115W/F150W/F200W images from JADES (blue/green/red channels), and an orange outline denotes HST ACS-F850LP/WFC3-F125W/WFC3-F160W images. The individual images are $1\farcs 0$ on a side, and are centred on the input coordinate of the target. North is up and East is to the left. 
}
    \label{fig:postage_mosaic2}
\end{figure*}

\begin{figure*}
    \centering
    \includegraphics*[width=0.95\textwidth]{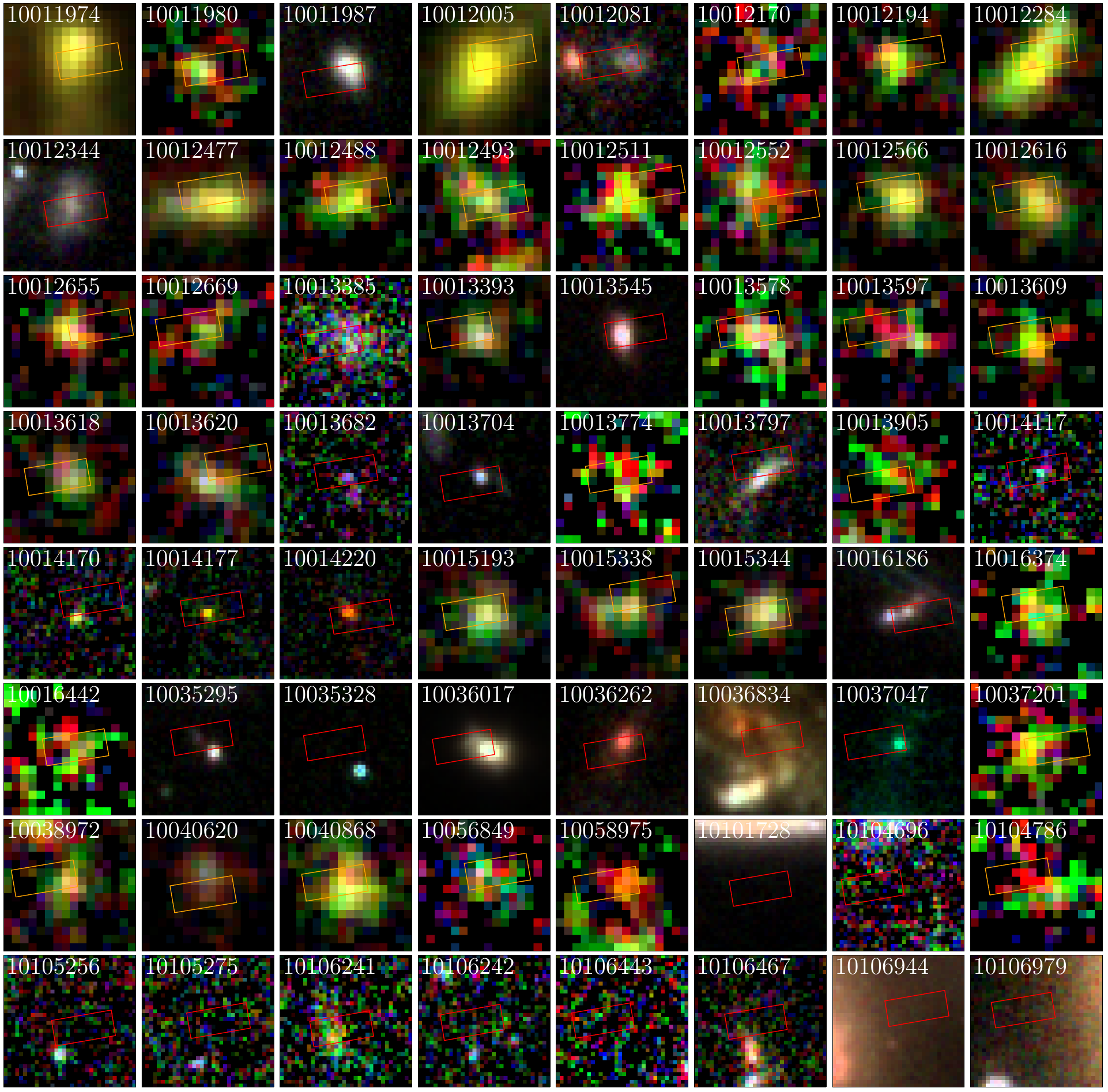}
    \caption{Overlay of target shutter positions onto the images for targets ID 10011974--10106979 sorted by NIRSpec ID number, starting at the top left. The illuminated shutter regions are outlined ($0\farcs46\times 0\farcs20$). A red 
    outline indicates that the image is derived from the JWST/NIRCam F115W/F150W/F200W images from JADES (blue/green/red channels), and an orange outline denotes HST ACS-F850LP/WFC3-F125W/WFC3-F160W images. The individual images are $1\farcs 0$ on a side, and are centred on the input coordinate of the target. North is up and East is to the left. 
}
    \label{fig:postage_mosaic3}
\end{figure*}

Figures~\ref{fig:postage_mosaic1}~-\ref{fig:postage_mosaic3}, continued from Figure~\ref{fig:postage_mosaic}, showing the positions MSA shutter overlaid on observed targtes.

\newpage

\section{Layout of Shutters in the HUDF/GOODS-South Field}
\label{sec:MSAquads}

\begin{figure*}
    \centering
    \includegraphics*[width=0.9\textwidth]{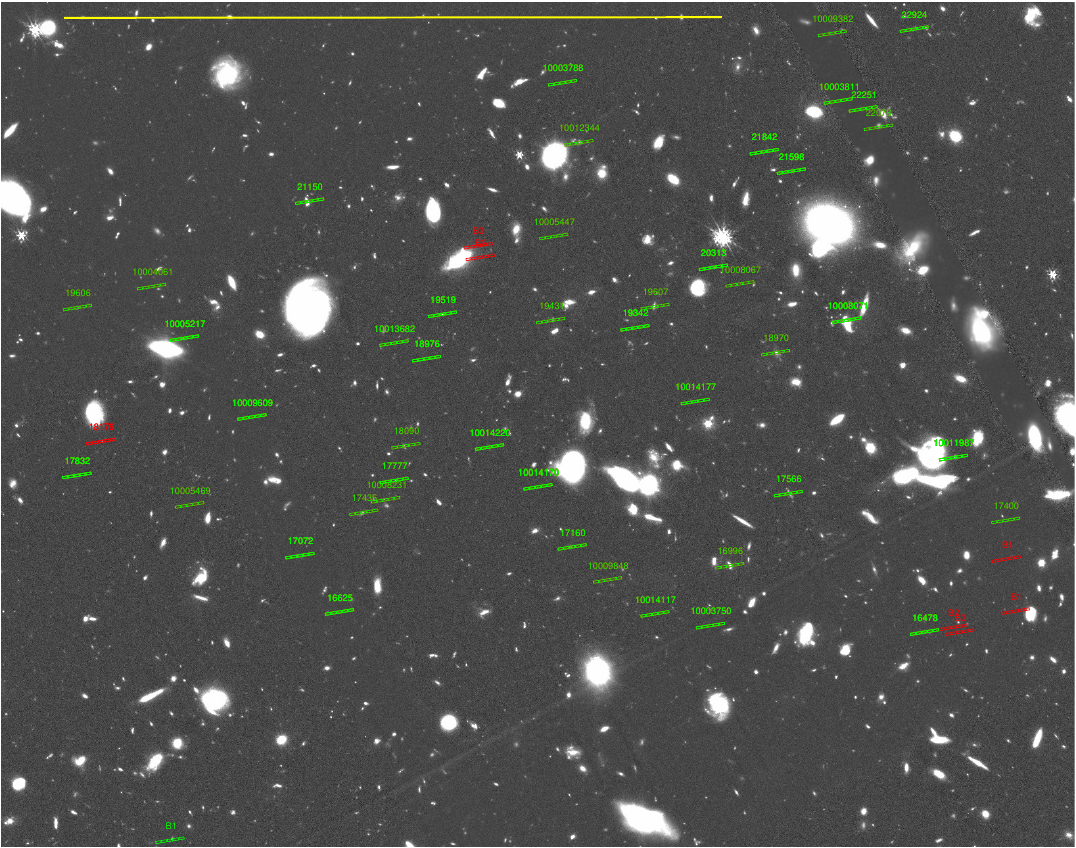}
    \caption{Quadrant 1 of the MSA, showing allocation of micro-shutters to targets. Those in green are covered by both the grating configurations and the low-dispersion prism. The red shutters are open only in the prism observations, as they would lead to overlapping spectra for our high priority targets in the grating configuration. Three micro-shutters are opened for each target, but the nodding by $\pm 1$ shutter means that spectra are obtained over the areas covered by five shutters (including background) which are displayed. The field displayed is  the NIRCam F200W image. The yellow scale bar denotes 1 arcmin. North is up and East is to the left. Shutters with the prefix `B' are empty sky background.}
    \label{fig:quad1}
\end{figure*}

\begin{figure*}
    \centering
    \includegraphics*[width=0.9\textwidth]{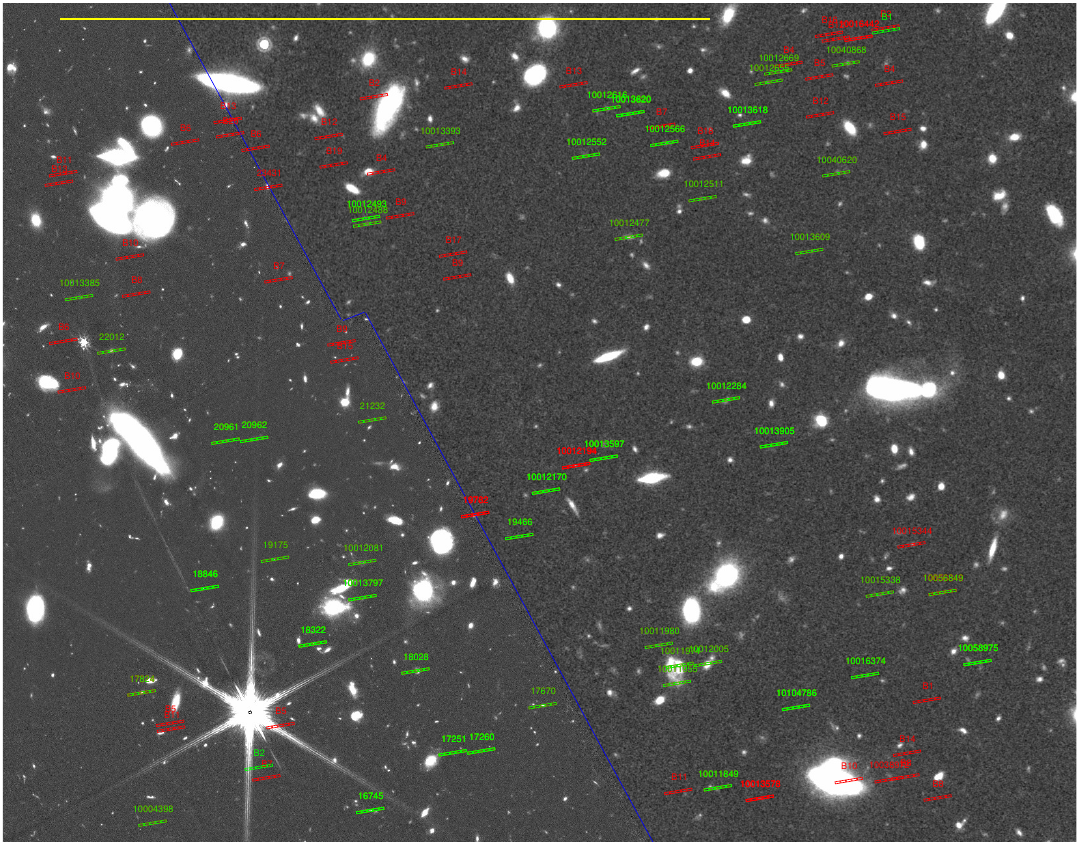}
    \caption{Quadrant 2 of the MSA, showing allocation of micro-shutters to targets. Those in green are covered by both the grating configurations and the low-dispersion prism. The red shutters are open only in the prism observations, as they would lead to overlapping spectra for our high priority targets in the grating configuration. Three micro-shutters are opened for each target, but the nodding by $\pm 1$ shutter means that spectra are obtained over the areas covered by five shutters (including background) which are displayed. North is up and East is to the left. The field displayed is the NIRCam F200W image, except for the area West of the blue line which had not yet been imaged by NIRCam and we show the HST F160W image. The yellow scale bar denotes 1 arcmin.  Shutters with the prefix `B' are empty sky background.}
    \label{fig:quad2}
\end{figure*}

\begin{figure*}
    \centering
    \includegraphics*[width=0.98\textwidth]{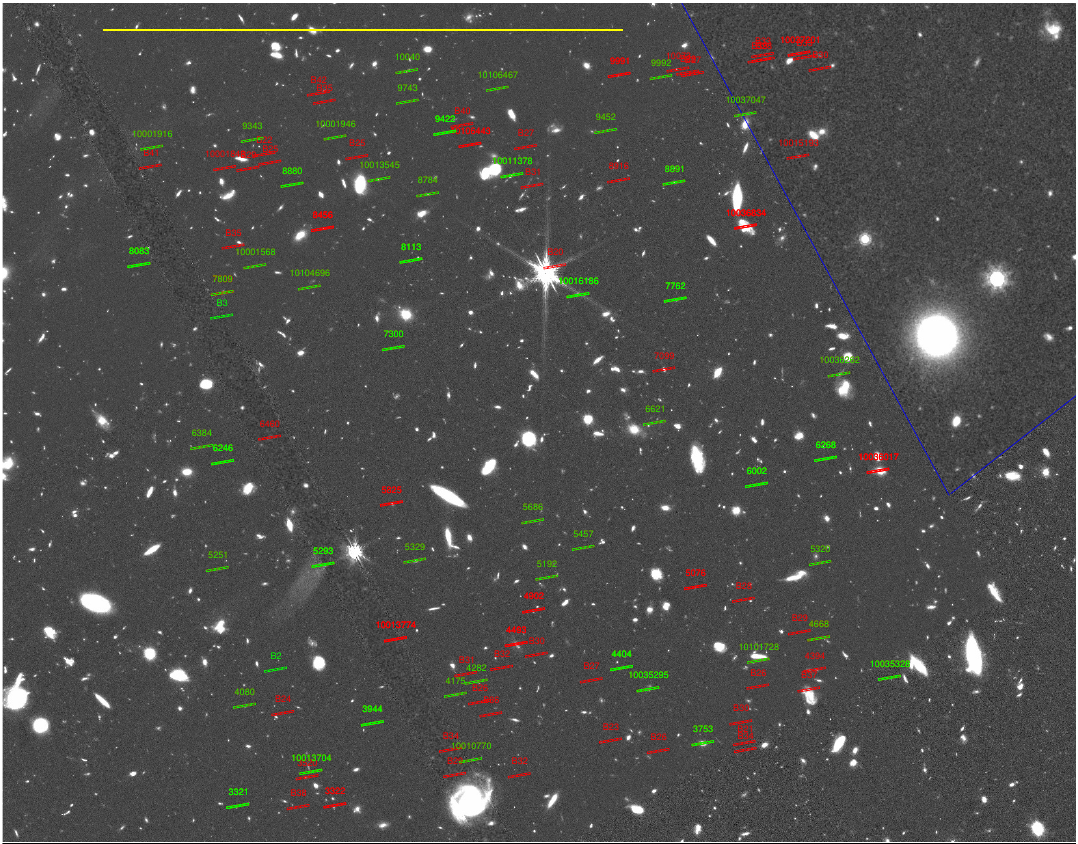}
    \caption{Quadrant 4 of the MSA, showing allocation of micro-shutters to targets. Those in green are covered by both the grating configurations and the low-dispersion prism. The red shutters are open only in the prism observations, as they would lead to overlapping spectra for our high priority targets in the grating configuration. Three micro-shutters are opened for each target, but the nodding by $\pm 1$ shutter means that spectra are obtained over the areas covered by five shutters (including background) which are displayed. North is up and East is to the left. The field displayed is the NIRCam F200W image, except for the area North-West of the blue line which had not yet been imaged by NIRCam and we show the HST F160W image. The yellow scale bar denotes 1 arcmin. Shutters with the prefix `B' are empty sky background.}
    \label{fig:quad4}
\end{figure*}

Figures~\ref{fig:quad1}~-\ref{fig:quad4} continue from Figure~\ref{fig:open_shutters}, showing the remaining three quadrants of the MSA positioned on the field.

\FloatBarrier

\clearpage\onecolumn

\section{Example Spectra}
\label{section:example_spectra}

\begin{figure*}[h]
        \centering
    \includegraphics[width=0.85\textwidth]{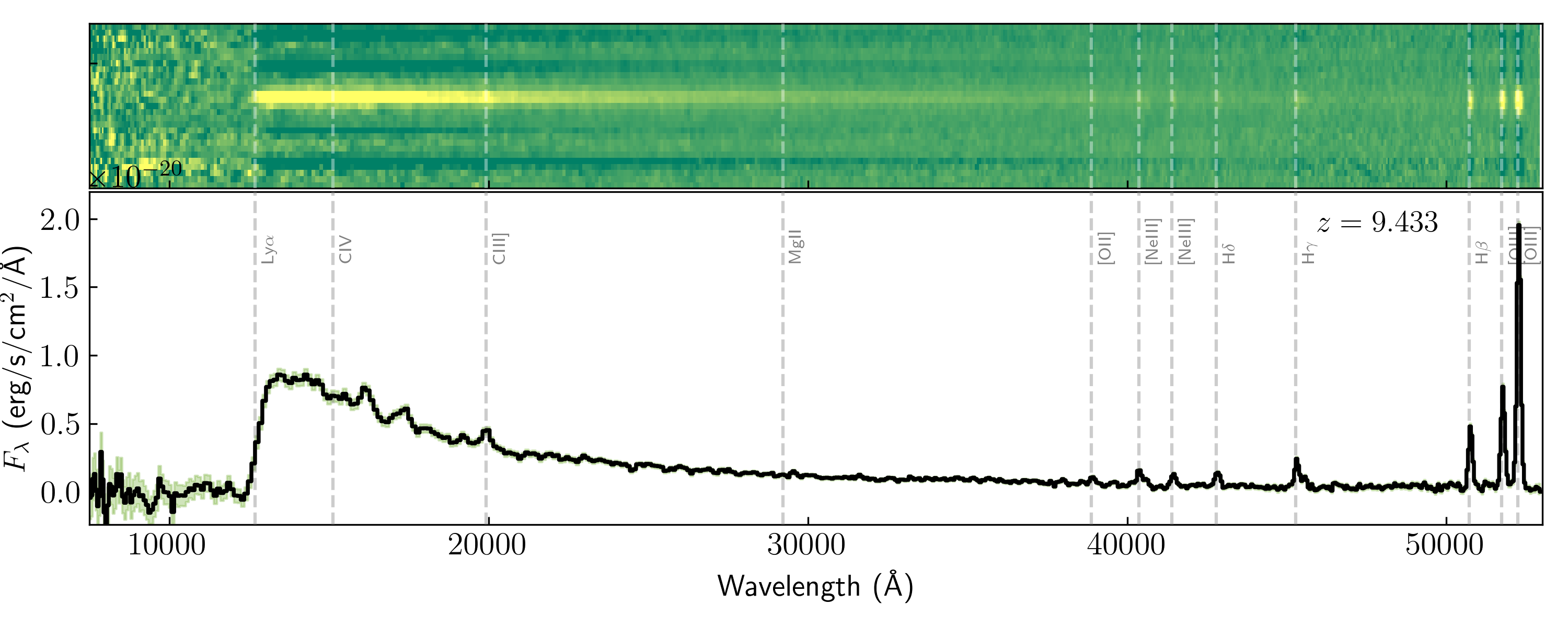}
    \includegraphics[width=0.6\textwidth]{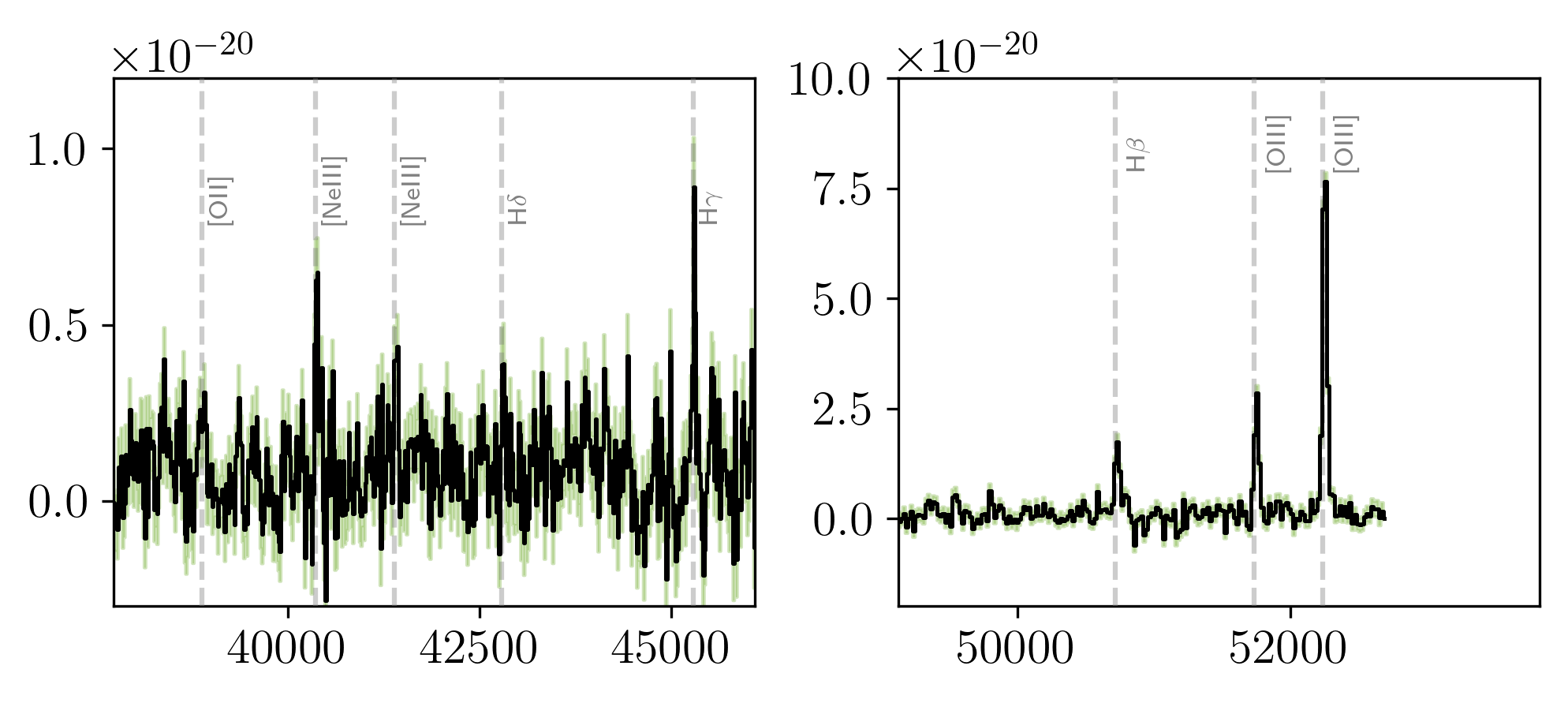}
    \caption{Low-dispersion prism spectra (1D and 2D) of 10058975 at $z=9.4327$, with the medium dispersion grating of prominent lines shown below. Green shaded regions on the 1D spectra denote the $1\sigma$ errors.
    The wavelengths of common emission lines are denoted by vertical lines.
    }
    \label{fig:example_spectra1}
\end{figure*}

\begin{figure*}
        \centering
   \includegraphics*[width=0.85\textwidth]{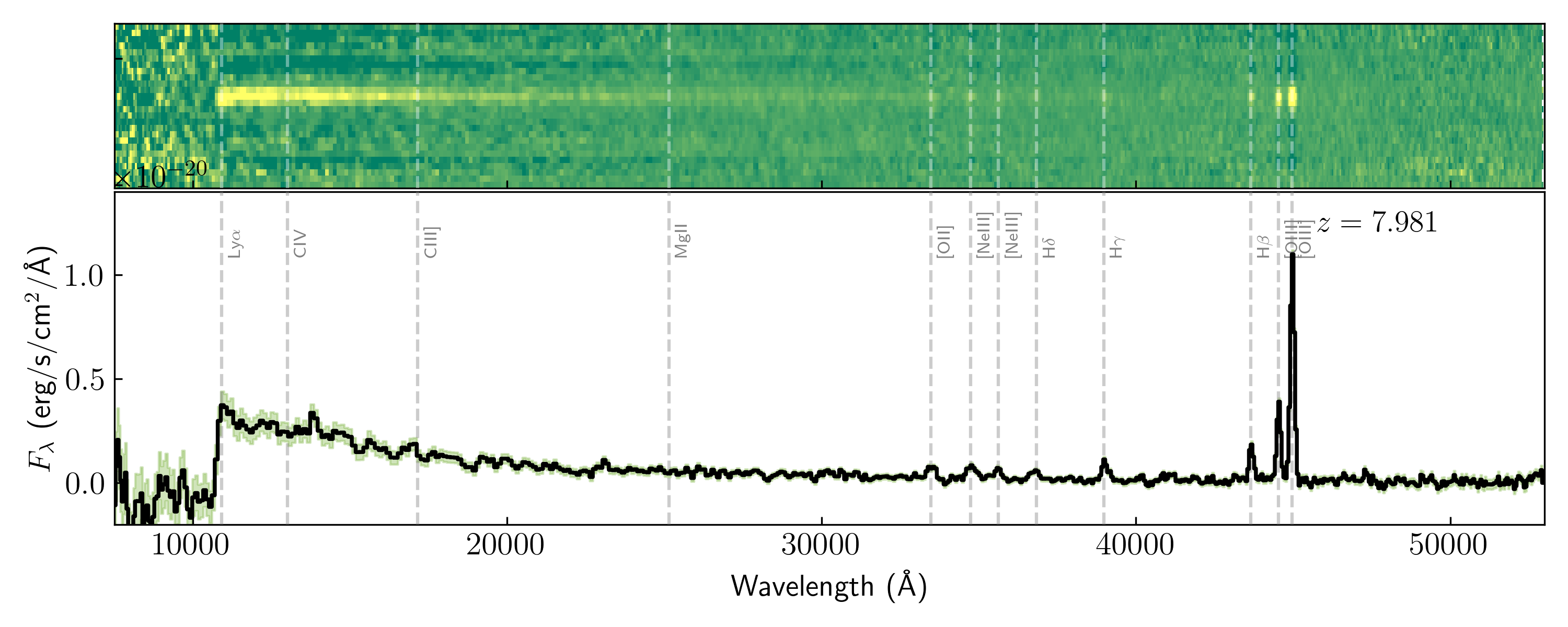}
    \includegraphics*[width=0.6\textwidth]{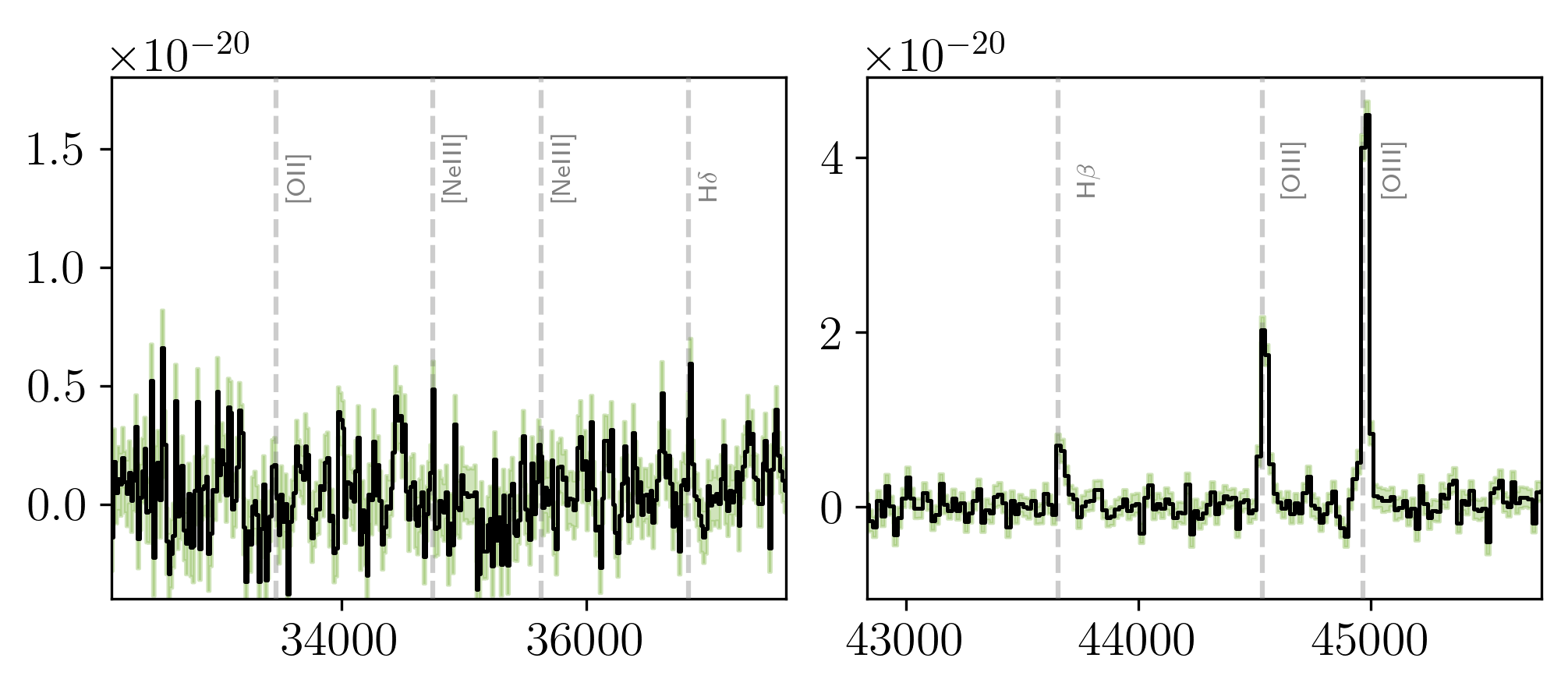}
    \caption{As for Figure~\ref{fig:example_spectra1}, but showing galaxy 021842 at $z=7.9806$.
    }
    \label{fig:example_spectra2}
\end{figure*}

\begin{figure*}
        \centering
   \includegraphics*[width=0.85\textwidth]{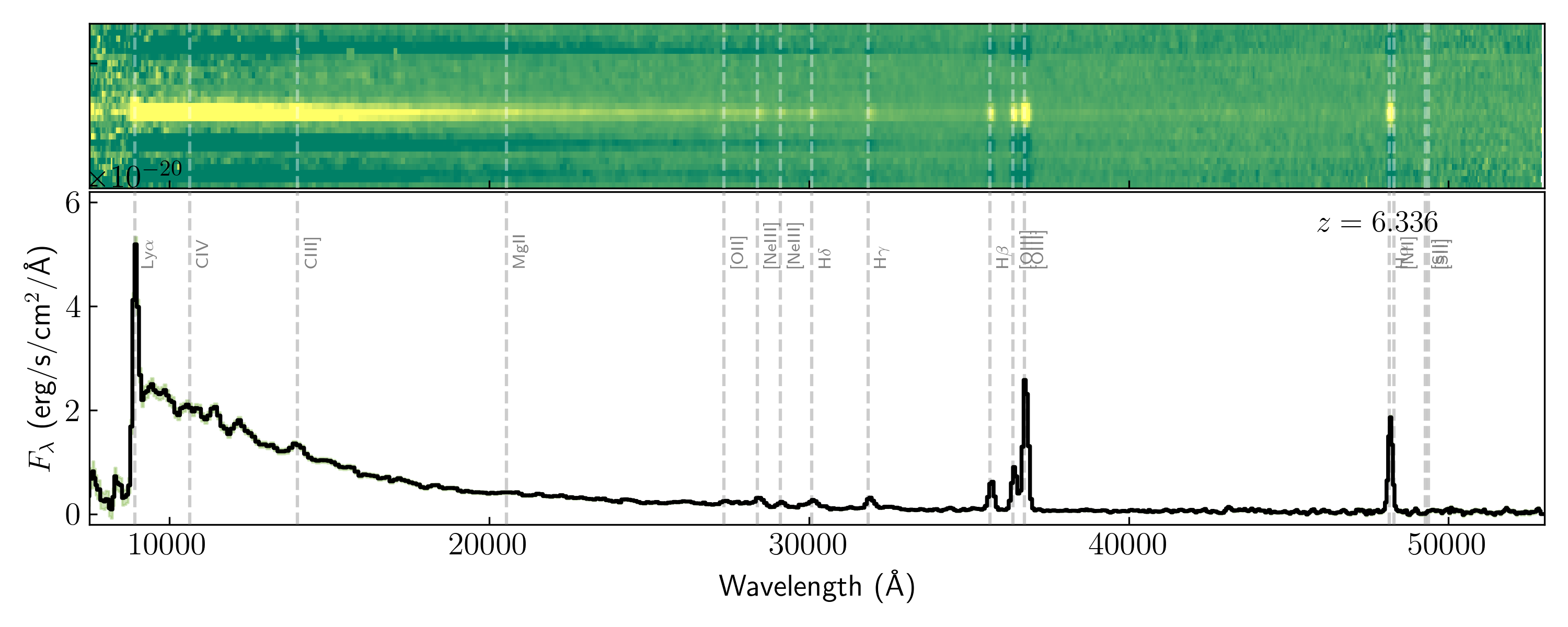}
    \includegraphics*[width=0.9\textwidth]{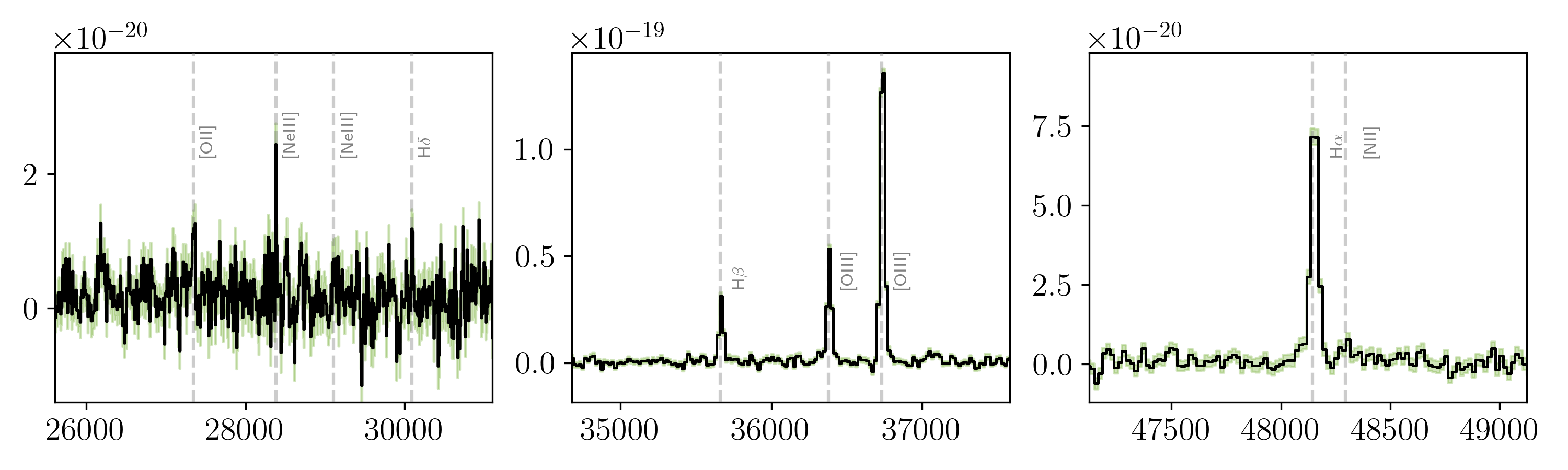}
    \caption{As for Figure~\ref{fig:example_spectra1}, but showing galaxy 018846 at $z=6.33$.
    }
    \label{fig:example_spectra3}
\end{figure*}

\begin{figure*}
        \centering
   \includegraphics*[width=0.85\textwidth]{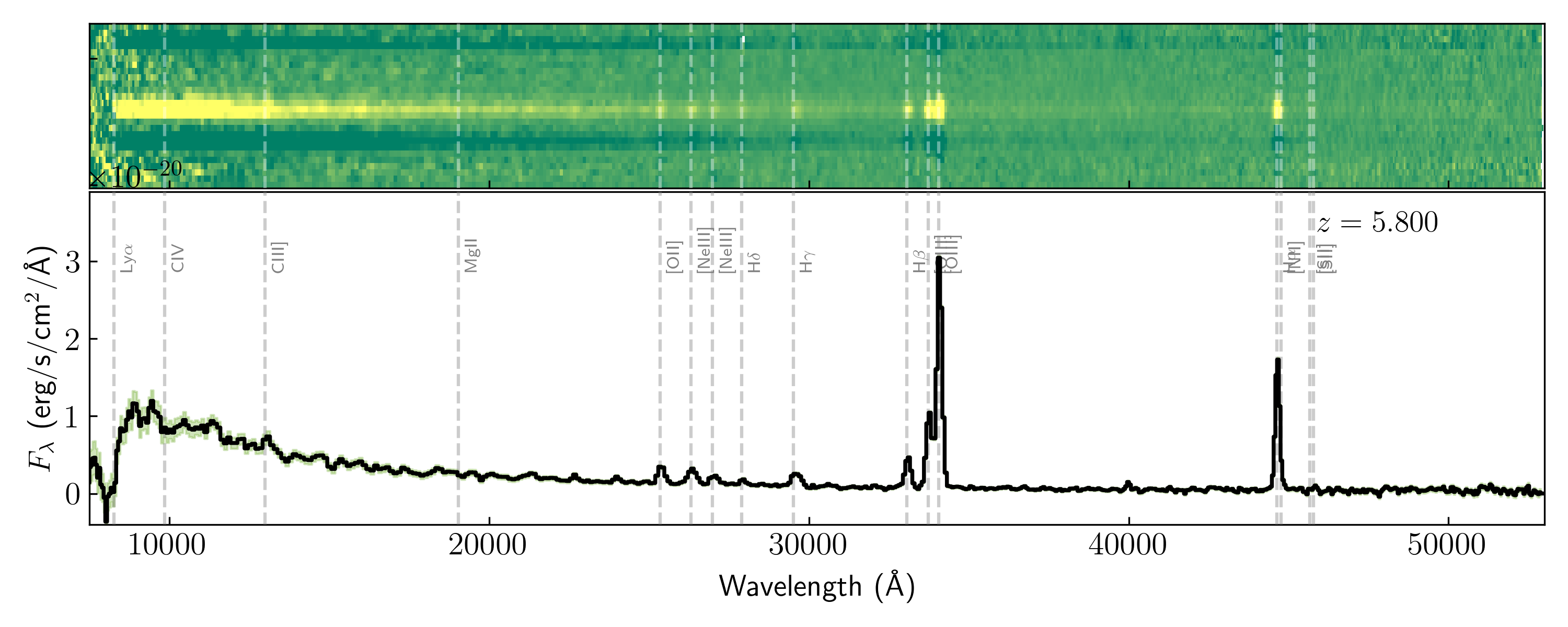}
    \includegraphics*[width=0.9\textwidth]{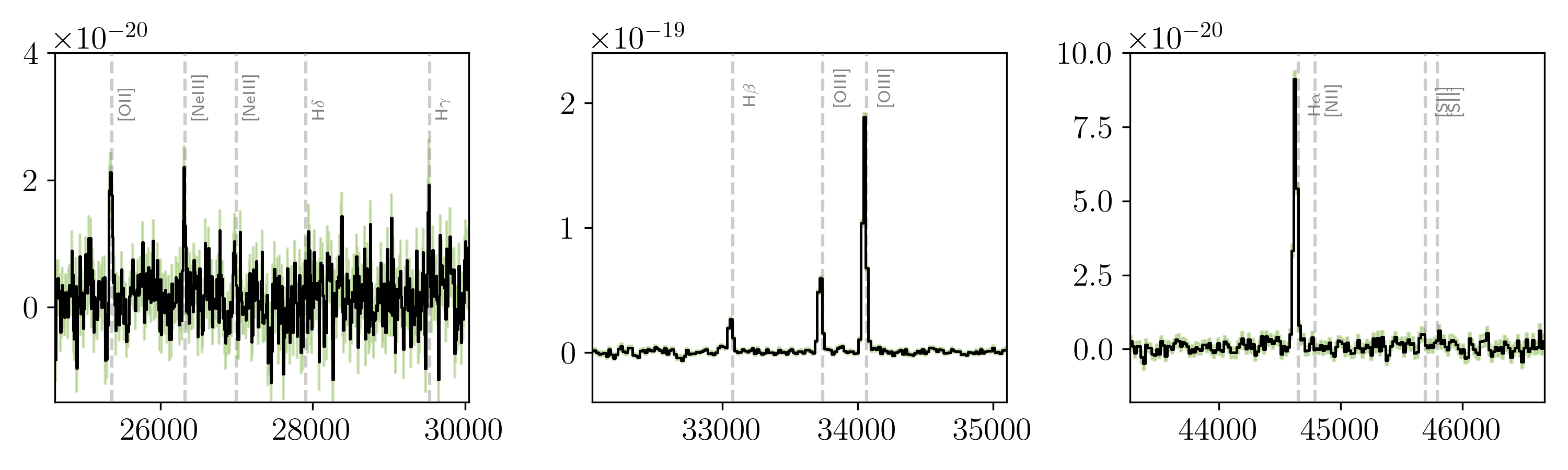}
    \caption{As for Figure~\ref{fig:example_spectra1}, but showing galaxy 022251 at $z=5.79$.
    }
    \label{fig:example_spectra4}
\end{figure*}

\begin{figure*}
        \centering
   \includegraphics*[width=0.85\textwidth]{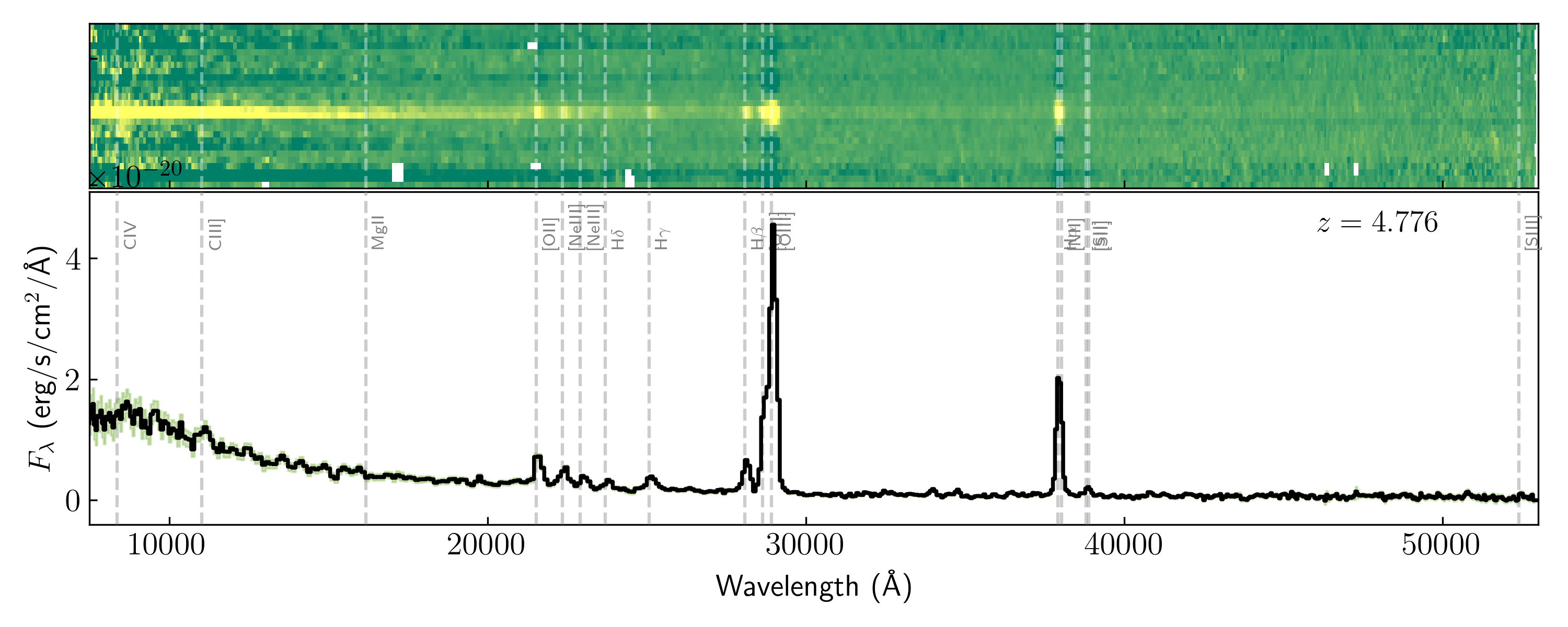}
    \includegraphics*[width=0.9\textwidth]{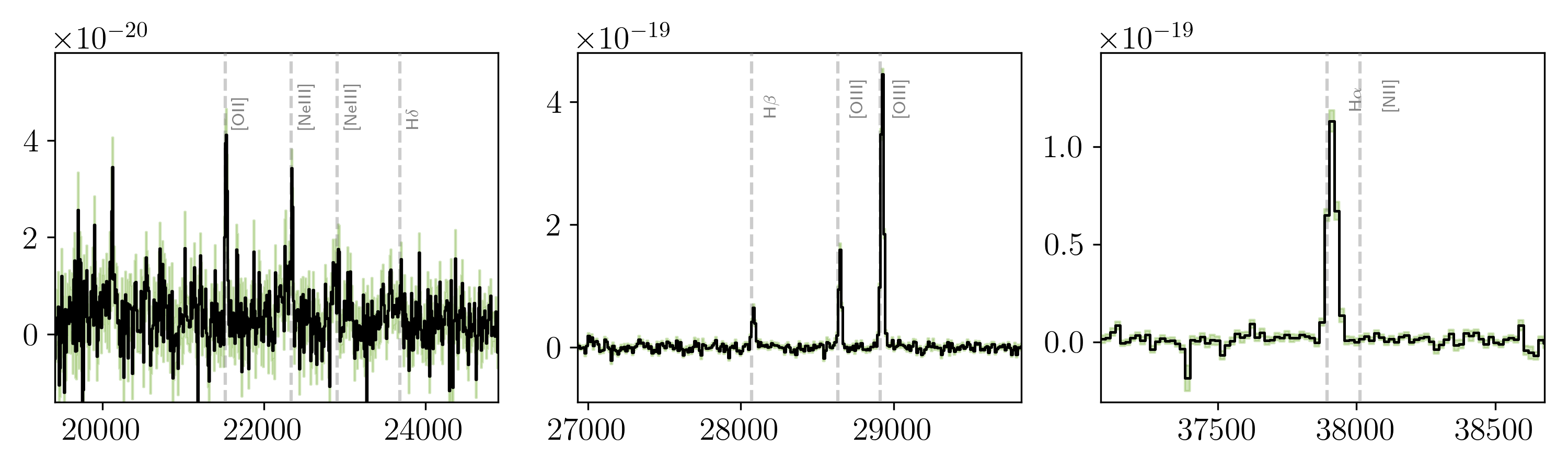}
    \caption{As for Figure~\ref{fig:example_spectra1}, but showing galaxy 018090 at $z=4.77$.
    }
    \label{fig:example_spectra5}
\end{figure*}

\begin{figure*}
        \centering
   \includegraphics*[width=0.85\textwidth]{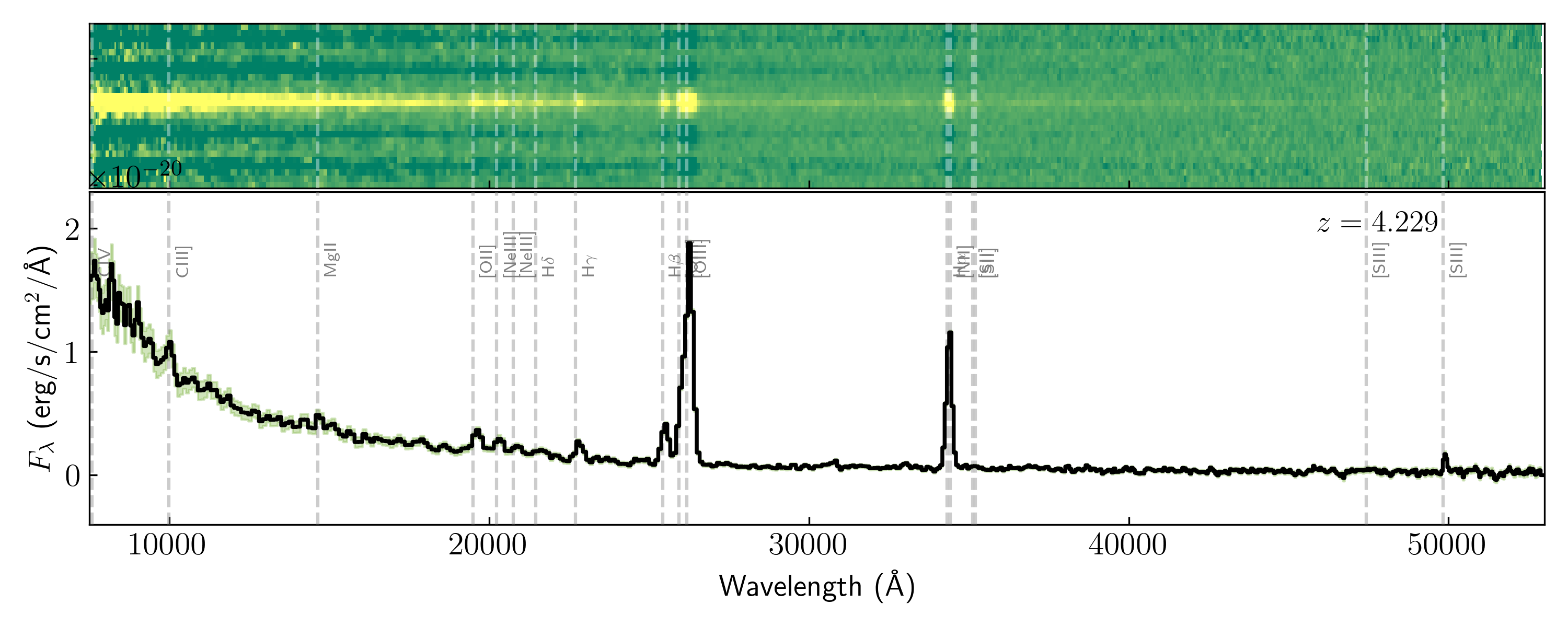}
    \includegraphics*[width=0.9\textwidth]{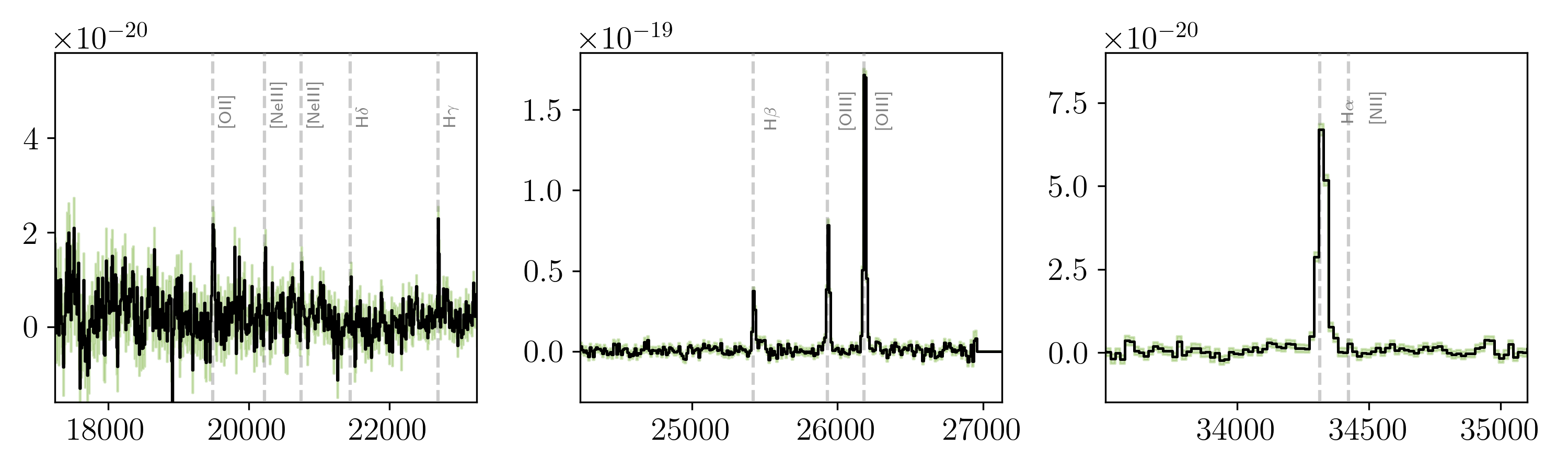}
    \caption{As for Figure~\ref{fig:example_spectra1}, but showing galaxy 007892 at $z=4.2287$. 
    }
    \label{fig:example_spectra6}
\end{figure*}

\begin{figure*}
        \centering
   \includegraphics*[width=0.85\textwidth]{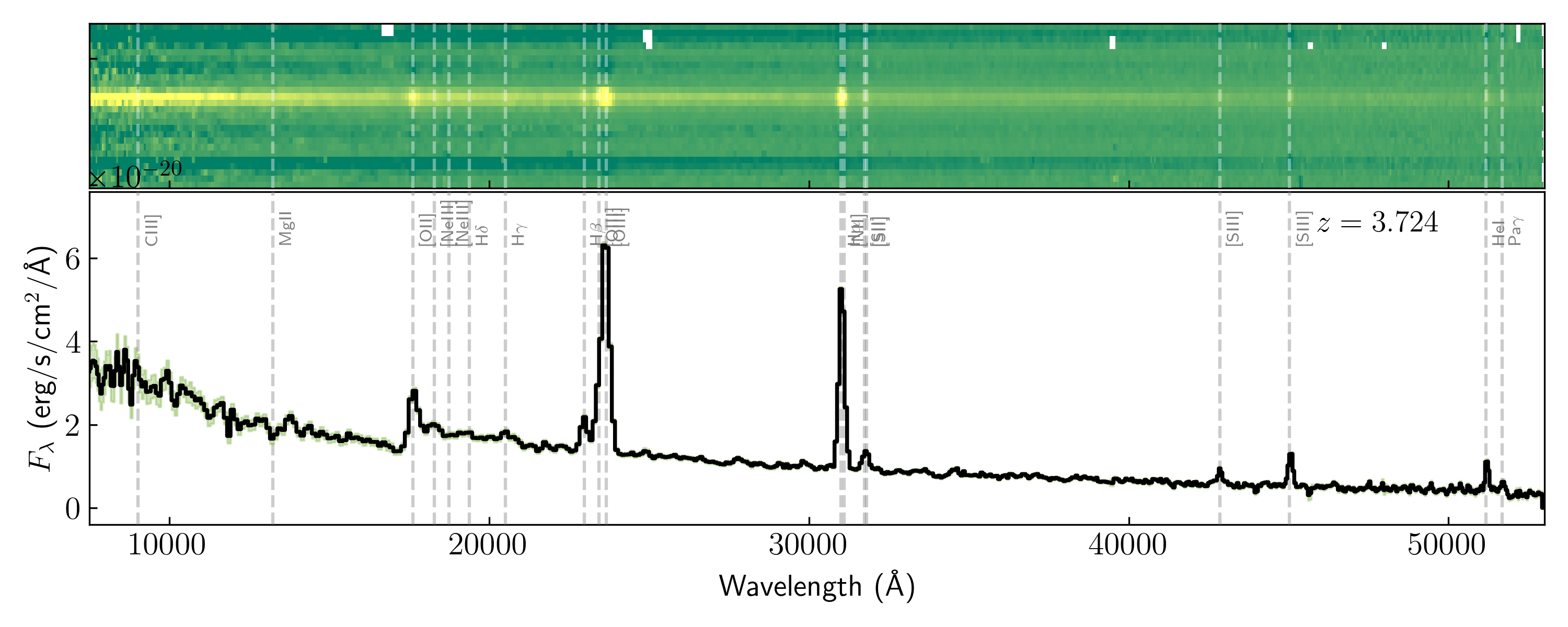}
    \includegraphics*[width=0.9\textwidth]{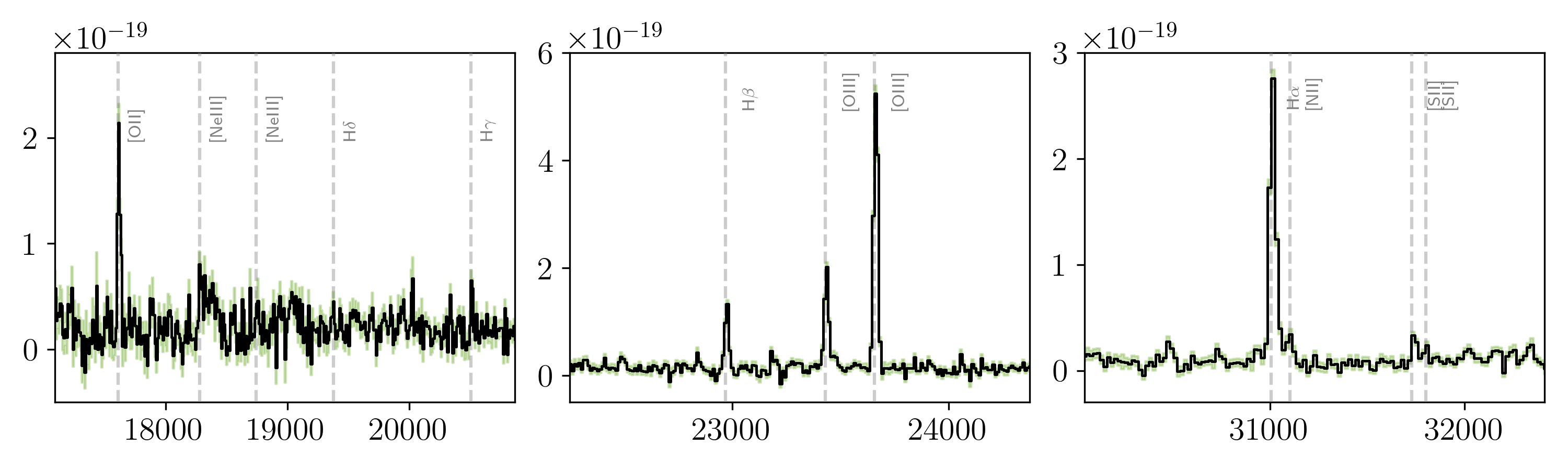}
    \caption{As for Figure~\ref{fig:example_spectra1}, but showing galaxy 018970 at $z=3.7245$. Note the Balmer break just below [OII]\,3727.
    }
    \label{fig:example_spectra7}
\end{figure*}

\begin{figure*}
        \centering
   \includegraphics*[width=0.85\textwidth]{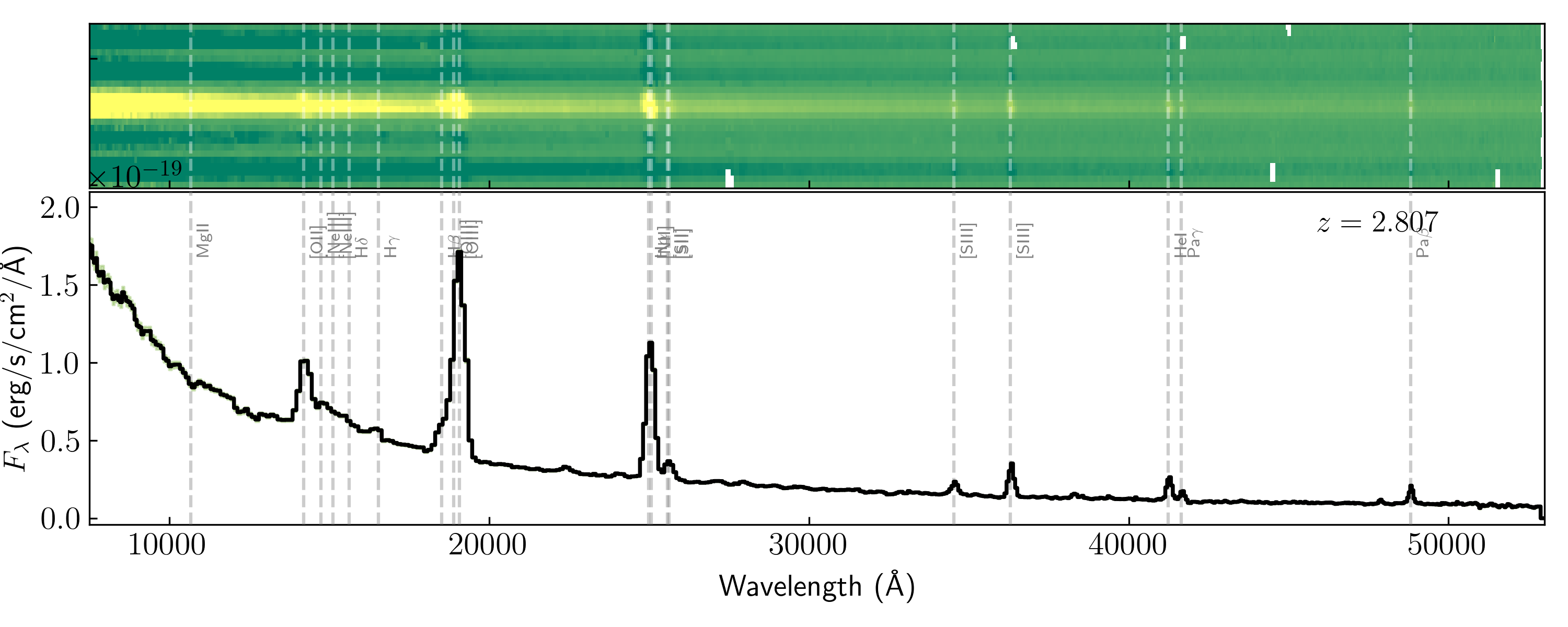}
    \includegraphics*[width=0.9\textwidth]{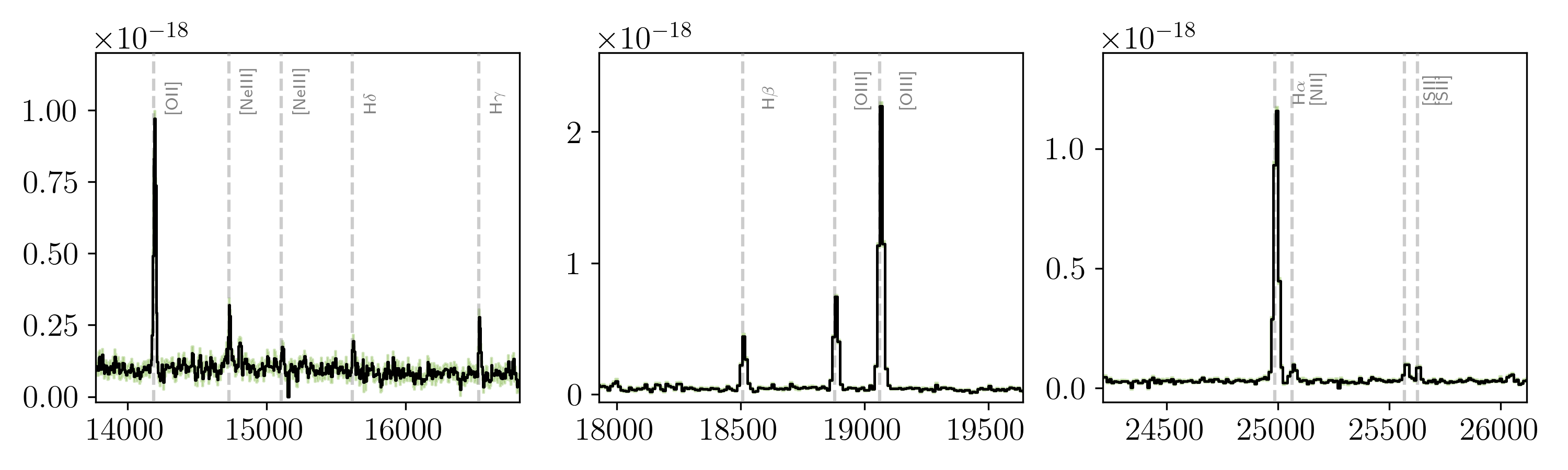}
    \caption{As for Figure~\ref{fig:example_spectra1}, but showing galaxy 003892 at $z=2.8072$. Note the Balmer break just below [OII]\,3727.
    }
    \label{fig:example_spectra8}
\end{figure*}

\begin{figure*}
        \centering
   \includegraphics*[width=0.85\textwidth]{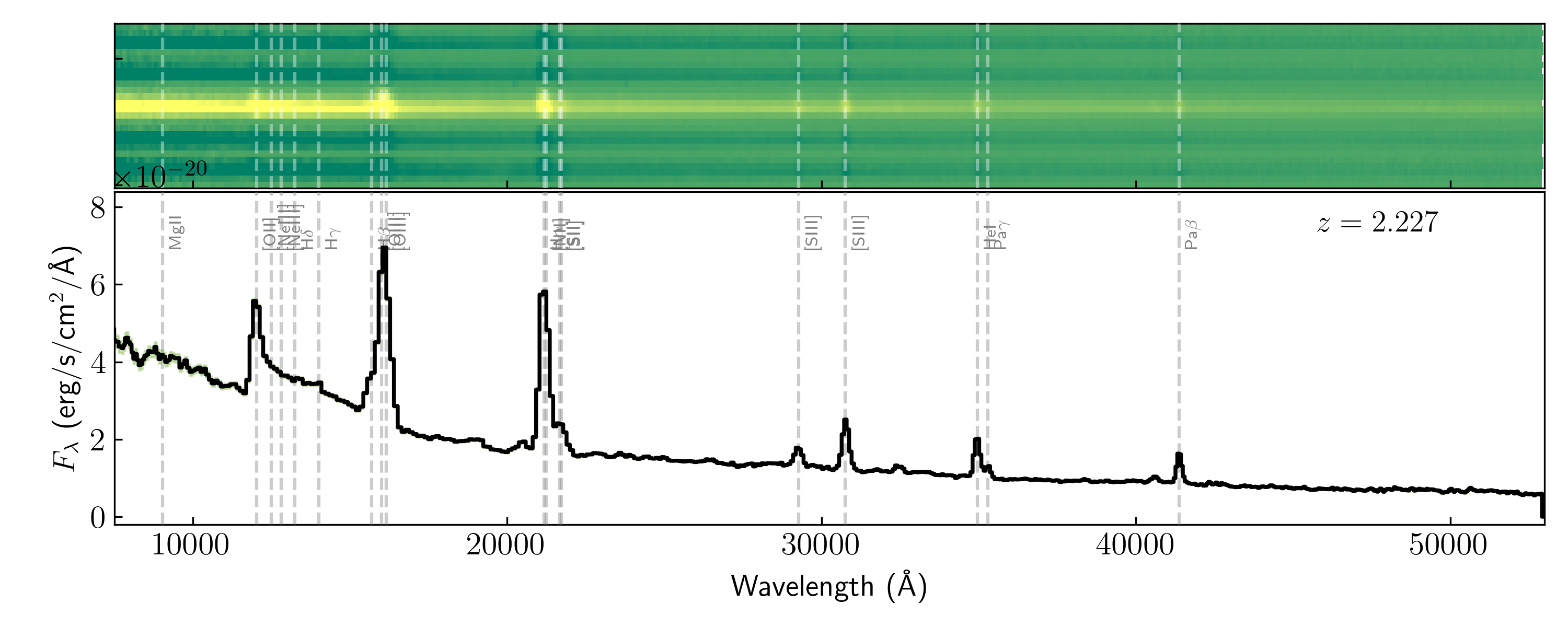}
    \includegraphics*[width=0.9\textwidth]{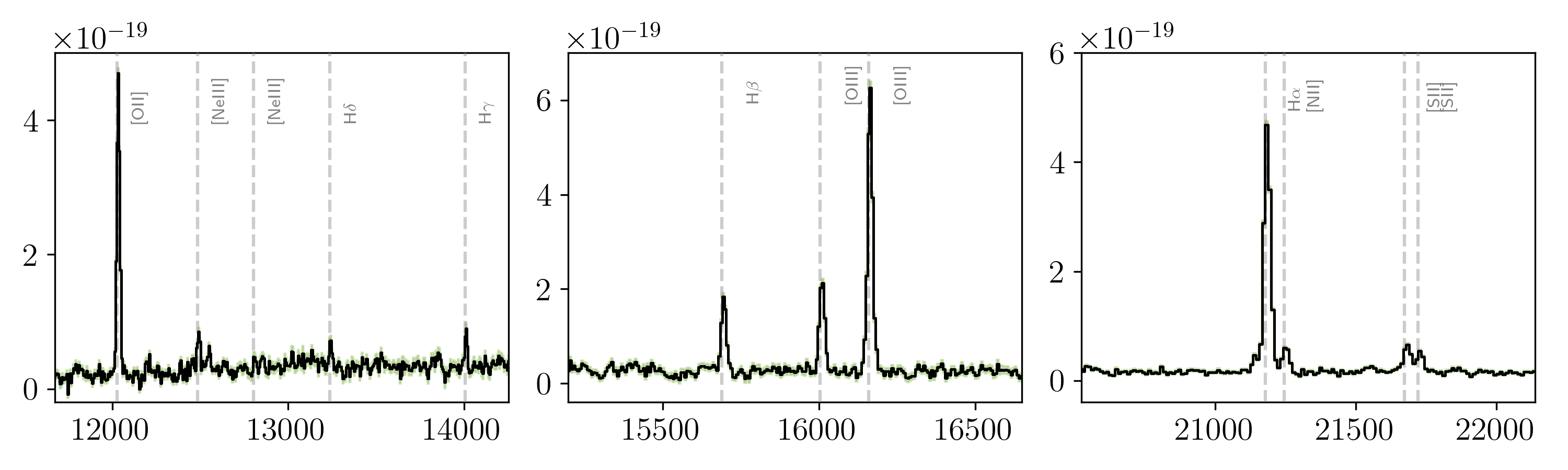}
    \caption{As for Figure~\ref{fig:example_spectra1}, but showing galaxy 003892 at $z=2.227$. Note the Balmer break just below [OII]\,3727.
    }
    \label{fig:example_spectra9}
\end{figure*}

\begin{figure*}
        \centering
    \includegraphics*[width=0.95\textwidth]{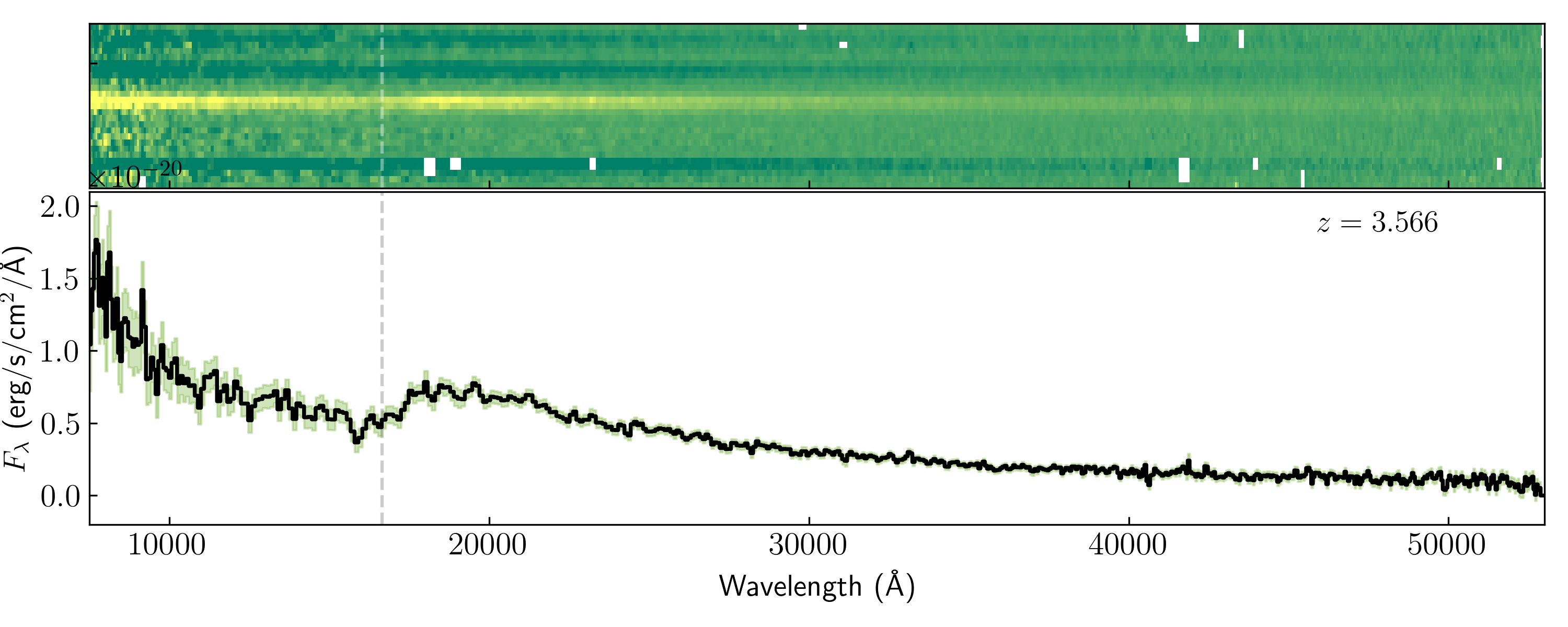}
    \caption{Prism spectra (1D and 2D) of 10036262, a galaxy which does not exhibit strong line emission but has a clear spectral break corresponding to a Balmer/4000\,\AA\ break at $z=3.566$.}
    \label{fig:example_spectra10}
    \end{figure*}

Example spectra covering a range of redshifts are shown in Figures~\ref{fig:example_spectra1}--\ref{fig:example_spectra10}. The low-dispersion prism spectrum is shown, along with the medium dispersion grating of emission prominent lines.

\FloatBarrier

\section{Tables of targets}
\label{sub:targets}

We present tables of the priority classes and positions for each object observed (Table~\ref{tab:targets}), and the fluxes of lines detected at $S/N>5$ in the medium dispersion grating spectra (Table~\ref{tab:grating_lines}) and the low-dispersion prism spectra (Table~\ref{tab:prism_lines}), along with the derived redshifts.

\clearpage\onecolumn

\begin{landscape}
\begin{longtable}{c c c c c c c c c c c c c}
\caption{Table of targeted sources, ranked in priority order for placement on the MSA (highest priority at the top).}
\label{tab:targets}
\\
\hline
\hline
  & & & & & & &\multicolumn{2}{c}{Intra-shutter offset ($''$)} & \multicolumn{2}{c}{Number of exposures} & pre-JWST &  \\
 NIRSpec\_ID & NIRCam\_ID & RA & Dec & priority & $z_\textrm{spec}$ & flag & x-offset & y-offset & R100 & Gratings & priority & references\\
\hline
\endfirsthead
\caption{continued} \\
\hline
  & & & & & & &\multicolumn{2}{c}{Intra-shutter offset ($''$)} & \multicolumn{2}{c}{Number of exposures} & pre-JWST &  \\
 NIRSpec\_ID & NIRCam\_ID & RA & Dec & priority & $z_\textrm{spec}$ & flag & x-offset & y-offset & R100 & Gratings & priority & references\\
\hline
\endhead
\hline
\endfoot
\hline
\endlastfoot
2773 & 96216 & 53.16634 & -27.82156 & 1 & 12.63 & C & 0.015 & 0.173 & 48 & 12 & -- & N \\
8013 & 110748 & 53.16446 & -27.80218 & 1 & 8.473 & A & -0.061 & 0.171 & 72 & 18 & 1 & 2,10,12,14,15,18,20 \\
10014170 & 129399 & 53.16339 & -27.77567 & 1 & -- & -- & 0.059 & -0.162 & 72 & 18 & 1 & 1,12,14,15,16,18$^*$ \\
10014220 & 130158 & 53.16477 & -27.77463 & 1 & 11.58 & C & -0.062 & -0.097 & 72 & 18 & 1 & 12,14$^*$ \\
10058975 & -1 & 53.11243 & -27.77461 & 1 & 9.433 & A & -0.012 & 0.143 & 72 & 18 & 1 & 1,17,18$^*$ \\
17400 & 128771 & 53.14988 & -27.7765 & 1 & 13.20 & C & -0.001 & 0.185 & 24 & 6 & -- & N \\
& & & & & & & & & & & & \\
6438 & 107324 & 53.16735 & -27.8075 & 2 & 9.689 & B & 0.053 & -0.175 & 72 & 18 & -- & N \\
\hline
\end{longtable}
\tablefoot{
The columns list the NIRSpec ID; the ID of the closest match within $0\farcs 2$ in the NIRCam data release catalogue (Rieke et al. 2023); RA and Dec supplied to the e\textsc{MPT} for target allocation, and for path loss determination; the priority class (see Table~\ref{tab:priorities}); the spectroscopic redshift, $z_\textrm{spec}$ and associated flag, as described here `A' - derived from S/N$>5$ emission lines detected in the R1000 grating, `B' - derived from S/N$>5$ emission lines detected in the prism observations, `C' - derived from spectral break and/or lower significance emission lines; the average intra-shutter offset of the source in arcseconds over the multiple pointings; the number of exposures (nods and dithers) for the prism and for each of the gratings (G140M, G235M, G395M, G395H), where a single exposure has duration 1400.5s; the priority from HST data only; and citations detailing where the source has been found before.
The citation key is as follows:
Targets marked with a `N' were NIRCam-based, and do not have a counterpart in our HST-based catalogue.
The majority of the remaining targets have counterparts in our HST-based catalogue drawn from one of \citet{Skelton2014_3dhst}, \citet{Guo2013}, \citet{Rafelski2015_UVUDF}, and/or \citet{Whitaker2019}. Targets which do {\em not} have a counterpart in any of these catalogues are marked with an asterisk, representing only a small number of targets.
Numbers indicate additional reference catalogues which were used for identifying and prioritising potential targets. Targets which are not NIRCam-based (not an `N'), flagged with an asterisk (i.e.\ not from one of the four large catalogues listed prior), and do not have a numerical reference listed were drawn from \citet{Coe2006}; this represents only six targets in Priority Class 9.
\tablebib{
[1]  \citet{Bouwens2015}, [2] \citet{Finkelstein2015}, [3] \citet{Harikane2016}, [4] \citet{Bunker2004}, [5] \citet{YanWindhorst2004}, [6] \citet{Bouwens2006}, [7] \citet{Bunker2010}, [8] \citet{Oesch2010}, [9] \citet{Yan2010}, [10] \citet{Bouwens2011}, [11] \citet{Wilkins2011}, [12] \citet{Ellis2013}, [13] \citet{Lorenzoni2013}, [14] \citet{McLure2013}, [15] \citet{Oesch2013}, [16] \citet{Schenker2013}, [17] \citet{Bouwens2016}, [18] \citet{Bouwens2021}, [19] \citet{Decarli2016}, [20] \citet{Inami2017}, [21] \citet{Popesso2009}.}
} 
\end{landscape}

\clearpage\onecolumn

\begin{longtable}{l c p{15cm} }
\caption{List of targets with one or more emission lines detected with $S/N>5$ in the $R\approx 1000$ grating data, and derived redshifts.}
\\
\hline       
\hline
ID & $z_{R\approx1000}$ & List of detected emission lines and flux in units of $10^{-19}$ erg s$^{-1}$ cm$^{-2}$ \\
\hline
\endfirsthead
\caption{continued} \\
\hline
ID & $z_{R\approx1000}$ & List of detected emission lines and flux in units of $10^{-19}$ erg s$^{-1}$ cm$^{-2}$ \\
\hline
\endhead
\hline
\endfoot
\hline
\endlastfoot
10058975 & 9.433 & [Ne{\sc iii}]~$\lambda$3869~$(3.1\pm0.6)$; H$\gamma$~$(3.0\pm0.6)$; H$\beta$~$(8.0\pm0.9)$; [O{\sc iii}]~$\lambda$4959~$(11.5\pm0.9)$; [O{\sc iii}]~$\lambda$5007~$(35.8\pm0.8)$\\
8013 & 8.473 & [O{\sc iii}]~$\lambda$5007~$(12.9\pm1.6)$\\
21842 & 7.980 & H$\beta$~$(3.1\pm0.5)$; [O{\sc iii}]~$\lambda$4959~$(8.7\pm0.5)$; [O{\sc iii}]~$\lambda$5007~$(17.9\pm0.7)$\\
10013682 & 7.275 & Ly-$\alpha$~$(23.8\pm3.3)$; [O{\sc iii}]~$\lambda$5007~$(7.2\pm0.6)$\\
10013905 & 7.197 & H$\beta$~$(7.4\pm0.9)$; [O{\sc iii}]~$\lambda$4959~$(5.8\pm0.9)$; [O{\sc iii}]~$\lambda$5007~$(20.6\pm0.8)$\\
20961 & 7.044 & H$\beta$~$(4.0\pm0.8)$; [O{\sc iii}]~$\lambda$5007~$(10.7\pm0.8)$\\
10013609 & 6.929 & H$\beta$~$(6.7\pm1.0)$; [O{\sc iii}]~$\lambda$4959~$(11.8\pm1.1)$; [O{\sc iii}]~$\lambda$5007~$(35.4\pm1.3)$; H$\alpha$~$(21.9\pm1.5)$\\
4297 & 6.713 & Ly-$\alpha$~$(30.9\pm5.2)$; H$\beta$~$(5.2\pm0.6)$; [O{\sc iii}]~$\lambda$4959~$(9.4\pm0.7)$; [O{\sc iii}]~$\lambda$5007~$(29.0\pm1.2)$; H$\alpha$~$(15.2\pm1.3)$\\
3334 & 6.706 & [O{\sc iii}]~$\lambda$4959~$(3.8\pm0.5)$; [O{\sc iii}]~$\lambda$5007~$(13.0\pm0.6)$\\
16625 & 6.631 & Ly-$\alpha$~$(21.1\pm3.8)$; H$\beta$~$(7.1\pm0.7)$; [O{\sc iii}]~$\lambda$4959~$(6.1\pm0.6)$; [O{\sc iii}]~$\lambda$5007~$(20.9\pm0.7)$; H$\alpha$~$(18.6\pm1.0)$\\
10005447 & 6.623 & [O{\sc iii}]~$\lambda$5007~$(5.4\pm0.9)$\\
18846 & 6.335 & Ly-$\alpha$~$(82.4\pm4.6)$; H$\gamma$~$(3.9\pm0.8)$; H$\beta$~$(10.3\pm0.6)$; [O{\sc iii}]~$\lambda$4959~$(19.8\pm0.7)$; [O{\sc iii}]~$\lambda$5007~$(58.4\pm1.1)$; H$\alpha$~$(36.0\pm1.2)$\\
18976 & 6.327 & H$\beta$~$(4.0\pm0.5)$; [O{\sc iii}]~$\lambda$4959~$(5.4\pm0.6)$; [O{\sc iii}]~$\lambda$5007~$(14.3\pm1.0)$; H$\alpha$~$(10.4\pm1.1)$\\
10009693 & 6.286 & [O{\sc iii}]~$\lambda$5007~$(5.9\pm0.8)$\\
17566 & 6.102 & [O{\sc ii}]~$\lambda\lambda$3727~$(10.0\pm1.5)$; H$\beta$~$(5.7\pm0.9)$; [O{\sc iii}]~$\lambda$4959~$(9.9\pm1.0)$; [O{\sc iii}]~$\lambda$5007~$(32.2\pm1.0)$; H$\alpha$~$(15.6\pm1.3)$\\
19342 & 5.974 & H$\beta$~$(5.0\pm0.7)$; [O{\sc iii}]~$\lambda$4959~$(5.7\pm0.6)$; [O{\sc iii}]~$\lambda$5007~$(19.5\pm0.8)$; H$\alpha$~$(13.8\pm0.9)$\\
10013618 & 5.944 & [O{\sc ii}]~$\lambda\lambda$3727~$(8.0\pm0.9)$; H$\beta$~$(1.9\pm0.4)$; [O{\sc iii}]~$\lambda$4959~$(5.1\pm0.6)$; [O{\sc iii}]~$\lambda$5007~$(13.1\pm0.5)$; H$\alpha$~$(7.5\pm0.7)$\\
6002 & 5.937 & Ly-$\alpha$~$(30.1\pm4.2)$; H$\beta$~$(2.6\pm0.4)$; [O{\sc iii}]~$\lambda$4959~$(6.1\pm0.6)$; [O{\sc iii}]~$\lambda$5007~$(16.3\pm0.5)$; H$\alpha$~$(10.2\pm0.5)$\\
9422 & 5.936 & Ly-$\alpha$~$(103.6\pm6.0)$; C{\sc iv}~$\lambda\lambda$1549~$(34.0\pm2.4)$; C{\sc iii}]~$\lambda\lambda$1909~$(7.4\pm0.9)$; [Ne{\sc iii}]~$\lambda$3869~$(6.5\pm0.3)$; H$\gamma$~$(8.1\pm0.7)$; H$\beta$~$(19.7\pm0.5)$; [O{\sc iii}]~$\lambda$4959~$(38.6\pm0.9)$; [O{\sc iii}]~$\lambda$5007~$(110.8\pm1.2)$\\
10013704\tablefootmark{a} & 5.920 & [Ne{\sc iii}]~$\lambda$3869~$(4.8\pm0.7)$; H$\gamma$~$(4.1\pm0.6)$; H$\beta$~$(9.0\pm0.7)$; [O{\sc iii}]~$\lambda$4959~$(17.5\pm0.6)$; [O{\sc iii}]~$\lambda$5007~$(48.3\pm0.9)$; H$\alpha$~$(63.5\pm2.0)$\\
10013620 & 5.918 & H$\beta$~$(4.7\pm0.8)$; [O{\sc iii}]~$\lambda$4959~$(8.6\pm0.8)$; [O{\sc iii}]~$\lambda$5007~$(23.1\pm0.9)$; H$\alpha$~$(13.8\pm1.3)$\\
19606 & 5.889 & H$\beta$~$(6.1\pm1.1)$; [O{\sc iii}]~$\lambda$4959~$(8.9\pm1.3)$; [O{\sc iii}]~$\lambda$5007~$(28.0\pm1.0)$; H$\alpha$~$(19.0\pm1.6)$\\
10005113 & 5.821 & H$\beta$~$(2.1\pm0.4)$; [O{\sc iii}]~$\lambda$4959~$(4.4\pm0.6)$; [O{\sc iii}]~$\lambda$5007~$(14.2\pm0.7)$\\
10056849 & 5.814 & H$\beta$~$(4.9\pm0.9)$; [O{\sc iii}]~$\lambda$4959~$(4.5\pm0.9)$; [O{\sc iii}]~$\lambda$5007~$(17.0\pm1.0)$; H$\alpha$~$(14.1\pm0.8)$\\
22251 & 5.798 & [O{\sc ii}]~$\lambda\lambda$3727~$(9.0\pm1.2)$; [Ne{\sc iii}]~$\lambda$3869~$(5.9\pm1.0)$; H$\beta$~$(10.9\pm0.8)$; [O{\sc iii}]~$\lambda$4959~$(23.4\pm0.7)$; [O{\sc iii}]~$\lambda$5007~$(67.1\pm1.2)$; H$\alpha$~$(33.8\pm0.9)$\\
3968 & 5.768 & [O{\sc iii}]~$\lambda$5007~$(7.9\pm0.8)$\\
4404 & 5.764 & [O{\sc ii}]~$\lambda\lambda$3727~$(5.5\pm1.0)$; [Ne{\sc iii}]~$\lambda$3869~$(6.0\pm1.0)$; H$\beta$~$(8.6\pm1.2)$; [O{\sc iii}]~$\lambda$4959~$(18.2\pm1.2)$; [O{\sc iii}]~$\lambda$5007~$(52.9\pm1.3)$; H$\alpha$~$(30.1\pm1.0)$\\
6384 & 5.615 & [O{\sc iii}]~$\lambda$5007~$(7.8\pm1.1)$\\
16745 & 5.567 & [O{\sc ii}]~$\lambda\lambda$3727~$(14.3\pm1.3)$; [Ne{\sc iii}]~$\lambda$3869~$(4.1\pm0.8)$; H$\beta$~$(4.1\pm0.6)$; H$\alpha$~$(20.8\pm0.7)$\\
6246 & 5.562 & H$\beta$~$(3.4\pm0.5)$; [O{\sc iii}]~$\lambda$4959~$(3.5\pm0.5)$; [O{\sc iii}]~$\lambda$5007~$(11.6\pm0.5)$; H$\alpha$~$(7.6\pm0.5)$\\
10016374 & 5.504 & [O{\sc ii}]~$\lambda\lambda$3727~$(5.4\pm1.0)$; H$\beta$~$(7.2\pm1.0)$; H$\alpha$~$(17.9\pm0.8)$\\
9343 & 5.443 & H$\beta$~$(4.4\pm0.7)$; [O{\sc iii}]~$\lambda$4959~$(7.7\pm1.0)$; [O{\sc iii}]~$\lambda$5007~$(20.1\pm1.2)$; H$\alpha$~$(10.6\pm0.8)$\\
9743 & 5.440 & [O{\sc iii}]~$\lambda$4959~$(11.3\pm1.8)$; [O{\sc iii}]~$\lambda$5007~$(24.6\pm2.7)$; H$\alpha$~$(26.9\pm2.0)$\\
9452 & 5.122 & [O{\sc iii}]~$\lambda$5007~$(47.7\pm3.8)$; H$\alpha$~$(18.8\pm2.1)$\\
10015338 & 5.077 & H$\gamma$~$(16.9\pm3.1)$; [O{\sc iii}]~$\lambda$4959~$(25.5\pm2.7)$; [O{\sc iii}]~$\lambda$5007~$(57.5\pm2.6)$; H$\alpha$~$(31.7\pm1.9)$\\
5759 & 5.052 & [O{\sc iii}]~$\lambda$4959~$(9.9\pm1.4)$; [O{\sc iii}]~$\lambda$5007~$(19.9\pm1.5)$; H$\alpha$~$(13.7\pm1.2)$\\
8113 & 4.903 & [O{\sc ii}]~$\lambda\lambda$3727~$(5.6\pm1.0)$; [Ne{\sc iii}]~$\lambda$3869~$(4.0\pm0.7)$; [O{\sc iii}]~$\lambda$4959~$(11.7\pm0.9)$; [O{\sc iii}]~$\lambda$5007~$(38.3\pm1.0)$; H$\alpha$~$(18.2\pm0.7)$\\
10005217 & 4.888 & H$\gamma$~$(4.5\pm0.6)$; H$\beta$~$(9.9\pm0.7)$; [O{\sc iii}]~$\lambda$4959~$(11.3\pm0.7)$; [O{\sc iii}]~$\lambda$5007~$(36.7\pm0.8)$; H$\alpha$~$(34.6\pm0.6)$\\
17260 & 4.885 & H$\alpha$~$(4.0\pm0.5)$\\
5457 & 4.864 & [O{\sc iii}]~$\lambda$5007~$(14.7\pm1.1)$; H$\alpha$~$(9.1\pm0.8)$\\
7938 & 4.806 & [O{\sc ii}]~$\lambda\lambda$3727~$(10.0\pm1.3)$; [Ne{\sc iii}]~$\lambda$3869~$(6.2\pm0.8)$; H$\beta$~$(10.6\pm1.1)$; [O{\sc iii}]~$\lambda$4959~$(24.7\pm1.4)$; [O{\sc iii}]~$\lambda$5007~$(73.2\pm1.4)$; H$\alpha$~$(32.2\pm0.9)$\\
18090 & 4.775 & [O{\sc ii}]~$\lambda\lambda$3727~$(13.3\pm1.6)$; [Ne{\sc iii}]~$\lambda$3869~$(7.4\pm1.1)$; H$\gamma$~$(6.7\pm1.2)$; H$\beta$~$(18.8\pm1.5)$; [O{\sc iii}]~$\lambda$4959~$(36.9\pm1.7)$; [O{\sc iii}]~$\lambda$5007~$(119.4\pm2.2)$; H$\alpha$~$(48.0\pm2.0)$\\
10001892 & 4.772 & H$\alpha$~$(2.1\pm0.4)$\\
17072 & 4.702 & [O{\sc iii}]~$\lambda$5007~$(8.2\pm0.9)$; H$\alpha$~$(5.3\pm0.5)$\\
8083\tablefootmark{a} & 4.649 & C{\sc iii}]~$\lambda\lambda$1909~$(7.4\pm1.4)$; [O{\sc ii}]~$\lambda\lambda$3727~$(10.9\pm1.3)$; [Ne{\sc iii}]~$\lambda$3869~$(14.7\pm1.1)$; [Ne{\sc iii}]~$\lambda$3967~$(8.4\pm0.8)$; H$\delta$~$(5.5\pm0.7)$; H$\gamma$~$(11.1\pm0.7)$; H$\beta$~$(26.3\pm1.1)$; H$\alpha$~$(120.9\pm1.7)$\\
10000626 & 4.464 & [O{\sc iii}]~$\lambda$5007~$(9.0\pm0.7)$; H$\alpha$~$(6.0\pm0.6)$\\
10001916 & 4.284 & [O{\sc iii}]~$\lambda$5007~$(4.7\pm0.8)$; H$\alpha$~$(3.1\pm0.6)$\\
7892 & 4.229 & [O{\sc ii}]~$\lambda\lambda$3727~$(7.7\pm1.2)$; H$\gamma$~$(4.5\pm0.6)$; H$\beta$~$(8.6\pm0.6)$; [O{\sc iii}]~$\lambda$4959~$(17.3\pm0.6)$; [O{\sc iii}]~$\lambda$5007~$(47.5\pm0.9)$; H$\alpha$~$(28.7\pm0.7)$\\
7762 & 4.149 & [O{\sc ii}]~$\lambda\lambda$3727~$(15.7\pm1.3)$; H$\gamma$~$(6.4\pm1.0)$; H$\beta$~$(10.1\pm0.5)$; [O{\sc iii}]~$\lambda$4959~$(19.9\pm0.5)$; [O{\sc iii}]~$\lambda$5007~$(57.3\pm1.0)$; H$\alpha$~$(36.8\pm0.7)$; [S{\sc ii}]~$\lambda$6716~$(3.2\pm0.4)$\\
17777 & 4.134 & [O{\sc iii}]~$\lambda$4959~$(6.2\pm0.7)$; [O{\sc iii}]~$\lambda$5007~$(16.5\pm0.7)$; H$\alpha$~$(6.3\pm0.6)$\\
7507 & 4.044 & H$\beta$~$(7.5\pm1.3)$; [O{\sc iii}]~$\lambda$4959~$(12.8\pm1.0)$; [O{\sc iii}]~$\lambda$5007~$(31.6\pm1.2)$; H$\alpha$~$(23.3\pm1.3)$\\
10013545 & 4.035 & [O{\sc ii}]~$\lambda\lambda$3727~$(19.3\pm2.0)$; H$\beta$~$(12.1\pm1.1)$; [O{\sc iii}]~$\lambda$4959~$(26.7\pm1.5)$; [O{\sc iii}]~$\lambda$5007~$(82.2\pm1.5)$; H$\alpha$~$(47.3\pm1.8)$\\
4270 & 4.023 & [O{\sc ii}]~$\lambda\lambda$3727~$(28.6\pm1.6)$; [Ne{\sc iii}]~$\lambda$3869~$(11.1\pm1.6)$; H$\delta$~$(6.3\pm1.2)$; H$\gamma$~$(12.4\pm1.1)$; H$\beta$~$(27.5\pm1.0)$; [O{\sc iii}]~$\lambda$4959~$(63.6\pm1.2)$; [O{\sc iii}]~$\lambda$5007~$(186.6\pm3.1)$; H$\alpha$~$(126.0\pm1.8)$; [S{\sc ii}]~$\lambda$6716~$(4.5\pm0.8)$; [S{\sc iii}]~$\lambda$9532~$(9.0\pm1.3)$\\
10016186 & 3.928 & [O{\sc ii}]~$\lambda\lambda$3727~$(15.9\pm1.7)$; H$\beta$~$(10.4\pm0.9)$; [O{\sc iii}]~$\lambda$4959~$(15.5\pm0.9)$; [O{\sc iii}]~$\lambda$5007~$(50.0\pm1.2)$; H$\alpha$~$(33.8\pm1.0)$\\
18028 & 3.731 & [O{\sc iii}]~$\lambda$5007~$(6.0\pm0.9)$; H$\alpha$~$(8.7\pm1.3)$\\
18970 & 3.725 & [O{\sc ii}]~$\lambda\lambda$3727~$(53.4\pm4.5)$; H$\beta$~$(27.7\pm2.4)$; [O{\sc iii}]~$\lambda$4959~$(49.6\pm3.1)$; [O{\sc iii}]~$\lambda$5007~$(139.4\pm3.7)$; H$\alpha$~$(105.4\pm3.3)$; [N{\sc ii}]~$\lambda$6583~$(10.2\pm1.9)$; [S{\sc ii}]~$\lambda$6716~$(9.4\pm1.4)$; [S{\sc iii}]~$\lambda$9532~$(11.2\pm2.1)$\\
19519 & 3.605 & [O{\sc ii}]~$\lambda\lambda$3727~$(16.5\pm1.3)$; H$\beta$~$(12.6\pm0.8)$; [O{\sc iii}]~$\lambda$4959~$(22.1\pm0.9)$; [O{\sc iii}]~$\lambda$5007~$(68.4\pm1.1)$; H$\alpha$~$(37.0\pm0.9)$\\
10009506 & 3.599 & [O{\sc ii}]~$\lambda\lambda$3727~$(18.4\pm2.4)$; [Ne{\sc iii}]~$\lambda$3869~$(9.9\pm1.7)$; H$\gamma$~$(6.8\pm1.0)$; H$\beta$~$(16.9\pm0.9)$; [O{\sc iii}]~$\lambda$4959~$(37.5\pm1.0)$; [O{\sc iii}]~$\lambda$5007~$(112.1\pm1.4)$; H$\alpha$~$(55.8\pm0.9)$; [S{\sc iii}]~$\lambda$9532~$(4.8\pm0.7)$\\
10035295 & 3.589 & H$\gamma$~$(6.0\pm1.2)$; [O{\sc iii}]~$\lambda$4363~$(8.6\pm1.5)$; H$\beta$~$(17.2\pm0.9)$; [O{\sc iii}]~$\lambda$4959~$(31.5\pm1.2)$; [O{\sc iii}]~$\lambda$5007~$(93.8\pm1.6)$; H$\alpha$~$(58.0\pm1.2)$; He{\sc i}~$\lambda$10830~$(15.2\pm1.3)$\\
5329 & 3.587 & [O{\sc iii}]~$\lambda$5007~$(9.9\pm1.6)$; H$\alpha$~$(10.8\pm1.2)$\\
7809 & 3.578 & [O{\sc iii}]~$\lambda$4959~$(24.4\pm2.4)$; [O{\sc iii}]~$\lambda$5007~$(62.4\pm2.0)$; H$\alpha$~$(36.3\pm2.2)$\\
3184 & 3.468 & [O{\sc ii}]~$\lambda\lambda$3727~$(13.0\pm1.7)$; H$\beta$~$(13.2\pm0.9)$; [O{\sc iii}]~$\lambda$4959~$(35.6\pm1.1)$; [O{\sc iii}]~$\lambda$5007~$(93.5\pm2.6)$; H$\alpha$~$(56.7\pm1.7)$; [S{\sc iii}]~$\lambda$9069~$(4.2\pm0.6)$; [S{\sc iii}]~$\lambda$9532~$(4.2\pm0.7)$; He{\sc i}~$\lambda$10830~$(7.0\pm1.1)$\\
10013597 & 3.323 & [O{\sc iii}]~$\lambda$5007~$(18.3\pm1.2)$; H$\alpha$~$(12.9\pm1.5)$\\
19431 & 3.320 & [O{\sc iii}]~$\lambda$4959~$(26.0\pm2.1)$; [O{\sc iii}]~$\lambda$5007~$(62.3\pm2.2)$; H$\alpha$~$(29.1\pm2.3)$\\
10040\tablefootmark{$\ddag$} & 3.154 & [O{\sc iii}]~$\lambda$5007~$(20.1\pm1.9)$\\
18322 & 3.152 & H$\beta$~$(12.4\pm1.1)$; [O{\sc iii}]~$\lambda$4959~$(22.8\pm1.0)$; [O{\sc iii}]~$\lambda$5007~$(65.5\pm1.4)$; H$\alpha$~$(34.2\pm1.0)$; He{\sc i}~$\lambda$10830~$(4.7\pm0.5)$\\
21150 & 3.088 & [O{\sc ii}]~$\lambda\lambda$3727~$(39.8\pm2.3)$; H$\gamma$~$(12.0\pm2.3)$; H$\beta$~$(23.0\pm1.9)$; [O{\sc iii}]~$\lambda$4959~$(50.4\pm2.3)$; [O{\sc iii}]~$\lambda$5007~$(176.2\pm4.5)$; H$\alpha$~$(79.7\pm2.0)$; [S{\sc iii}]~$\lambda$9532~$(8.9\pm0.8)$; He{\sc i}~$\lambda$10830~$(9.1\pm1.3)$\\
8245 & 3.065 & [O{\sc iii}]~$\lambda$5007~$(18.4\pm1.5)$\\
10004721 & 3.064 & [O{\sc iii}]~$\lambda$5007~$(11.0\pm1.2)$; H$\alpha$~$(6.3\pm1.1)$\\
2923 & 3.015 & [O{\sc iii}]~$\lambda$4959~$(7.7\pm1.3)$; [O{\sc iii}]~$\lambda$5007~$(18.1\pm1.7)$; H$\alpha$~$(10.9\pm1.0)$\\
10004819 & 2.990 & H$\beta$~$(11.9\pm2.0)$; [O{\sc iii}]~$\lambda$4959~$(16.0\pm2.0)$; [O{\sc iii}]~$\lambda$5007~$(36.9\pm2.4)$; H$\alpha$~$(25.2\pm2.2)$\\
10011980 & 2.867 & H$\beta$~$(14.8\pm2.6)$; H$\alpha$~$(47.2\pm1.6)$; He{\sc i}~$\lambda$10830~$(5.5\pm0.9)$\\
20313 & 2.845 & H$\beta$~$(8.3\pm1.4)$; [O{\sc iii}]~$\lambda$4959~$(10.4\pm1.3)$; [O{\sc iii}]~$\lambda$5007~$(27.8\pm1.8)$; H$\alpha$~$(20.3\pm1.0)$\\
8102 & 2.840 & [O{\sc iii}]~$\lambda$5007~$(5.8\pm0.9)$; H$\alpha$~$(6.3\pm0.7)$\\
5686 & 2.819 & [O{\sc iii}]~$\lambda$5007~$(13.8\pm2.3)$; H$\alpha$~$(10.8\pm2.1)$\\
3892 & 2.807 & [O{\sc ii}]~$\lambda\lambda$3727~$(195.6\pm4.0)$; [Ne{\sc iii}]~$\lambda$3869~$(45.4\pm5.1)$; [Ne{\sc iii}]~$\lambda$3967~$(18.0\pm3.3)$; H$\delta$~$(17.5\pm2.9)$; H$\gamma$~$(31.5\pm3.4)$; H$\beta$~$(91.7\pm4.0)$; [O{\sc iii}]~$\lambda$4959~$(165.7\pm3.2)$; [O{\sc iii}]~$\lambda$5007~$(498.8\pm5.8)$; He{\sc i}~$\lambda$5875~$(12.1\pm1.6)$; H$\alpha$~$(299.7\pm3.4)$; [N{\sc ii}]~$\lambda$6583~$(18.1\pm1.3)$; [S{\sc ii}]~$\lambda$6716~$(20.9\pm1.3)$; [S{\sc ii}]~$\lambda$6731~$(16.1\pm1.3)$; [S{\sc iii}]~$\lambda$9069~$(18.9\pm1.3)$; [S{\sc iii}]~$\lambda$9532~$(49.6\pm1.8)$; He{\sc i}~$\lambda$10830~$(32.4\pm1.6)$; Pa-$\gamma$~$(11.2\pm1.4)$; Pa-$\beta$~$(18.7\pm1.6)$\\
16996 & 2.776 & [O{\sc ii}]~$\lambda\lambda$3727~$(45.4\pm3.8)$; [O{\sc iii}]~$\lambda$4959~$(18.4\pm2.7)$; H$\alpha$~$(54.8\pm2.4)$; [S{\sc iii}]~$\lambda$9532~$(4.4\pm0.9)$; He{\sc i}~$\lambda$10830~$(7.3\pm1.4)$\\
10011849 & 2.694 & [O{\sc ii}]~$\lambda\lambda$3727~$(45.5\pm2.8)$; [Ne{\sc iii}]~$\lambda$3869~$(19.6\pm2.4)$; H$\gamma$~$(12.0\pm2.2)$; H$\beta$~$(25.5\pm1.8)$; [O{\sc iii}]~$\lambda$4959~$(54.3\pm2.0)$; [O{\sc iii}]~$\lambda$5007~$(188.9\pm2.6)$; H$\alpha$~$(97.3\pm2.1)$; He{\sc i}~$\lambda$10830~$(16.1\pm1.5)$\\
17824 & 2.693 & [O{\sc ii}]~$\lambda\lambda$3727~$(41.2\pm4.5)$; H$\beta$~$(20.8\pm3.4)$; [O{\sc iii}]~$\lambda$4959~$(49.1\pm3.1)$; [O{\sc iii}]~$\lambda$5007~$(121.5\pm3.4)$; H$\alpha$~$(63.4\pm2.7)$\\
10011378 & 2.577 & [O{\sc ii}]~$\lambda\lambda$3727~$(27.2\pm1.6)$; [O{\sc iii}]~$\lambda$4959~$(12.7\pm1.1)$; [O{\sc iii}]~$\lambda$5007~$(45.1\pm1.6)$; H$\alpha$~$(28.3\pm0.9)$; [S{\sc iii}]~$\lambda$9532~$(3.9\pm0.5)$\\
17670 & 2.343 & [O{\sc ii}]~$\lambda\lambda$3727~$(33.5\pm2.6)$; H$\beta$~$(14.2\pm2.7)$; [O{\sc iii}]~$\lambda$4959~$(26.6\pm3.8)$; [O{\sc iii}]~$\lambda$5007~$(102.6\pm3.0)$; H$\alpha$~$(52.7\pm2.4)$; He{\sc i}~$\lambda$10830~$(6.4\pm1.2)$\\
8784 & 2.343 & [O{\sc iii}]~$\lambda$4959~$(11.4\pm1.6)$; [O{\sc iii}]~$\lambda$5007~$(35.0\pm2.7)$; H$\alpha$~$(19.1\pm1.5)$\\
8880 & 2.342 & H$\beta$~$(10.0\pm1.1)$; [O{\sc iii}]~$\lambda$5007~$(21.2\pm1.7)$; H$\alpha$~$(25.6\pm1.0)$\\
10011955 & 2.228 & H$\beta$~$(14.3\pm2.7)$; [O{\sc iii}]~$\lambda$5007~$(26.1\pm3.6)$; H$\alpha$~$(45.8\pm2.4)$; [N{\sc ii}]~$\lambda$6583~$(6.9\pm1.3)$; [S{\sc iii}]~$\lambda$9532~$(14.2\pm1.7)$\\
10008071 & 2.227 & [O{\sc ii}]~$\lambda\lambda$3727~$(94.8\pm2.1)$; [Ne{\sc iii}]~$\lambda$3869~$(11.2\pm2.0)$; H$\delta$~$(8.5\pm1.4)$; H$\gamma$~$(9.6\pm1.1)$; H$\beta$~$(30.7\pm1.4)$; [O{\sc iii}]~$\lambda$4959~$(40.4\pm1.0)$; [O{\sc iii}]~$\lambda$5007~$(119.3\pm2.0)$; H$\alpha$~$(133.5\pm2.3)$; [N{\sc ii}]~$\lambda$6583~$(14.8\pm1.1)$; [S{\sc ii}]~$\lambda$6716~$(15.7\pm1.1)$; [S{\sc ii}]~$\lambda$6731~$(11.5\pm1.1)$; [S{\sc iii}]~$\lambda$9069~$(13.9\pm1.3)$; [S{\sc iii}]~$\lambda$9532~$(31.5\pm1.5)$; Pa-$\delta$~$(4.9\pm0.8)$; He{\sc i}~$\lambda$10830~$(26.0\pm0.7)$; Pa-$\gamma$~$(6.9\pm0.5)$; Pa-$\beta$~$(16.8\pm0.7)$\\
10011974 & 2.225 & H$\gamma$~$(23.4\pm2.3)$; H$\beta$~$(56.1\pm3.7)$; [O{\sc iii}]~$\lambda$4959~$(33.5\pm2.8)$; [O{\sc iii}]~$\lambda$5007~$(104.6\pm2.9)$; H$\alpha$~$(204.0\pm2.9)$; [N{\sc ii}]~$\lambda$6583~$(36.5\pm1.7)$; [S{\sc ii}]~$\lambda$6716~$(18.4\pm1.9)$; [S{\sc ii}]~$\lambda$6731~$(16.1\pm1.5)$; [S{\sc iii}]~$\lambda$9069~$(17.0\pm1.6)$; [S{\sc iii}]~$\lambda$9532~$(33.0\pm2.0)$; He{\sc i}~$\lambda$10830~$(20.4\pm1.5)$; Pa-$\gamma$~$(9.2\pm1.1)$; Pa-$\beta$~$(15.6\pm1.4)$\\
6268 & 2.024 & [O{\sc iii}]~$\lambda$4959~$(6.5\pm1.0)$; [O{\sc iii}]~$\lambda$5007~$(16.2\pm1.1)$; H$\alpha$~$(12.5\pm0.7)$\\
8891 & 1.997 & [O{\sc ii}]~$\lambda\lambda$3727~$(35.3\pm2.0)$; [Ne{\sc iii}]~$\lambda$3869~$(7.6\pm1.5)$; H$\beta$~$(17.2\pm1.4)$; [O{\sc iii}]~$\lambda$4959~$(24.5\pm1.8)$; [O{\sc iii}]~$\lambda$5007~$(62.1\pm1.3)$; H$\alpha$~$(44.2\pm1.8)$; He{\sc i}~$\lambda$10830~$(4.5\pm0.6)$\\
3137 & 1.904 & H$\alpha$~$(4.7\pm0.9)$\\
10010642 & 1.902 & [O{\sc ii}]~$\lambda\lambda$3727~$(29.1\pm3.5)$; [O{\sc iii}]~$\lambda$4959~$(12.0\pm1.7)$; [O{\sc iii}]~$\lambda$5007~$(21.2\pm2.7)$; H$\alpha$~$(19.5\pm1.8)$\\
10012005 & 1.864 & [O{\sc ii}]~$\lambda\lambda$3727~$(75.7\pm3.7)$; H$\gamma$~$(14.3\pm2.8)$; H$\beta$~$(30.0\pm4.0)$; [O{\sc iii}]~$\lambda$4959~$(23.3\pm2.3)$; [O{\sc iii}]~$\lambda$5007~$(74.1\pm2.5)$; [S{\sc iii}]~$\lambda$9069~$(8.4\pm1.6)$; [S{\sc iii}]~$\lambda$9532~$(21.3\pm2.0)$; Pa-$\beta$~$(6.7\pm1.3)$\\
17160 & 1.850 & [O{\sc iii}]~$\lambda$4959~$(7.7\pm1.4)$; [O{\sc iii}]~$\lambda$5007~$(22.2\pm1.9)$; H$\alpha$~$(13.6\pm1.3)$\\
19607 & 1.847 & [O{\sc ii}]~$\lambda\lambda$3727~$(54.0\pm3.9)$; H$\beta$~$(32.1\pm2.5)$; [O{\sc iii}]~$\lambda$4959~$(53.5\pm3.5)$; [O{\sc iii}]~$\lambda$5007~$(159.3\pm3.1)$; H$\alpha$~$(103.9\pm3.2)$; [S{\sc iii}]~$\lambda$9532~$(9.2\pm1.2)$; He{\sc i}~$\lambda$10830~$(9.4\pm1.3)$\\
20962 & 1.806 & [O{\sc iii}]~$\lambda$5007~$(5.7\pm1.1)$\\
4950 & 1.781 & [O{\sc iii}]~$\lambda$5007~$(18.0\pm2.0)$; H$\alpha$~$(11.0\pm2.2)$\\
3753 & 1.768 & [O{\sc ii}]~$\lambda\lambda$3727~$(237.2\pm2.8)$; [Ne{\sc iii}]~$\lambda$3869~$(32.8\pm2.1)$; H$\delta$~$(19.3\pm2.1)$; H$\gamma$~$(34.7\pm1.7)$; [O{\sc iii}]~$\lambda$4363~$(10.3\pm1.9)$; H$\beta$~$(88.1\pm1.6)$; [O{\sc iii}]~$\lambda$4959~$(111.9\pm2.2)$; [O{\sc iii}]~$\lambda$5007~$(355.9\pm3.3)$; He{\sc i}~$\lambda$5875~$(15.9\pm2.6)$; H$\alpha$~$(322.1\pm3.9)$; [N{\sc ii}]~$\lambda$6583~$(28.9\pm2.0)$; [S{\sc ii}]~$\lambda$6716~$(24.4\pm2.9)$; [S{\sc ii}]~$\lambda$6731~$(20.8\pm1.9)$; [S{\sc iii}]~$\lambda$9069~$(21.7\pm1.2)$; [S{\sc iii}]~$\lambda$9532~$(52.1\pm1.1)$; Pa-$\delta$~$(6.1\pm1.0)$; He{\sc i}~$\lambda$10830~$(34.2\pm1.7)$; Pa-$\gamma$~$(10.3\pm1.5)$; Pa-$\beta$~$(25.7\pm0.9)$; Pa-$\alpha$~$(54.4\pm2.1)$\\
17435 & 1.728 & H$\beta$~$(10.8\pm1.8)$; [O{\sc iii}]~$\lambda$4959~$(21.8\pm1.6)$; [O{\sc iii}]~$\lambda$5007~$(51.8\pm2.2)$; H$\alpha$~$(26.7\pm2.0)$\\
21598 & 1.715 & H$\gamma$~$(8.4\pm1.4)$; H$\beta$~$(18.0\pm1.1)$; [O{\sc iii}]~$\lambda$4959~$(17.7\pm1.2)$; [O{\sc iii}]~$\lambda$5007~$(52.3\pm1.5)$; H$\alpha$~$(40.2\pm1.3)$; Pa-$\alpha$~$(4.7\pm0.7)$\\
22078 & 1.611 & [O{\sc ii}]~$\lambda\lambda$3727~$(119.5\pm8.9)$; H$\beta$~$(38.5\pm4.0)$; [O{\sc iii}]~$\lambda$4959~$(87.7\pm4.4)$; [O{\sc iii}]~$\lambda$5007~$(272.6\pm6.5)$; H$\alpha$~$(148.9\pm6.0)$; [S{\sc iii}]~$\lambda$9069~$(10.9\pm2.0)$; [S{\sc iii}]~$\lambda$9532~$(23.3\pm2.8)$; He{\sc i}~$\lambda$10830~$(23.4\pm1.9)$; Pa-$\alpha$~$(26.2\pm1.6)$\\
10012284 & 1.306 & [O{\sc iii}]~$\lambda$4959~$(14.2\pm2.6)$; [O{\sc iii}]~$\lambda$5007~$(27.1\pm2.6)$; H$\alpha$~$(22.9\pm2.6)$\\
10010770 & 1.295 & [O{\sc iii}]~$\lambda$5007~$(21.8\pm3.3)$; H$\alpha$~$(20.9\pm4.1)$\\
10012616 & 1.097 & [O{\sc iii}]~$\lambda$5007~$(29.8\pm5.0)$; H$\alpha$~$(20.5\pm1.8)$\\
10012477 & 0.670 & [O{\sc iii}]~$\lambda$5007~$(82.8\pm11.1)$; H$\alpha$~$(93.5\pm4.8)$; He{\sc i}~$\lambda$10830~$(20.0\pm3.4)$; Pa-$\alpha$~$(13.2\pm2.3)$\\
\label{tab:grating_lines}
\end{longtable}
\tablefoot{
Details of the emission line fitting can be found in Section~\ref{sec:visual_redshifts}. The last column gives a list of emission lines detected with $S/N>5$, and the flux measured for that line. 
Targets marked \tablefoottext{$\ddag$}{were identified as likely having multiple objects in the shutter (see Section~\ref{sub:individual_targets}).}
Targets marked \tablefoottext{a}{showed a broad component under H$\alpha$ and reported flux was obtained from direct integration rather than a single component fit (Section~\ref{sec:visual_redshifts}).}
These measurements are also available in a machine-readable format on The Mikulski Archive for Space Telescopes: \url{https://archive.stsci.edu/hlsp/jades}.}

\clearpage
\begin{longtable}{l c p{15cm} }
\caption{List of targets with one or more emission lines detected at $S/N>5$ in the Prism/Clear data.}
\\
\hline\hline  
ID & $z_{\rm PRISM}$ & List of detected emission lines and flux in units of $10^{-19}$ erg s$^{-1}$ cm$^{-2}$ \\
\hline
\\
\endfirsthead
\caption{continued}\\
\hline
ID & $z_{\rm PRISM}$ & List of detected emission lines and flux in units of $10^{-19}$ erg s$^{-1}$ cm$^{-2}$ \\
\hline
\\
\endhead
\endfoot
\hline
\endlastfoot
6438 & 9.689 & H$\gamma$~$(1.5\pm0.2)$; H$\beta$~$(2.0\pm0.2)$\\
10058975 & 9.438 & C{\sc iv}~$\lambda\lambda$1549~$(4.6\pm0.8)$; He{\sc ii}~+~O{\sc iii}]~$(7.3\pm1.0)$; C{\sc iii}]~$\lambda\lambda$1909~$(4.6\pm0.3)$; [Ne{\sc iii}]~$\lambda$3869~$(2.6\pm0.3)$; H$\delta$~$(2.1\pm0.0)$; H$\gamma$~$(3.7\pm0.1)$; [O{\sc iii}]~$\lambda$4363~$(1.5\pm0.1)$; H$\beta$~$(6.7\pm0.2)$; [O{\sc iii}]~$\lambda$4959~$(10.7\pm0.2)$; [O{\sc iii}]~$\lambda$5007~$(29.4\pm0.3)$\\
8013 & 8.479 & H$\beta$~$(1.8\pm0.2)$; [O{\sc iii}]~$\lambda$4959~$(3.2\pm0.2)$; [O{\sc iii}]~$\lambda$5007~$(9.1\pm0.3)$\\
21842 & 7.981 & [O{\sc ii}]~$\lambda\lambda$3727~$(1.9\pm0.2)$; [Ne{\sc iii}]~$\lambda$3869~$(2.0\pm0.2)$; H$\delta$~$(1.5\pm0.2)$; H$\gamma$~$(1.9\pm0.1)$; [O{\sc iii}]~$\lambda$4363~$(0.6\pm0.1)$; H$\beta$~$(2.8\pm0.2)$; [O{\sc iii}]~$\lambda$4959~$(5.9\pm0.2)$; [O{\sc iii}]~$\lambda$5007~$(17.7\pm0.2)$\\
10013682 & 7.277 & H$\beta$~$(1.0\pm0.2)$; [O{\sc iii}]~$\lambda$4959~$(2.4\pm0.2)$; [O{\sc iii}]~$\lambda$5007~$(6.0\pm0.2)$\\
8079 & 7.260 & [O{\sc iii}]~$\lambda$5007~$(1.7\pm0.2)$\\
10013905 & 7.206 & [O{\sc ii}]~$\lambda\lambda$3727~$(1.3\pm0.2)$; H$\gamma$~+~[O{\sc iii}]~$\lambda$4363~$(2.7\pm0.4)$; H$\beta$~$(4.0\pm0.2)$; [O{\sc iii}]~$\lambda$4959~$(6.6\pm0.2)$; [O{\sc iii}]~$\lambda$5007~$(19.5\pm0.3)$\\
20961 & 7.051 & H$\beta$~$(2.5\pm0.2)$; [O{\sc iii}]~$\lambda$4959~$(3.1\pm0.2)$; [O{\sc iii}]~$\lambda$5007~$(9.6\pm0.3)$\\
\end{longtable}
\tablefoot{
Details of the emission line fitting can be found in Section~\ref{sec:visual_redshifts}.
Note that some lines (notably, H$\beta$, [O {\sc iii}] $\lambda$4959 and [O {\sc iii}] $\lambda$5007) are reported independently at high-redshift, but are reported as blends at lower redshift due to the reduced spectral resolution of the prism at shorter wavelengths. As a visual guide, horizontal rules demarcate points where the reporting of the H$\beta$ + [O {\sc iii}] $\lambda\lambda$4959, 5007 complex changes. We also caution that some fluxes reported here as being for individual lines may feature non-negligible contributions from fainter lines (e.g.\ H$\alpha$ from [N {\sc ii}] $\lambda$6583, or [Ne {\sc iii}] $\lambda$3869 from He {\sc i} $\lambda$3889). In these cases, the reported flux will represent the total flux of the observed emission feature. 
Targets marked \tablefoottext{$\ddag$} were identified as likely having multiple objects in the shutter (see Section~\ref{sub:individual_targets}).
These measurements are available in a machine-readable format on The Mikulski Archive for Space Telescopes: \url{https://archive.stsci.edu/hlsp/jades}.
}
\twocolumn

\begin{table}
\caption{Targets not meeting the $S/N>5$ emission line threshhold required to appear in Tables~\ref{tab:grating_lines} -- \ref{tab:prism_lines}, but for which a secure redshift could be identified visually based on spectral breaks or low $S/N$ emission features.}            
\label{tab:marginal}      
\centering          
\begin{tabular}{l c }    
\hline\hline       
ID & $z_{\rm Visual}$ \\
\hline
17400 & 13.20 \\
2773 & 12.63 \\
10014220 & 11.58 \\ 
10014177 & 10.38 \\ 
8115 & 7.30 \\
17251 & 5.043 \\
6460 & 3.573 \\
10036262 & 3.566 \\
6855 & 2.868 \\
17832 & 2.832 \\
10006084 & 2.815 \\
10000081 & 2.712 \\
4668 & 2.649 \\
10010639 & 2.621 \\
19175 & 2.556 \\
2528 & 2.345 \\
5320 & 2.305 \\
6710 & 2.249 \\
10037047 & 2.058 \\
10010691 & 1.968 \\
9992 & 1.962 \\
10008722 & 1.742 \\
10040868 & 1.115 \\
\hline
\hline
\end{tabular}
\end{table}

\end{appendix}

\end{document}